%% file: main.tex
\documentclass[aps,twocolumn,showpacs,showkeys,preprintnumbers,superscriptaddress,nobibnotes,floatfix,longbibliography,notitlepage,nofootinbib]{revtex4-2}

\usepackage[T1]{fontenc} 
\usepackage{ae,aecompl}
\usepackage{graphicx}
\usepackage{tabularx}
\usepackage{placeins}
\usepackage{gensymb}
\usepackage{rotating}
\usepackage{upgreek}
\usepackage{natbib}
\usepackage{amsmath}
\usepackage[dvipsnames]{xcolor}
\usepackage[all]{nowidow}
\usepackage[colorlinks=true,allcolors=blue]{hyperref}
\usepackage{array}
\usepackage{lipsum}
\usepackage{cleveref}
\usepackage{ragged2e}
\usepackage[justification=justified]{caption}

\usepackage{graphicx}
\usepackage{bm}
\usepackage{tabularx}
\usepackage{epsf}
\usepackage{rotating}
\usepackage{epsfig,graphics,rotate,color}
\usepackage{wrapfig}
\usepackage{amssymb}
\usepackage{amsmath}
\usepackage{amsfonts}
\usepackage{array,hhline,dcolumn}
\usepackage{gensymb}
\usepackage{placeins}
\usepackage{xr}

\makeatletter

\newcommand*{\addFileDependency}[1]{
\typeout{(#1)}
\@addtofilelist{#1}
\IfFileExists{#1}{}{\typeout{No file #1.}}
}\makeatother

\newcommand*{\myexternaldocument}[1]{%
\externaldocument{#1}%
\addFileDependency{#1.tex}%
\addFileDependency{#1.aux}%
}

\myexternaldocument{appendix}
\myexternaldocument{jdefs}
\def\orcid#1{\href{https://orcid.org/#1}{\includegraphics[keepaspectratio,width=1.1em]{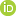}}}

\usepackage{tikz}
\def\checkmark{\tikz\fill[scale=0.4](0,.35) -- (.25,0) -- (1,.7) -- (.25,.15) -- cycle;} 
\usepackage{multirow}

\begin{document}

\include{jdefs}
\title{Dark Matter Decay to Neutrinos}


\author{Carlos A. Arg{\"u}elles~\orcid{0000-0003-4186-4182}}
\affiliation{Department of Physics \& Laboratory for Particle Physics and Cosmology, Harvard University, Cambridge, MA 02138, USA}

\author{Diyaselis Delgado~\orcid{0000-0002-4306-8828}}
\email[Corresponding author: ]{ddelgado@g.harvard.edu}
\affiliation{Department of Physics \& Laboratory for Particle Physics and Cosmology, Harvard University, Cambridge, MA 02138, USA}

\author{Avi Friedlander~\orcid{0000-0002-5152-8898}}
\affiliation{Department of Physics, Engineering Physics and Astronomy, Queen's University, Kingston ON K7L 3N6, Canada}
\affiliation{Arthur B. McDonald Canadian Astroparticle Physics Research Institute, Kingston ON K7L 3N6, Canada}

\author{\\ Ali Kheirandish~\orcid{0000-0001-7074-0539}}
\affiliation{Department of Physics \& Astronomy, University of Nevada, Las Vegas, NV, 89154, USA}
\affiliation{Nevada Center for Astrophysics, University of Nevada, Las Vegas, NV 89154, USA}

\author{Ibrahim Safa~\orcid{0000-0001-8737-6825}}
\affiliation{Department of Physics, Columbia University, New York, NY, 10027, USA}
\affiliation{Department of Physics \& Wisconsin IceCube Particle Astrophysics Center, University of Wisconsin, Madison, WI 53706, USA}
\affiliation{Department of Physics \& Laboratory for Particle Physics and Cosmology, Harvard University, Cambridge, MA 02138, USA}

\author{Aaron C. Vincent~\orcid{0000-0003-3872-0743}}
\affiliation{Department of Physics, Engineering Physics and Astronomy, Queen's University, Kingston ON K7L 3N6, Canada}
\affiliation{Arthur B. McDonald Canadian Astroparticle Physics Research Institute, Kingston ON K7L 3N6, Canada}
\affiliation{Perimeter Institute for Theoretical Physics, Waterloo ON N2L 2Y5, Canada}

\author{Henry White~\orcid{0000-0003-1771-2218}}
\affiliation{Department of Physics, Engineering Physics and Astronomy, Queen's University, Kingston ON K7L 3N6, Canada}
\affiliation{Arthur B. McDonald Canadian Astroparticle Physics Research Institute, Kingston ON K7L 3N6, Canada}

\begin{abstract}
It is possible that the strongest interactions between dark matter and the Standard Model occur via the neutrino sector.
Unlike gamma rays and charged particles, neutrinos provide a unique avenue to probe for astrophysical sources of dark matter, since they arrive unimpeded and undeflected from their sources.
Previously, we reported on annihilations of dark matter to neutrinos; here, we review constraints on the decay of dark matter into neutrinos over a range of dark matter masses from MeV to ZeV, compiling previously reported limits, exploring new electroweak corrections and computing constraints where none have been computed before.
We examine the expected contributions to the neutrino flux at current and upcoming neutrino experiments as well as photons from electroweak emission expected at gamma-ray telescopes, leading to constraints on the dark matter decay lifetime, which ranges from $\tau \sim 1.2\times10^{21}$ s at 10~MeV to $1.5\times10^{29}$~s at 1~PeV.
\end{abstract}

\newcommand{\refeq}[1]{Eq.~(\ref{#1})}
\newcommand{\refeqs}[2]{Eqs.~(\ref{#1})~and~(\ref{#2})}
\newcommand{\refeqss}[3]{Eqs.~(\ref{#1}), (\ref{#2})~and~(\ref{#3})}
\newcommand{\reffig}[1]{Fig.~\ref{#1}}
\newcommand{\reffigs}[2]{Figs.~\ref{#1}~and~\ref{#2}}
\newcommand{\refsec}[1]{Section~\ref{#1}}
\newcommand{\refsecs}[5]{Sections~\ref{#1},~\ref{#2},~\ref{#3},~\ref{#4},~and~\ref{#5}}
\newcommand{\refappsec}[1]{Appendix Section~\ref{#1}}
\newcommand{\refapp}[1]{Appendix~\ref{#1}}
\newcommand{\reftab}[1]{Table~\ref{#1}}
\newcommand{\refref}[1]{~\cite{#1}}
\newcommand{\refrefs}[2]{~\cite{#1}~and~\cite{#2}}

\maketitle

\section{Introduction\label{sec:intro}}
The angular power spectrum of the cosmic microwave background (CMB) as well as the matter distribution on large scales, the clustering of galaxies, and the measured kinematics of stars and gas within those galaxies all point to a large component of weakly-interacting \textit{dark matter} (DM), constituting 85\% of all matter in the Universe~\cite{Lisanti:2016jxe,Aghanim:2018eyx}.
While these observations imply an equation of state consistent with a cold, collisionless fluid, no microphysical connection has yet been found between DM and the Standard Model (SM) of particle physics.
Numerical coincidences such as the one-to-five ratio of dark-to-ordinary matter sustain our hope that DM decoupled late enough in the history of the Universe to require a coupling well below the Planck scale and thus be describable in the language of particle physics.

The parameter space of such nongravitational interactions is immense, and myriad portals are potentially available.
Traditional searches for electroweak and supersymmetry-inspired WIMPs in the GeV-TeV mass range that scatter with or annihilate to quarks have expanded in the past decades to encompass light axionlike~\cite{Adams:2018bvi} and minicharged particles~\cite{Magill:2018tbb,Stebbins:2019xjr,Harnik:2020ugb,ArguellesDelgado:2021lek,Montero:2022jrc}, sub-GeV nonthermal DM candidates~\cite{Allahverdi:2012wb,Arguelles:2022fqq}, primordial black holes~\cite{Green:2020jor}, and other exotic objects.
In some of these scenarios, dark matter can be unstable and decay to Standard Model particles. 

Direct searches for such DM rely on elastic scattering with electrons or nuclei, while indirect searches look for signatures of decay or annihilation into SM particles. 
Products of DM decay (or annihilation) into SM particles eventually create a flux of stable particles, i.e., protons, electrons, photons, or neutrinos.
Here, we focus on the latter.
A direct neutrino portal would render direct detection impracticable, and indirect detection very difficult, owing to the minuscule cross section of neutrinos at low energies.
However, at high energies, the neutrino cross section grows and is no longer suppressed by the mass of the heavy bosons but by the momentum transfer as is the case of photon-nucleon interactions~\cite{Gandhi:1995tf,Arguelles:2015wba,Garcia:2020jwr}.
Additionally, high-energy gamma rays can be attenuated as they travel from their sources of production to Earth, while neutrinos voyage unimpeded.
Therefore, the study of neutrinos represents a final frontier in the search for indirect signatures of DM.
The study of this channel is further motivated by connections between the dark sector and neutrinos.
These have been proposed in a variety of different contexts, including the scotogenic scenarios where neutrinos gain their mass via interacting with the dark sector~\cite{Alvey_2019,Baumholzer_2020,Boehm:2006mi,Escudero:2016tzx,Farzan:2012sa,Farzan:2014gza,Patel:2019zky}, the Majoron scenario~\cite{Garcia-Cely:2017oco} or see-saw models~\cite{Coy:2020wxp}.
In many of these models, UV physics can destabilize the DM, leading to a decay to $\bar \nu \nu$, which may dominate over other channels.
These models have motivated numerous dedicated studies, mainly in the context of discovering heavy DM using neutrino line searches~\cite{Beacom:2006tt,Esmaili:2012us,PhysRevD.92.123515,ElAisati:2017ppn,Bhattacharya:2019ucd,Dutta:2022wuc,PhysRevD.101.115029,PhysRevD.105.L041301,Murase:2015gea,Murase:2012xs,Chianese:2021htv,Chianese:2021jke}, and many neutrino experiments have hunted for DM signatures in their observations~\cite{Super-Kamiokande:2004pou, Super-Kamiokande:2020sgt, IceCube:2014rqf,IceCube:2017rdn, IceCube:2021xzo,IceCube:2022vtr,PhysRevD.99.095014}, so far yielding null results.

Previously, we presented an updated compendium of constraints on particle DM annihilation to neutrinos~\cite{Arguelles:2019ouk}.
Here, we turn our attention to the production of neutrinos from DM decay, directing special attention to the higher-mass region.
At masses greater than $\sim$ TeV, electroweak (EW) corrections can ``decloak'' the DM, producing high-energy photons, even when directly decaying to neutrinos that can be detectable at current and future gamma-ray observatories~\cite{Bauer_2021}.
We thus present limits from current measurements and sensitivities for upcoming experiments covering a DM mass range starting at 20~MeV and spanning well into the ultra-heavy domain up to $10^{11}$~GeV.
Although recent LHAASO results are on par with IceCube's recent analyses, we will find that current and future neutrino telescopes retain superior sensitivity across nearly the entire range of masses that we consider.
This is due to three factors: the loop suppression of the gamma-ray production rate, the growth of the electroweak cross section with energy, and the loss of high-energy gamma rays to interactions in the interstellar and intergalactic medium.

We begin by briefly describing the signature of DM decaying to neutrinos at both neutrino telescopes and gamma-ray observatories.
We present our new results in~\Cref{sec:results} and offer parting words of wisdom in~\Cref{sec:conclusion}.

\section{Dark Matter decays to neutrinos}

The expected flux per neutrino flavor at Earth from decay of DM with mass $m_\chi$ and lifetime $\tau_\chi$ is
\begin{equation}
    \frac{d\Phi_{\nu+ \bar{\nu}}}{dE_\nu} = \frac{1}{4 \pi}
    \frac{1}{ \tau_\chi m_\chi}  \frac{1}{3}  \frac{dN_\nu}{dE_\nu} D(\Omega).
    \label{eq:galaxyDecRate}
\end{equation}
Below the electroweak scale, the neutrino spectrum per decay is ${dN_\nu}/{dE_\nu} = 2\delta(1 - 2E/m_\chi) m_\chi/2E^2$; at higher masses a low-energy tail arises as discussed in Ref.~\cite{Bauer_2021}.
Because relevant backgrounds follow a power-law distribution, only the delta contribution is relevant for neutrino constraints.
The so-called $D$-factor, $D(\Omega)$, is an integral of the DM distribution $\rho(x)$ along the line of sight and solid angle $\Delta \Omega$:
\begin{equation}
     D \equiv  \int d\Omega \int_{\mathrm{l.o.s.}}  \rho_\chi(x) dx. 
 \label{eq:Dfactordef}
\end{equation}
We assume the Galactic DM spatial distribution is modeled by an Navarro-Frenk-White (NFW) profile with a slope parameter $\gamma = 1.2$ and a scale radius $r_s = 20~{\rm kpc}$, and we set the local DM density to $\rho_0 = 0.4~{\rm GeV~cm}^{-3}$.
These parameters are consistent with the results of, e.g., Ref.~\cite{deSalas:2019pee}, which point out a strong dependence on how the baryonic potential is modeled. 
We take the distance to the galactic centre to be $R_0 = 8.127~{\rm kpc}$~\cite{Abuter:2018drb}.
Here, we mainly strive to make results as self-consistent as possible by using common halo parameters in all of our analyses. 
We have assumed equal production of each flavor, which leads to equal flavors at Earth.
Due to neutrino oscillation, this will remain approximately true regardless of the initial flavor composition.

The $D$-factor depends on the field of view of each experiment. Effective areas are reported as a function of elevation (or equivalently, zenith angle). Given each experiment's latitude and altitude, we integrate these acceptances over a period of 24 hours, where the solid angle integral is weighted by the fractional acceptance.
This defines an effective $D$-factor:
\begin{equation}
    D_{\rm eff} = \int dt \int d\Omega \int_{l.o.s.} \rho_\chi (x) F (\Omega,t) \, dx,
    \label{eq:Deff}
\end{equation}
where $F(\Omega , t)$ is the fractional acceptance in equatorial coordinates. This procedure is simplified for experiments at the South Pole (IceCube, ANITA), where elevation and declination are equivalent.

In computing the yield of dark matter in a given experiment, we convolve the experimental efficiency with flux from neutrinos from a given direction. 
For our background agnostic constraints, the flux obtained by this procedure is then compared to the unfolded neutrino fluxes. 
When published experimental results on dark matter annihilation use a directional analysis, such as the case of ANTARES, we rescale the result by the efficiency-weighted ratio of the dark matter $J$ to $D$ factors; namely the ratio of expected signal yield in the annihilation to decay scenarios.
Efficiencies used for each experiment are given in Suppl. Table \ref{tab:Jtable}.

\begin{table}[th]
\begin{center}
\caption{\textbf{\textit{Neutrino (top) and gamma-ray (bottom) observatories considered in this work.}}
Here, ``All Flavors'' denotes both neutrinos and antineutrinos of electron, muon, and tau flavor. 
The experiments given in {\em italic} font are upcoming or proposed detectors.}
\label{tab:tableofexperiments}
{

\begin{tabular}{  l m{7em} c c m{8em}} 
 \hline
 \textbf{Energy (GeV)} & \textbf{Experiment} & \textbf{Dir.} & \textbf{Particles} \\ \hline \hline \vspace*{-2.4mm} \\

 $\left(2.5-15\right) \times 10^{-3} $ & Borexino~\cite{Bellini:2010gn}& $\boldsymbol{\times}$ &  $\bar{\nu}_e$ \\ 
 $\left(8.3-18.3\right)\times 10^{-3}$ & KamLAND~\cite{Collaboration:2011jza} &  \checkmark &  $\bar{\nu}_e$ \\
 $\left(10-40\right)\times 10^{-3}$ & \textit{JUNO}~\cite{An:2015jdp} &  \checkmark &  $\bar{\nu}_e$ \\

 $\left(1.5-100\right)\times 10^{-2}$ & SK~\cite{2018PhRvD..97g5039O}& $\boldsymbol{\times}$ &  $\bar{\nu}_e$ \\
 
 $0.1-30$ & \textit{DUNE}~\cite{Abi:2020evt}& $\boldsymbol{\times}$ &  $\nu_e, \bar{\nu}_e, \nu_{\tau}, \bar{\nu}_{\tau}$  \\ 
 
 $0.1-30$ & \textit{HK} \cite{Abi:2020evt}& $\boldsymbol{\times}$ &  $\nu_e, \bar{\nu}_e, \nu_{\tau}, \bar{\nu}_{\tau}$  \\ 

  $1-10^4$ & SK \cite{Frankiewicz:2015zma, Abe:2020sbr} & \checkmark &  All Flavors \\ 
    $20-10^4$ & IceCube \cite{Aartsen:2016pfc} & \checkmark &  All Flavors\\ 
     $50-10^5$ & ANTARES \cite{Adrian-Martinez:2015wey} & \checkmark &  $\nu_\mu,\,\bar{\nu}_\mu$ \\ 
          $10^3 - 10^7$  & \textit{P-ONE}~\cite{P-ONE:2020ljt} &  \checkmark &  All Flavors \\ 
          $10^4 - 10^7$  & \textit{KM3NeT}~\cite{Adrian-Martinez:2016fdl} &  \checkmark &  All Flavors \\ 
          $10^6 - 10^8$ & \textit{TAMBO}~\cite{Wissel:2019alx} &  \checkmark &  $\nu_\tau,\,\bar{\nu}_\tau $  \\ 
          $> 10^7$ & \textit{IceCube-Gen2}~\cite{Aartsen:2014njl} &  \checkmark &  All Flavors \\ 
          $> 10^8$ & \textit{RNO-G}~\cite{Aguilar:2019jay} &  \checkmark &  All Flavors \\ 
          $> 10^8$ & \textit{GRAND}~\cite{Alvarez-Muniz:2018bhp} &  \checkmark &  $\nu_\tau,\,\bar{\nu}_\tau $  \\  
          $10^8 - 10^{11}$ & Auger~\cite{PierreAuger:2015eyc}& \checkmark & All Flavors \vspace*{0.5mm}
          \\ \hline \hline \vspace*{-2.4mm} \\ 
          
          
           $10^{-1} - 10^2$ & Fermi-LAT~\cite{Atwood_2009} & & $\gamma$\\
           $10^3 - 10^9$ & \textit{CTA}~\cite{Bigongiari:2016amk} & &  $\gamma$ \\
           $10^4 -  10^9$  & HAWC~\cite{HAWC:2013htd} & & $\gamma$ \\
           $10^5 - 10^9$ & LHAASO~\cite{LHAASO:2019qtb} & & $\gamma$ \\
           $10^6 - 10^9$ & IceTop~\cite{Abbasi_2013} & & $\gamma$ \\
           $10^7 - 2\times10^9$ & KASCADE~\cite{KLAGES199792} & & $\gamma$ \\
           $10^8 - 2\times10^{10}$ & CASA-MIA~\cite{GIBBS198867} & & $\gamma$\\
           $10^9 - 2\times10^{12}$ & EAS-MSU~\cite{EAS-MSU} & & $\gamma$\\
           $10^{11.5} - 10^{14}$& TA-SD~\cite{TASD} & & $\gamma$  \\
           $>10^{12}$& Auger-SD~\cite{PierreAuger:2007kus} & & $\gamma$ \vspace*{0.5mm} \vspace*{0.5mm} \\
           \hline 
    \hline
\end{tabular}
}
\end{center}
\end{table}

Based on Eq.~\ref{eq:galaxyDecRate}, we produce limits on the DM decay rate, using results from different analyses of existing data~\cite{Arguelles:2019ouk,Zas:2017xdj,Agostini:2019yuq,Alvarez-Muniz:2018bhp,Aartsen:2016pfc,Aartsen:2017ulx,iovine2021indirect,Aartsen:2015xup,Aartsen:2018vtx,Bhattacharya:2019ucd,KamLAND:2021gvi,Agostini:2018uly,WanLinyan:2018,2018PhRvD..97g5039O,Frankiewicz:2017trk,Richard:2015aua} or forecasted sensitivities~\cite{Bell:2020rkw,Akita:2022lit,gozzini2019search,Aguilar_2021,Agostini:2020aar,Romero-Wolf:2020pzh}.
The full list of neutrino experiments is given in the top section of~\Cref{tab:tableofexperiments}.
We also list the neutrino energy range covered by each experiment, spanning from 10~MeV at Borexino to $> 10^{11}$~GeV at IceCube and AUGER, as well as each experiment's neutrino flavor sensitivity.
For a detailed description of each experiment and its sensitivity, we point the reader to Ref.~\cite{Arguelles:2019ouk}.
The $D$-factors for each experiment are computed by integrating the exposure of each telescope over 24 hours.
The resulting exposures and $D$-factors are tabulated in Table~\ref{tab:Jtable} in the Supplementary Material. %

The decay lifetime constraints result from  a comparison between the flux sensitivities from each experiment and the expected neutrino flux from DM decay. This approach assumes a branching ratio of 100\% of DM decay to neutrinos can describe the total neutrino flux measurements in the Galactic Center region. Our forecasts assume five years of exposure for each of the following experiments: JUNO~\cite{An:2015jdp}, DUNE~\cite{Abi:2020evt}, Hyper-Kamiokande (HK)~\cite{Abi:2020evt}, RNO-G~\cite{Aguilar:2019jay}, IceCube-Gen2~\cite{Aartsen:2014njl}, KM3NeT~\cite{Adrian-Martinez:2016fdl}, P-ONE~\cite{P-ONE:2020ljt}, TAMBO~\cite{Wissel:2019alx}, and GRAND~\cite{Alvarez-Muniz:2018bhp}.
Constraints that are not derived by us, but are reported by experiments or other groups, are rescaled to match the $D$-factors used in this work.
This enables a fair comparison between different experimental constraints.

Data are available at various stages of the analysis pipeline. The closer to event-level, the stronger the constraining power. Depending on how data are reported, we are able to compute lifetime limits with varying precision. 
The methods by which these different data sets are converted into a lower bound on DM lifetime is outlined below. Further, Table \ref{tab:lifetime_method} outlines the type of data used to calculate all lifetime limits within this work. Below we discuss the approach applied to each individual data set according to the exposure time and neutrino flavor detected by the experiment.  

The full list of references for each experiment is provided in Tab. \ref{tab:tableofexperiments}.

\begin{enumerate}
\item \textbf{Lifetime limit}

Constraints labeled IceCube (Bhattacharya) are based on Ref.~\cite{Bhattacharya:2019ucd}. They performed an event-level calculation of limits on dark matter decay and annihilation. They present separate constraints for decays to electron, mu, and tau flavor. We take the least constraining (most conservative) limit for each energy bin, and divide by three to account for our assumption of equal decay to all flavors. 
These limits were not scaled by the ratio of $D$ factors due to matching NFW halo profile assumptions.

Similarly, for the sensitivities of JUNO, based on Ref. \cite{Akita:2022lit}, the $D$-factor definition matches the halo parameters used for this analysis and does not require rescaling.


\item \textbf{Rescaled annihilation cross section limits}

A number of experiments have presented constraints on dark matter annihilation cross section $\langle \sigma v \rangle$, but not decay.  These annihilation limits can be converted into a limit on DM lifetime by rescaling by the appropriate ratio of $D$-factor to $J$-factor, i.e. 
\begin{equation}
\tau_\chi^{limit} = \frac{ 2 m_\chi}{DJ\langle \sigma v \rangle^{limit}}.    
\label{eq:DJ}
\end{equation}
Note that the $J$-factor refers to the annihilation equivalent of Eq. \eqref{eq:Dfactordef} with $\rho_\chi^2$.

This is done for IceCube, IceCube-DeepCore, and SuperKamiokande. This procedure is used for sensitivity forecasts for DUNE, HyperKamiokande and KM3NeT. For our ANTARES constraint, based on \cite{Albert:2016emp}, effective areas were presented in Ref. \cite{2005foap.conf..573R} for three different nadir angle bins, as a function of neutrino energy. This allows us to perform a more accurate rescaling of the effective $\langle \sigma v \rangle$ to $\tau_\chi$ conversion, by combining Eqs. \eqref{eq:DJ} and \eqref{eq:Deff} to take into account the acceptance for each mass for a given time of day.



\item \textbf{Upper limit on neutrino flux}

Borexino, KamLand, SuperKamiokande ($\bar \nu_e$ search) have presented energy-binned neutrino flux limits. We translate these to limits on the DM lifetime  using Eq. \eqref{eq:galaxyDecRate}. Limits on only neutrinos or antineutrinos were scaled by an additional factor of to translate them to a limit on $\Phi_{\nu + \bar \nu}$. 
These limits are not derived using angular information in the data, and are thus less sensitive than a dedicated analysis. 

\item \textbf{Diffuse Neutrino Flux}

Limits on the diffuse neutrino flux are typically presented under the assumption of a power-law spectrum. We label the size of the logarithmic energy bins, $\Delta \equiv \log_{10} E_i - \log_{10} E_{i+1}$ for the $i^\mathrm{th}$ bin, and the power law, $E^{-\alpha}$, such that the limit on the diffuse flux can be written as
\begin{equation}
    \frac{d\phi}{d E}\bigg|_{lim} = f_0 E^{-\alpha},
\end{equation}
where $f_0$ is constant.  This first needs to be integrated to turn this limit into a limit on the total  flux within each energy bin, and then compared with the integral of Eq. \eqref{eq:galaxyDecRate} within the bin, for a given $m_\chi$. 
Schematically,
\begin{equation}
    \phi_{lim}(\bar E) = 4\pi \int_{\bar E 10^{- \Delta/2}}^{\bar E 10^{+ \Delta/2}} f_0 E^{-\alpha} dE,
    \label{eq:philim}
\end{equation}


Equating Eq. \eqref{eq:philim} and the integral of Eq. \eqref{eq:galaxyDecRate} yields
\begin{equation}
    \tau = \frac{ 2D (\alpha-1)}{3m_\chi^2(4\pi)^2} \left( \left(10^{\Delta/2} - 10^{-\Delta/2}\right) \frac{d\phi}{d E}\bigg|_{lim} \right)^{-1}.
    \label{taulimitalpha1}
\end{equation}
for $\alpha = 1$, and
\begin{equation}
    \tau = \frac{2 D}{3m_\chi^2(4\pi)^2} \left(\Delta \ln{(10)\frac{d\phi}{d E}\bigg|_{lim} } \right)^{-1}.
    \label{taulimitalpha2}
\end{equation}
for $\alpha = 2$. Eq. \ref{taulimitalpha2} was applied for the analyses done on Auger data, SuperKamiokande atmospheric neutrino data, and IceCube atmospheric neutrino data. Eq. \ref{taulimitalpha1} was applied for GRAND and IceCube HE analyses.

\item \textbf{Projected Effective Areas}

Finally, some of our projections require a complete derivation of sensitivities. We proceed as in Ref. \cite{Arguelles:2019ouk}.

 For IceCube Gen-2, P-ONE, and TAMBO we assume an atmospheric background \cite{IceCube:2015mgt} $\phi_{atm,0} (E/10^4 \, \mathrm{GeV})^{-3.39}$, with $\phi_{atm,0} = 1.1749\times 10^{-14}$ GeV$^{-1}$ s$^{-1}$ cm$^{-2}$ sr$^{-1}$ and  a power law astrophysical neutrino flux $\phi_{astro,0} (E/10^5 \, \mathrm{GeV})^{-2.28}$, 
with $\phi_{astro,0} =1.44\times 10^{-18}$ GeV$^{-1}$ s$^{-1}$ cm$^{-2}$ sr$^{-1}$ \cite{IceCube-Gen2:2020qha}. We use projected effective areas, presented in elevation bins, from Refs.~\cite{Aartsen:2019swn}, \cite{Agostini:2020aar} and \cite{Romero-Wolf:2020pzh} respectively for Gen-2, P-ONE and TAMBO. These are convolved with the expected flux in Eq. \eqref{eq:Deff}. We derive an upper limit sensitivity based on 5 years of exposure using a binned Poisson likelihood (see Ref. \cite{Arguelles:2019ouk}). 

For the lower energies of DUNE and JUNO, we do not have effective areas in terms of the top-of-the-atmosphere neutrino flux. We use the predictions of \cite{PhysRevD.92.023004} to model the atmospheric background at SURF, and nuSQuIDS to account for oscillation ~\cite{Delgado:2014kpa,Arguelles:2020hss,nusquids}. We focus on $e$- and $\tau$-flavored charged-current interactions, comparing the expected energy distribution. We do not consider event-by-event directional information. For the charged lepton energy resolution, we assume a fractional resolution of $2\% + 15\%/\sqrt{E/{\rm GeV}}$~\cite{Acciarri:2015uup} and assume $100\%$ efficiency.

We assume that charged-current electron-neutrino interactions deposit all their energy in the detector. However, for tau-neutrino charged-current interactions, the visible energy is lower due to invisible neutrinos produced in the prompt $\tau$ decay. To reduce the atmospheric neutrino background, which is mainly contributed by muon-neutrino charged-current processes, we exclude them from the analysis, considering that DUNE morphological identification can effectively identify them.
\end{enumerate}

\subsection{Gamma rays from electroweak corrections}

Above the TeV scale, electroweak corrections can lead to the production of photons. 
These result in two distinct gamma-ray fluxes.
First, ``prompt'' high-energy flux consisting of primary photons emitted during the dark matter decay to neutrinos.
Second, lower-energy (GeV--TeV) photon signal due to scattering of primary photons with the CMB and extragalactic background light (EBL).
The prompt gamma-ray spectra can be obtained via the \texttt{HDMSpectra} package~\cite{Bauer_2021}, which solves the Dokshitzer-Gribo-Lipatov-Altarelli-Parisi (DGLAP) equations above the electroweak scale with initial conditions given by the DM decay channel.
At the edge of this scale, \texttt{HDMSpectra} matches its solution with \texttt{Pythia-8.2}~\cite{Sjostrand:2014zea}, which calculates the effects of particle showers, hadronization, and light particle decays.

We use the gamma-ray distribution from this package to derive constraints using gamma-ray data sets in a similar way to how we proceed with neutrinos.
One important difference is the inclusion of an additional factor of $\exp(-\tau_{\gamma\gamma})$ in Eq.~\ref{eq:galaxyDecRate}.
This factor accounts for attenuation due to pair-production from scattering of high-energy gamma-rays with the background light \cite{Stecker:2006eh}, for which CMB photons provide the dominant attenuation channel.
For the galactic component, we conservatively include this as a constant factor by taking the average attenuation rate over a distance of 10 kpc.
We will present limits from {\em Fermi}-LAT~\cite{Atwood_2009}, HAWC~\cite{HAWC:2013htd}, LHAASO~\cite{LHAASO:2019qtb}, IceTop~\cite{Abbasi_2013}, KASCADE-Grande~\cite{KLAGES199792}, CASA-MIA~\cite{GIBBS198867}, EAS-MSU~\cite{EAS-MSU}, TA-SD~\cite{TASD}, and Auger-SD~\cite{PierreAuger:2007kus} observations as well as a projected sensitivity for CTA~\cite{Bigongiari:2016amk}.
For CTA, we consider the differential sensitivity from~\cite{cta_sens} and convert it to an upper limit on the total flux per decade of energy, which is defined as the minimum flux required to obtain a $5\sigma$ point source detection from CTA Southern array for a total observation time of 50 hours.

In the case of IceTop, we use the differential upper limits at 90\% C.L. reported in~\cite{IceCube:2019scr}.
These are then converted to total integrated emission per energy decade.
For KASCADE, KASCADE-Grande, CASA-MIA, EAS-MSU, TA-SD, and Auger-SD, we use the integral gamma-ray flux upper limits reported in Refs.~\cite{Apel_2017,PhysRevLett.79.1805,PhysRevD.95.123011,TelescopeArray:2018rbt,PierreAuger:2022aty}.
For HAWC, we follow the same procedure for the flux upper limit in each declination band~\citep{HAWC:2017udy} and then further select the most stringent constraint among all bands.
All integrated gamma-ray fluxes are then compared to the expected total flux from DM decay to neutrinos with the photon spectrum from electroweak corrections.
This comparison then yields our constraints on the DM decay lifetime. 
Limits for decaying DM to all neutrino flavors from LHAASO were taken from~\cite{Cao:2022myt}. 

\begin{figure}
    \includegraphics[width =\columnwidth]{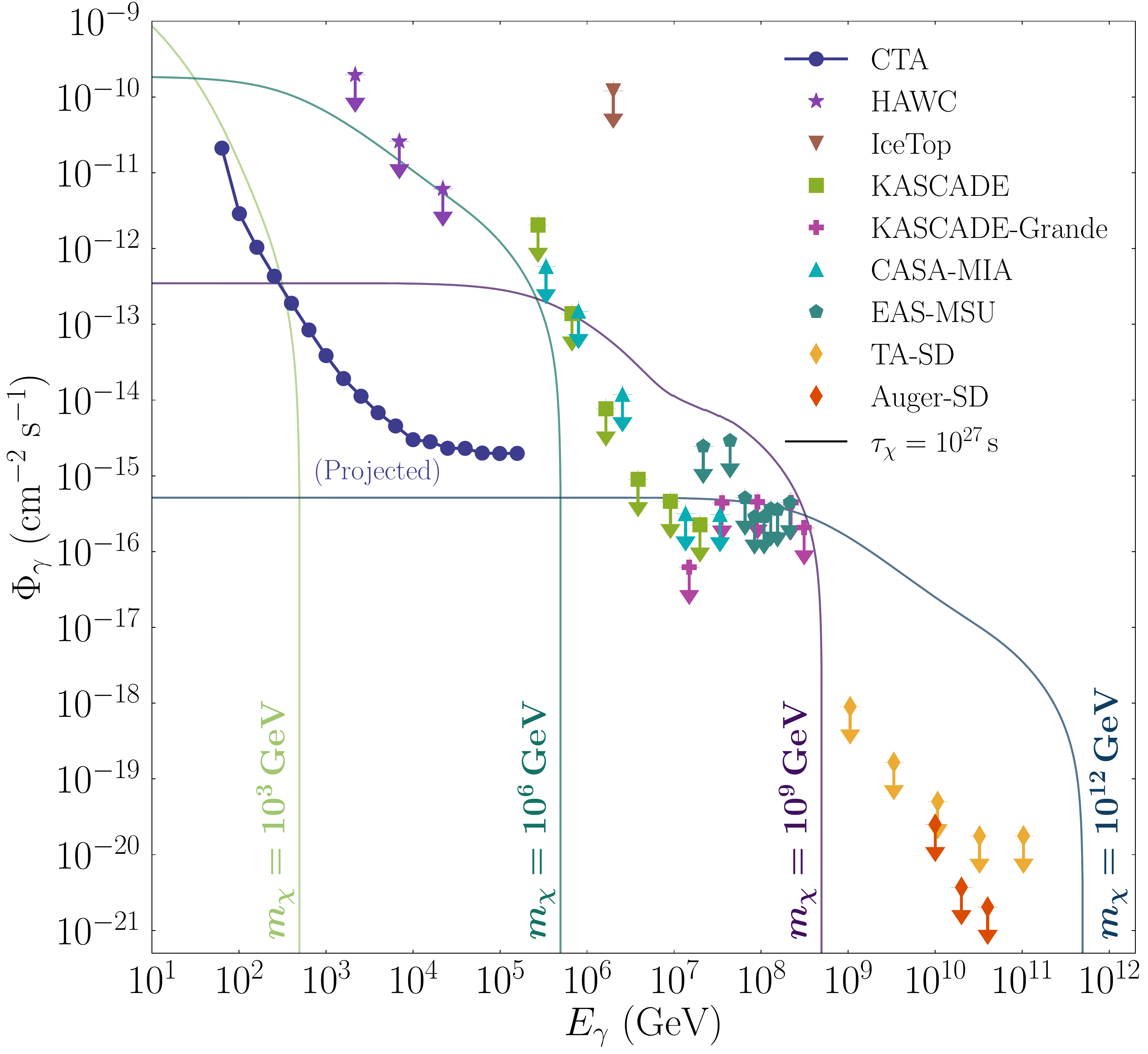}
    \caption{\textbf{\textit{Expected integral gamma-ray fluxes produced by electroweak corrections to dark matter decay to neutrinos overlaid on the observed gamma-ray distributions.}}
    Integral fluxes, defined as the integral of the flux from $E_\gamma$ to infinity, for four different dark matter masses and lifetime of $\tau_\chi = 10^{27}{\rm s}$, are shown as solid lines.
    Colored symbols indicate observations detailed in the bottom half of~\Cref{tab:tableofexperiments}.}
    \label{fig:GammaFluxwithData}
\end{figure}

At sufficiently large masses, gamma rays produced from decays outside our galaxy can scatter down to produce a signal that is observable at lower energies in experiments such as \textit{Fermi}-LAT. 
High-energy gamma rays traversing the intergalactic medium (IGM) are absorbed and scattered by photons from the CMB and EBL, attenuating the signal~\cite{Franceschini:2017iwq}; see Ref.~\cite{Skrzypek:2022hpy} for a recent detailed discussion.
Scattering and absorption of gamma rays result in cascades that transform any sufficiently high-energy gamma-ray source into a universal spectrum~\cite{Murase:2012df} that peaks within the \textit{Fermi} telescope's sensitivity range.
In what follows we take advantage of this universality to extend gamma-ray limits to higher dark matter masses, and convert them to limits on decay to neutrinos.

The experimental upper-limits are given in terms of the gamma-ray integral flux, which is defined as:
\begin{equation}
    \Phi_\gamma(E_\gamma) = \int_{E_\gamma}^{\infty} d\tilde E_\gamma \frac{d\Phi_\gamma}{d\tilde E_\gamma} = \int_{E_\gamma}^{\infty} d\tilde E_\gamma \; \Phi_0 \Big(\frac{\tilde E_\gamma}{E_0}\Big)^{-\Gamma},
    \label{eq:integralFlux}
\end{equation}
where the power-law index, $\Gamma$ is conventionally chosen to be 2. 
Given a null experimental observation, the gamma-ray integral flux upper limit is obtained by find the value of $\Phi_0$, such that the number of expected events, obtained from the convolution of the detector effective area with the power-law flux. 
Namely, it saturates the following equation:
\begin{equation}
    N^{90\%}_{\rm upper \, limit} = \int_{E_\gamma}^{\infty} d\tilde E_\gamma \; \Phi_0^{90\%~u.l.} \Big(\frac{\tilde E_\gamma}{E_0}\Big)^{-\Gamma} A_{\rm eff}(E_\gamma) \, T.
    \label{eq:integralFluxLimit}
\end{equation}

Ref.~\cite{Cohen:2016uyg} sets constraints on the lifetime of DM decay to SM particles using \textit{Fermi} observations of the isotropic gamma-ray background.
We use the limits presented there for DM decays to neutrino pairs that extend up to $m_\chi = 10^{7}$~GeV.
The limits presented in Ref.~\cite{Cohen:2016uyg} constrain the channel $\chi \rightarrow \bar b b$ up to $10^{10}$ GeV.
We use these to obtain corresponding limits in the channel of interest, DM decay to neutrino pairs.
The idea of the \textit{universal spectrum} means that regardless of the initial gamma-ray spectral shape, the spectrum arriving at Earth is universal. 
Therefore limits on dark matter decay to neutrinos are related to the $\chi \rightarrow \bar b b$ limits by a factor of $F^\gamma_{\chi\rightarrow\bar \nu \nu}/F^\gamma_{\chi \rightarrow \bar b b} = 0.06$, where $F^\gamma_{\chi \rightarrow \bar X X}$ is the fraction of energy per decay to species $X$ going into photons.
Suppl. Fig.~\ref{fig:FermiLimits} shows that this rescaling yields the published $\bar \nu \nu$ results below $10^{7}$ to a reasonable accuracy, allowing us to confidently extend these limits up to $10^{10}$ GeV.
We find that the \textit{Fermi}-LAT constraints remain subdominant over the full range of masses considered.

\begin{figure*}[ht!]
    \centering
    \includegraphics[width=\textwidth]{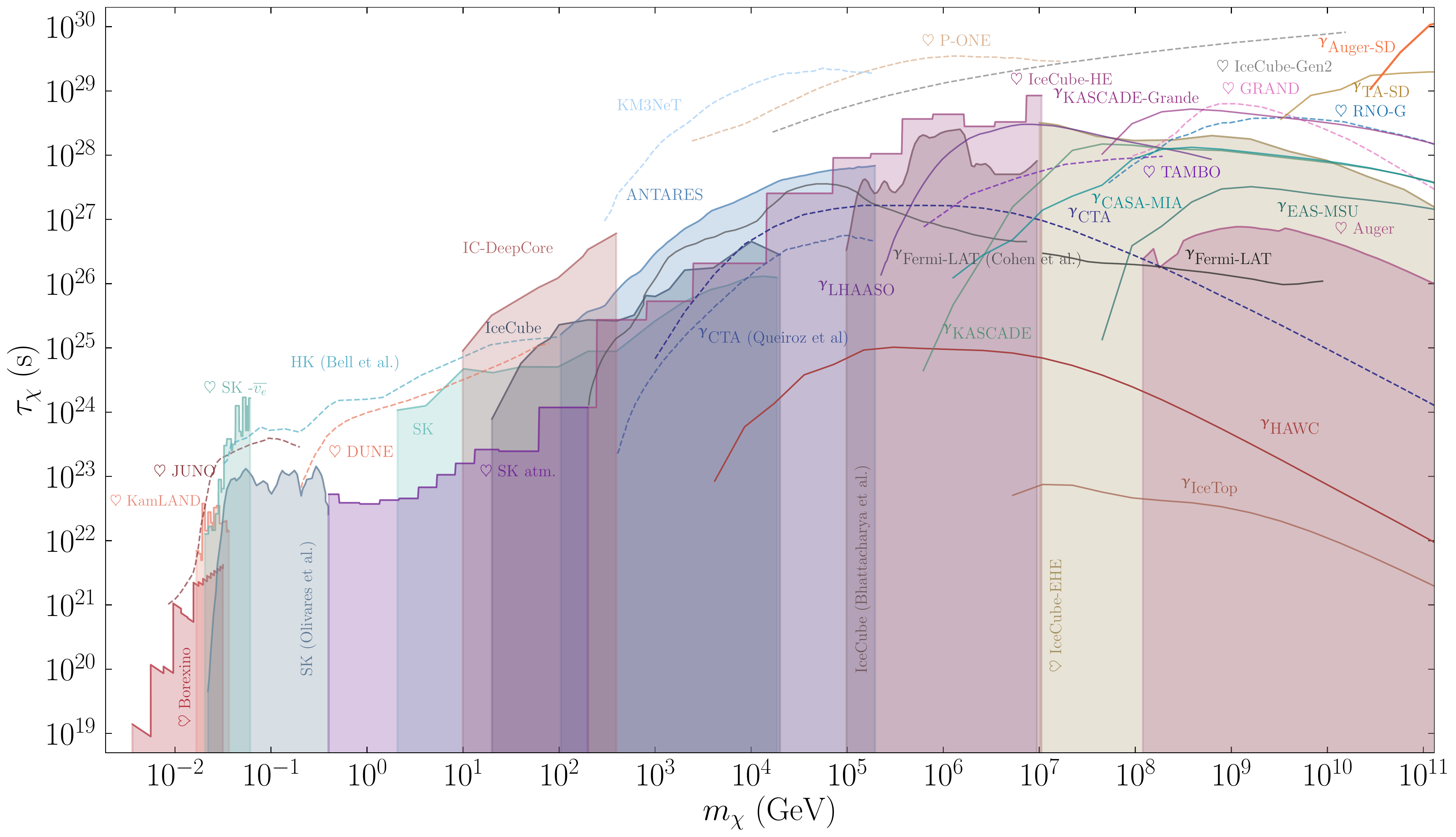}
    \caption{\textbf{\textit{Constraints on the lifetime of dark matter decaying to neutrinos $\boldsymbol{\chi \rightarrow \bar \nu \nu}$.}}
    Solid lines bordering shaded regions represent limits from existing neutrino telescope data, solid lines without shading correspond to limits from existing gamma-ray observatories (as shown in Fig.~\ref{fig:GammaDecayLimits}), and dashed lines show the reach of future experiments. Labels with a heart symbol ($\heartsuit$) correspond to limits derived for this work.}
    \label{fig:DecayLimits}
    \end{figure*}

\section{Results\label{sec:results}}

Using the methods outlined above, we present constraints on the DM decay lifetime in Fig.~\ref{fig:DecayLimits}.
We label the results derived for this work with a heart ($\heartsuit$).

Constraints from neutrino telescopes are shown as shaded regions bordered by solid lines.
The overlap in experimental sensitivities yields continuous constraints on the DM lifetime that are much greater than the age of the Universe, ranging from $\tau > 10^{19}$~s at $m_\chi\sim 50$~MeV to $\tau > 10^{27}$~s  for $m_\chi \sim 10^{11}$~GeV.
The expected neutrino flux at Earth from DM decay is independent of mass.
Below $\sim 10^7$~GeV, this is reflected by sensitivity closely following the growth of the electroweak cross section with energy, with some scaling between experiments owing to differences in effective volumes.
Above $\sim 10^7$~GeV energies, the Earth becomes opaque to neutrinos, and detection technologies become sensitive to a much smaller solid angle, usually restricted to an area just around the horizon.

Estimated sensitivities of future observatories are shown as dashed lines; these assume five years of data taking.
JUNO, Hyper-K, DUNE, KM3NeT, P-ONE, and IceCube-Gen2 should each lead to an improvement of one to two orders of magnitude over current bounds, mainly owing to much larger effective detector volumes.
Projected improvements from future radio (GRAND, RNO-G) and modular Cherenkov arrays (TAMBO) are more modest, which we mainly attribute to restricted fields of view.
 
\begin{figure}
    \includegraphics[width=\columnwidth]{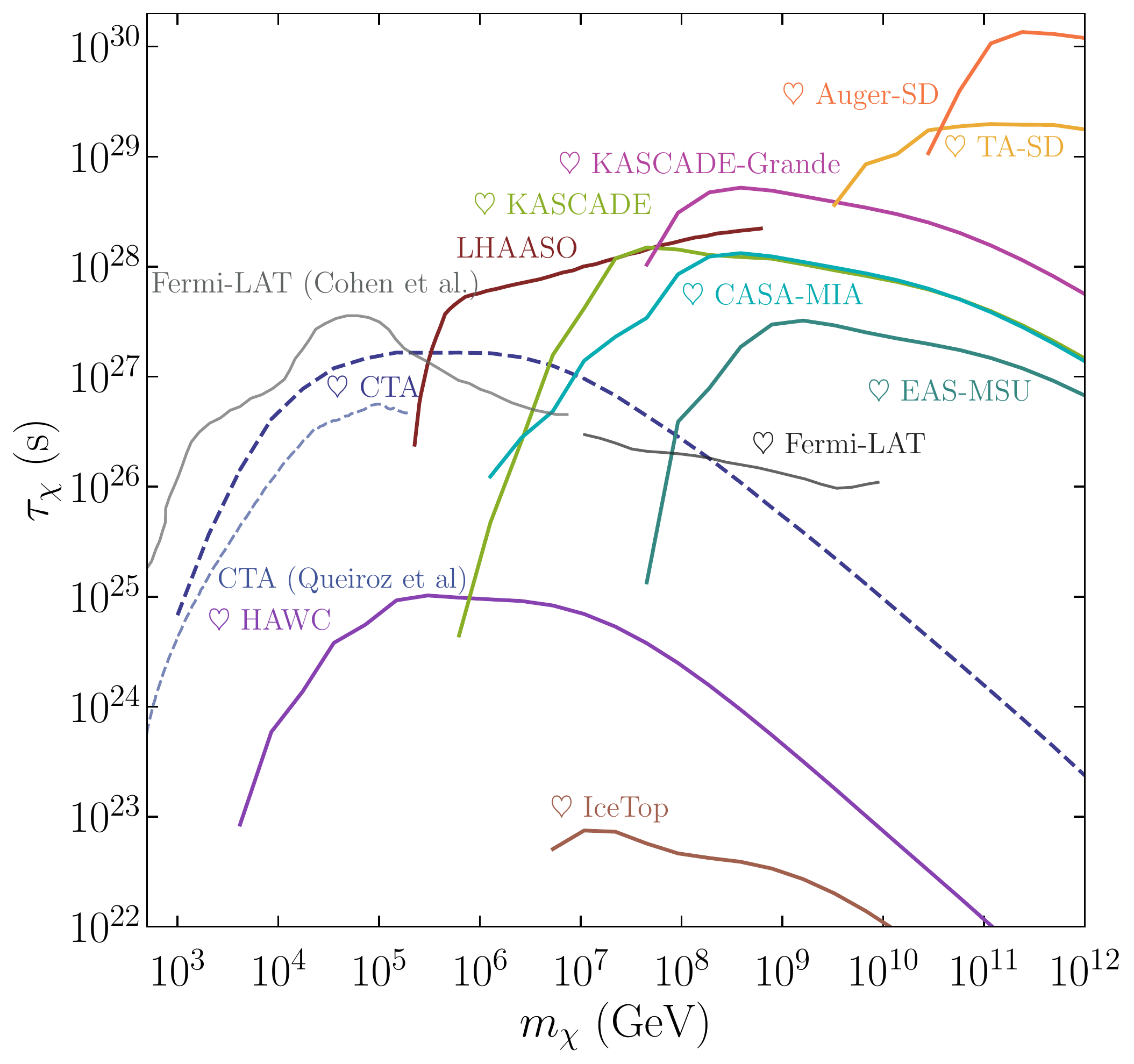}
    \caption{\textbf{\textit{Gamma-ray constraints on dark matter decay lifetime $\boldsymbol{\chi \rightarrow \bar \nu \nu}$ due to $\boldsymbol{\gamma}$ emission from electroweak processes.}}
    Solid lines correspond to existing constraints, while the dashed line is a projection for a future experiment.
    Hearts indicate the new constraints derived in this work.
    Gamma-ray emission below the electroweak scale is suppressed by powers of $M_W$~\cite{Bauer_2021}.
    }
    \label{fig:GammaDecayLimits}
\end{figure}

Limits from gamma-ray observatories are marked with a $\gamma$ superscript.
These are also shown separately in Fig.~\ref{fig:GammaDecayLimits}.
Four experiments dominate the constraints at three different energy ranges. 
At masses below $\sim 10^5$~GeV, the flux of extragalactic gamma-rays produced by interactions with the IGM is probed by \textit{Fermi}-LAT, yielding the dominant source of gamma-ray constraints in this mass range.
At masses between $10^6$ and $10^7$~GeV, recent measurements by LHAASO supersede prior experiments and improve constraints by nearly four orders of magnitude compared to HAWC. 
At masses above $10^{7.5}$~GeV, KASCADE-Grande measurements establish the most competitive constraints on the DM decay lifetime limits, outperforming existing neutrino telescopes; at $m_\chi \gtrsim 10^{10}$~GeV, Auger-SD supersedes all other experiments thanks to its monumental effective area.
Other experiments considered, such as HAWC or IceTop, remain subdominant over the entire mass range probed here.

\section{Future Prospects \& Conclusions\label{sec:conclusion}}

As shown in Fig.~\ref{fig:DecayLimits}, existing neutrino telescopes are able to constrain the lifetime of DM decay to neutrino pairs to values ranging from $10^3$ to $10^{12}$ times the age of the Universe.
Upcoming neutrino telescopes will make improvements of one to two orders of magnitude: DUNE and Hyper-Kamiokande will fill in the gap around $m_\chi \sim$~GeV, while the strongest improvements will take place for the next generation of large-volume water and ice Cherenkov telescopes: KM3NeT, P-ONE, and IceCube-Gen2.
Though not included here, the ongoing scintillator phase of SNO+~\cite{Andringa:2015tza} may also shore up constraints on the low-mass end, depending on the timeline for tellurium filling; the inclusion of directional information in, e.g., Borexino~\cite{BOREXINO:2021efb} or KamLAND analyses could also yield a modest improvement in reach~\cite{Klein:2022tqr,McCormick:2022xjx}.

Above the $\sim$ TeV range, the electroweak emission of gamma rays opens a new opportunity for discovery, and above $10^8$ GeV, gamma rays become the dominant source of information, thanks to the large telescope areas and unsuppressed electroweak emission of photons. What's more the observation of an electromagnetic counterpart will be key in the event of a discovery.
Intriguingly, the Square Kilometre Array (SKA) will be sensitive to $\chi\rightarrow \bar \nu \nu$~\cite{Dutta:2022wuc} in nearby dwarf galaxies for DM masses above a few hundred GeV. Taken together, these observations highlight the importance of multimessenger observations when it comes to elucidating the nature of dark matter.

\acknowledgements{}

We thank Matheus Hostert, Alejandro Diaz, Nicholas Rodd, Marco Chianese, and Damiano Fiorillo for useful discussions.
We thank Barbara (Basia) Skrzypek for providing the code for gamma-ray attenuation.
We thank Thomas Hambye for useful discussion on the model building of dark matter and its connections to neutrinos.
We thank Jean DeMerit for carefully proofreading this manuscript.
CAA, IS, and DD are supported by the Faculty of Arts and Sciences of Harvard University and the Alfred P. Sloan Foundation.
AK work is partially supported by the Nevada Center for Astrophysics.
IS is supported by NSF under grants PLR-1600823 and PHY-1607644 and by the University of Wisconsin Research Council with funds granted by the Wisconsin Alumni Research Foundation.
AF is supported by an Ontario Graduate Scholarship.
ACV is supported by the Arthur B. McDonald Canadian Astroparticle Physics Research Institute and NSERC, with equipment funded by the Canada Foundation for Innovation and the Ontario Government.
Research at Perimeter Institute is supported by the Government of Canada through the Department of Innovation, Science, and Economic Development, and by the Province of Ontario.

\bibliographystyle{apsrev4-1}
\bibliography{nudm}

\include{nudm.bib}

\include{appendix}

\end{document}

%% file: jdefs.tex
\def\pdu{Phys.\ Dark Univ.}
\def\lnp{Lec.\ Notes in Physics}
\def\cpc{Comp.\ Phys.\ Comm.}
\def\jpg{J. Phys. G}
\def\ijmpa{Int.\ J.\ Mod.\ Phys.\ A}
\def\epjc{Eur.\ Phys.\ J.\ C}
\def\nima{Nuc.\ Inst.\ Methods A}
\def\nimb{Nuc.\ Inst.\ Methods B}
\def\njp{New J.\ Phys.}
\def\rmp{Rev.\ Mod.\ Phys.}
\def\app{Astropart.\ Phys.}
\def\aj{AJ}%
\def\actaa{Acta Astron.}%
\def\araa{ARA\&A}%
\def\arnps{Ann.~Rev.~Nucl.~\& Part.~Sci.}%
\def\apj{ApJ}%
\def\apjl{ApJ}%
\def\apjs{ApJS}%
\def\ao{Appl.\ Opt.}%
\def\apss{Ap\&SS}%
\def\aap{A\&A}%
\def\aapr{A\&A~Rev.}%
\def\aaps{A\&AS}%
\def\azh{AZh}%
\def\pos{PoS}%
\def\baas{BAAS}%
\def\bac{Bull.\ Astr.\ Inst.\ Czechosl.}%
\def\caa{Chinese Astron.\ Astrophys.}%
\def\cjaa{Chinese J.\ Astron.\ Astrophys.}%
\def\icarus{Icarus}%
\def\jhep{JHEP}%
\def\jcap{JCAP}%
\def\jinst{JINST}%
\def\jpsj{J.\ Phys.\ Soc.\ Japan}%
\def\jrasc{JRASC}%
\def\canjphys{Can.~J.~Phys.}
\def\apphys{Astropart.~Phys.}
\def\mnras{MNRAS}%
\def\memras{MmRAS}%
\def\na{New A}%
\def\nar{New A Rev.}%
\def\pasa{PASA}%
\def\pra{Phys.\ Rev.\ A}%
\def\prb{Phys.\ Rev.\ B}%
\def\prc{Phys.\ Rev.\ C}%
\def\prd{Phys.\ Rev.\ D}%
\def\pre{Phys.\ Rev.\ E}%
\def\prx{Phys.\ Rev.\ X}%
\def\prl{Phys.\ Rev.\ Lett.}%
\def\pasp{PASP}%
\def\pasj{PASJ}%
\def\qjras{QJRAS}%
\def\rmxaa{Rev. Mexicana Astron. Astrofis.}%
\def\skytel{S\&T}%
\def\solphys{Sol.\ Phys.}%
\def\sovast{Soviet~Ast.}%
\def\ssr{Space~Sci.\ Rev.}%
\def\zap{ZAp}%
\def\nat{Nature}%
\def\science{Science}%
\def\sci{\science}%
\def\iaucirc{IAU~Circ.}%
\def\aplett{Astrophys.\ Lett.}%
\def\apspr{Astrophys.\ Space~Phys.\ Res.}%
\def\bain{Bull.\ Astron.\ Inst.\ Netherlands}%
\def\fcp{Fund.\ Cosmic~Phys.}%
\def\gca{Geochim.\ Cosmochim.\ Acta}%
\def\grl{Geophys.\ Res.\ Lett.}%
\def\jcp{J.\ Chem.\ Phys.}%
\def\jgr{J.\ Geophys.\ Res.}%
\def\jqsrt{J.\ Quant.\ Spec.\ Radiat.\ Transf.}%
\def\memsai{Mem.\ Soc.\ Astron.\ Italiana}%
\def\nphysa{Nucl.\ Phys.\ A}%
\def\nphysb{Nucl.\ Phys.\ B}%
\def\physrep{Phys.\ Rep.}%
\def\physscr{Phys.\ Scr}%
\def\planss{Planet.\ Space~Sci.}%
\def\procspie{Proc.\ SPIE}%
\def\repprogphys{Rep.\ Prog.\ Phys.}%
\def\jpcrd{J. Phys. Chem. Ref. Data}%
\def\jphysb{J. Phys. B}%
\def\jphysd{J. Phys. D}%
\def\jphysconfseries{J. Phys. Conf. Series}%
\def\physrev{\pr}
\def\pr{Phys. Rev.}%
\def\josa{J. Opt. Soc. Amer. (1917-1983)}%
\def\josab{J. Opt. Soc. Amer. B}%
\def\pla{Phys. Lett. A}%
\def\plb{Phys. Lett. B}%
\def\os{Opt. Spectrosc. (Russ.)}%
\def\jas{J. Appl. Spectrosc.}%
\def\annp{Ann. Phys.}%
\def\sa{Spectrochim. Acta}%
\def\prsoca{Proc. R. Soc. London Ser. A}%
\def\zphysa{Z. Phys. A}%
\def\zphysb{Z. Phys. B}%
\def\zphysc{Z. Phys. C}%
\def\zphysd{Z. Phys. D}%
\def\zphyse{Z. Phys. E}%
\def\zphys{Z. Phys.}%
\def\adndt{Atom. Data Nuc. Data Tables}%
\def\jmolspec{J. Mol. Spectrosc.}%
\def\aphysb{Appl. Phys. B}%
\def\nim{Nuc. Inst. Meth.}%
\def\jphysique{J. Phys. (Paris)}%
\def\epjp{Eur.~Phys.~J.~Plus}%
\def\epjc{Eur.~Phys.~J.~C}%
\def\epl{Europhys.~Lett}%
\def\njp{New J.~Phys.}
\def\statsci{StatSci}
\def\an{AstNach}
\let\astap=\aap
\let\apjlett=\apjl
\let\apjsupp=\apjs
\let\applopt=\ao

%% file: nudm.bib

@article{Stecker:2006eh,
    author = "Stecker, Floyd W. and Malkan, M. A. and Scully, S. T.",
    title = "{Corrected Table for the Parametric Coefficients for the Optical Depth of the Universe to Gamma-rays at Various Redshifts}",
    eprint = "astro-ph/0612048",
    archivePrefix = "arXiv",
    doi = "10.1086/511738",
    journal = "Astrophys. J.",
    volume = "658",
    pages = "1392",
    year = "2007"
}
@article{IceCube:2020wum,
    author = "Abbasi, R. and others",
    collaboration = "IceCube",
    title = "{The IceCube high-energy starting event sample: Description and flux characterization with 7.5 years of data}",
    eprint = "2011.03545",
    archivePrefix = "arXiv",
    primaryClass = "astro-ph.HE",
    doi = "10.1103/PhysRevD.104.022002",
    journal = "Phys. Rev. D",
    volume = "104",
    pages = "022002",
    year = "2021"
}

@article{McCormick:2022xjx,
    author = "McCormick, Katie",
    title = "{Solar-Neutrino Detection Gets a Boost}",
    doi = "10.1103/Physics.15.s25",
    journal = "APS Physics",
    volume = "15",
    pages = "s25",
    year = "2022"
}

@INPROCEEDINGS{2005foap.conf..573R,
       author = {{Racca}, C. and {ANTARES Collaboration}},
        title = "{The ANTARES neutrino telescope: Status report}",
    booktitle = {Frontier Objects in Astrophysics and Particle Physics},
         year = 2005,
        month = jan,
        pages = {573},
       adsurl = {https://ui.adsabs.harvard.edu/abs/2005foap.conf..573R},
      adsnote = {Provided by the SAO/NASA Astrophysics Data System}
}

@article{IceCube:2015mgt,
    author = "Aartsen, M. G. and others",
    collaboration = "IceCube",
    title = "{Measurement of the Atmospheric $\nu_e$ Spectrum with IceCube}",
    eprint = "1504.03753",
    archivePrefix = "arXiv",
    primaryClass = "astro-ph.HE",
    doi = "10.1103/PhysRevD.91.122004",
    journal = "Phys. Rev. D",
    volume = "91",
    pages = "122004",
    year = "2015"
}

@article{Klein:2022tqr,
    author = "Klein, Joshua R. and others",
    title = "{Future Advances in Photon-Based Neutrino Detectors: A SNOWMASS White Paper}",
    eprint = "2203.07479",
    archivePrefix = "arXiv",
    primaryClass = "physics.ins-det",
    reportNumber = "FERMILAB-PUB-22-193-ND-V",
    month = "3",
    year = "2022"
}

@article{BOREXINO:2021efb,
    author = "Agostini, M. and others",
    collaboration = "BOREXINO",
    title = "{First Directional Measurement of Sub-MeV Solar Neutrinos with Borexino}",
    eprint = "2112.11816",
    archivePrefix = "arXiv",
    primaryClass = "hep-ex",
    doi = "10.1103/PhysRevLett.128.091803",
    journal = "Phys. Rev. Lett.",
    volume = "128",
    number = "9",
    pages = "091803",
    year = "2022"
}

@article{Skrzypek:2022hpy,
    author = {Skrzypek, Barbara and Chianese, Marco and Delgado Arg\"uelles, Carlos},
    title = "{Multi-messenger High-Energy Signatures of Decaying Dark Matter and the Effect of Background Light}",
    eprint = "2205.03416",
    archivePrefix = "arXiv",
    primaryClass = "astro-ph.HE",
    month = "5",
    year = "2022"
}

@article{Arguelles:2015wba,
    author = {Arg\"uelles, Carlos A. and Halzen, Francis and Wille, Logan and Kroll, Mike and Reno, Mary Hall},
    title = "{High-energy behavior of photon, neutrino, and proton cross sections}",
    eprint = "1504.06639",
    archivePrefix = "arXiv",
    primaryClass = "hep-ph",
    doi = "10.1103/PhysRevD.92.074040",
    journal = "Phys. Rev. D",
    volume = "92",
    number = "7",
    pages = "074040",
    year = "2015"
}

@article{Garcia:2020jwr,
    author = "Garcia, Alfonso and Gauld, Rhorry and Heijboer, Aart and Rojo, Juan",
    title = "{Complete predictions for high-energy neutrino propagation in matter}",
    eprint = "2004.04756",
    archivePrefix = "arXiv",
    primaryClass = "hep-ph",
    doi = "10.1088/1475-7516/2020/09/025",
    journal = "JCAP",
    volume = "09",
    pages = "025",
    year = "2020"
}

@article{Gandhi:1995tf,
    author = "Gandhi, Raj and Quigg, Chris and Reno, Mary Hall and Sarcevic, Ina",
    title = "{Ultrahigh-energy neutrino interactions}",
    eprint = "hep-ph/9512364",
    archivePrefix = "arXiv",
    reportNumber = "FERMILAB-PUB-95-221-T, CLNS-95-1357, MRI-PHY-16-95, UIOWA-95-06, AZPH-TH-95-15",
    doi = "10.1016/0927-6505(96)00008-4",
    journal = "Astropart. Phys.",
    volume = "5",
    pages = "81--110",
    year = "1996"
}

@article{Allahverdi:2012wb,
    author = "Allahverdi, Rouzbeh and Dutta, Bhaskar and Sinha, Kuver",
    title = "{Non-thermal Higgsino Dark Matter: Cosmological Motivations and Implications for a 125 GeV Higgs}",
    eprint = "1208.0115",
    archivePrefix = "arXiv",
    primaryClass = "hep-ph",
    doi = "10.1103/PhysRevD.86.095016",
    journal = "Phys. Rev. D",
    volume = "86",
    pages = "095016",
    year = "2012"
}

@article{Arguelles:2022fqq,
    author = {Arg\"uelles, Carlos A. and Mu\~noz, V\'\i{}ctor and Shoemaker, Ian M. and Takhistov, Volodymyr},
    title = "{Hadrophilic light dark matter from the atmosphere}",
    eprint = "2203.12630",
    archivePrefix = "arXiv",
    primaryClass = "hep-ph",
    reportNumber = "IPMU22-0010",
    doi = "10.1016/j.physletb.2022.137363",
    journal = "Phys. Lett. B",
    volume = "833",
    pages = "137363",
    year = "2022"
}

@article{Magill:2018tbb,
    author = "Magill, Gabriel and Plestid, Ryan and Pospelov, Maxim and Tsai, Yu-Dai",
    title = "{Millicharged particles in neutrino experiments}",
    eprint = "1806.03310",
    archivePrefix = "arXiv",
    primaryClass = "hep-ph",
    reportNumber = "FERMILAB-PUB-18-631-A",
    doi = "10.1103/PhysRevLett.122.071801",
    journal = "Phys. Rev. Lett.",
    volume = "122",
    number = "7",
    pages = "071801",
    year = "2019"
}

@article{ArguellesDelgado:2021lek,
    author = {Arg\"uelles Delgado, Carlos Alberto and Kelly, Kevin James and Mu\~noz Albornoz, V\'\i{}ctor},
    title = "{Millicharged particles from the heavens: single- and multiple-scattering signatures}",
    eprint = "2104.13924",
    archivePrefix = "arXiv",
    primaryClass = "hep-ph",
    reportNumber = "FERMILAB-PUB-21-214-T",
    doi = "10.1007/JHEP11(2021)099",
    journal = "JHEP",
    volume = "11",
    pages = "099",
    year = "2021"
}

@article{Harnik:2020ugb,
    author = "Harnik, Roni and Plestid, Ryan and Pospelov, Maxim and Ramani, Harikrishnan",
    title = "{Millicharged cosmic rays and low recoil detectors}",
    eprint = "2010.11190",
    archivePrefix = "arXiv",
    primaryClass = "hep-ph",
    reportNumber = "FERMILAB-PUB-20-523-T",
    doi = "10.1103/PhysRevD.103.075029",
    journal = "Phys. Rev. D",
    volume = "103",
    number = "7",
    pages = "075029",
    year = "2021"
}

@article{Montero:2022jrc,
    author = "Montero, Miguel and Mu\~noz, Julian B. and Obied, Georges",
    title = "{Swampland Bounds on Dark Sectors}",
    eprint = "2207.09448",
    archivePrefix = "arXiv",
    primaryClass = "hep-ph",
    month = "7",
    year = "2022"
}

@article{Stebbins:2019xjr,
    author = "Stebbins, Albert and Krnjaic, Gordan",
    title = "{New Limits on Charged Dark Matter from Large-Scale Coherent Magnetic Fields}",
    eprint = "1908.05275",
    archivePrefix = "arXiv",
    primaryClass = "astro-ph.CO",
    reportNumber = "FERMILAB-PUB-19-271-A",
    doi = "10.1088/1475-7516/2019/12/003",
    journal = "JCAP",
    volume = "12",
    pages = "003",
    year = "2019"
}

@article{Super-Kamiokande:2004pou,
    author = "Desai, S. and others",
    collaboration = "Super-Kamiokande",
    title = "{Search for dark matter WIMPs using upward through-going muons in Super-Kamiokande}",
    eprint = "hep-ex/0404025",
    archivePrefix = "arXiv",
    doi = "10.1103/PhysRevD.70.083523",
    journal = "Phys. Rev. D",
    volume = "70",
    pages = "083523",
    year = "2004",
    note = "[Erratum: Phys.Rev.D 70, 109901 (2004)]"
}

@article{Super-Kamiokande:2020sgt,
    author = "Abe, K. and others",
    collaboration = "Super-Kamiokande",
    title = "{Indirect search for dark matter from the Galactic Center and halo with the Super-Kamiokande detector}",
    eprint = "2005.05109",
    archivePrefix = "arXiv",
    primaryClass = "hep-ex",
    doi = "10.1103/PhysRevD.102.072002",
    journal = "Phys. Rev. D",
    volume = "102",
    number = "7",
    pages = "072002",
    year = "2020"
}

@article{IceCube:2014rqf,
    author = "Aartsen, M. G. and others",
    collaboration = "IceCube",
    title = "{Multipole analysis of IceCube data to search for dark matter accumulated in the Galactic halo}",
    eprint = "1406.6868",
    archivePrefix = "arXiv",
    primaryClass = "astro-ph.HE",
    doi = "10.1140/epjc/s10052-014-3224-5",
    journal = "Eur. Phys. J. C",
    volume = "75",
    number = "99",
    pages = "20",
    year = "2015"
}

@article{IceCube:2017rdn,
    author = "Aartsen, M. G. and others",
    collaboration = "IceCube",
    title = "{Search for Neutrinos from Dark Matter Self-Annihilations in the center of the Milky Way with 3 years of IceCube/DeepCore}",
    eprint = "1705.08103",
    archivePrefix = "arXiv",
    primaryClass = "hep-ex",
    doi = "10.1140/epjc/s10052-017-5213-y",
    journal = "Eur. Phys. J. C",
    volume = "77",
    number = "9",
    pages = "627",
    year = "2017"
}

@article{IceCube:2021xzo,
    author = "Abbasi, R. and others",
    collaboration = "IceCube",
    title = "{Search for GeV-scale dark matter annihilation in the Sun with IceCube DeepCore}",
    eprint = "2111.09970",
    archivePrefix = "arXiv",
    primaryClass = "astro-ph.HE",
    doi = "10.1103/PhysRevD.105.062004",
    journal = "Phys. Rev. D",
    volume = "105",
    number = "6",
    pages = "062004",
    year = "2022"
}

@article{IceCube:2022vtr,
    author = "Abbasi, R. and others",
    collaboration = "IceCube",
    title = "{Searches for Connections between Dark Matter and High-Energy Neutrinos with IceCube}",
    eprint = "2205.12950",
    archivePrefix = "arXiv",
    primaryClass = "hep-ex",
    month = "5",
    year = "2022"
}

@article{Arguelles:2019ouk,
    author = {Arg\"uelles, Carlos A. and Diaz, Alejandro and Kheirandish, Ali and Olivares-Del-Campo, Andr\'es and Safa, Ibrahim and Vincent, Aaron C.},
    title = "{Dark matter annihilation to neutrinos}",
    eprint = "1912.09486",
    archivePrefix = "arXiv",
    primaryClass = "hep-ph",
    doi = "10.1103/RevModPhys.93.035007",
    journal = "Rev. Mod. Phys.",
    volume = "93",
    number = "3",
    pages = "035007",
    year = "2021"
}

@article{Garcia-Cely:2017oco,
    author = "Garcia-Cely, Camilo and Heeck, Julian",
    title = "{Neutrino Lines from Majoron Dark Matter}",
    eprint = "1701.07209",
    archivePrefix = "arXiv",
    primaryClass = "hep-ph",
    reportNumber = "ULB-TH-17-01",
    doi = "10.1007/JHEP05(2017)102",
    journal = "JHEP",
    volume = "05",
    pages = "102",
    year = "2017"
}

@article{Coy:2020wxp,
    author = "Coy, Rupert and Hambye, Thomas",
    title = "{Neutrino lines from DM decay induced by high-scale seesaw interactions}",
    eprint = "2012.05276",
    archivePrefix = "arXiv",
    primaryClass = "hep-ph",
    reportNumber = "ULB-TH/20-19",
    doi = "10.1007/JHEP05(2021)101",
    journal = "JHEP",
    volume = "05",
    pages = "101",
    year = "2021"
}

@article{Esmaili:2012us,
    author = "Esmaili, Arman and Ibarra, Alejandro and Peres, Orlando L. G.",
    title = "{Probing the stability of superheavy dark matter particles with high-energy neutrinos}",
    eprint = "1205.5281",
    archivePrefix = "arXiv",
    primaryClass = "hep-ph",
    doi = "10.1088/1475-7516/2012/11/034",
    journal = "JCAP",
    volume = "11",
    pages = "034",
    year = "2012"
}

@article{PhysRevD.92.123515,
  title = {New search for monochromatic neutrinos from dark matter decay},
  author = {Aisati, Cha\"{\i}mae El and Gustafsson, Michael and Hambye, Thomas},
  journal = {Phys. Rev. D},
  volume = {92},
  issue = {12},
  pages = {123515},
  numpages = {12},
  year = {2015},
  month = {Dec},
  publisher = {American Physical Society},
  doi = {10.1103/PhysRevD.92.123515},
  url = {https://link.aps.org/doi/10.1103/PhysRevD.92.123515}
}

@article{KamLAND:2021gvi,
    author = "Abe, S. and others",
    collaboration = "KamLAND",
    title = "{Limits on Astrophysical Antineutrinos with the KamLAND Experiment}",
    eprint = "2108.08527",
    archivePrefix = "arXiv",
    primaryClass = "astro-ph.HE",
    doi = "10.3847/1538-4357/ac32c1",
    journal = "Astrophys. J.",
    volume = "925",
    number = "1",
    pages = "14",
    year = "2022"
}

@article{IceCube-Gen2:2020qha,
    author = "Aartsen, M. G. and others",
    collaboration = "IceCube-Gen2",
    title = "{IceCube-Gen2: the window to the extreme Universe}",
    eprint = "2008.04323",
    archivePrefix = "arXiv",
    primaryClass = "astro-ph.HE",
    doi = "10.1088/1361-6471/abbd48",
    journal = "J. Phys. G",
    volume = "48",
    number = "6",
    pages = "060501",
    year = "2021"
}

@article{Abbasi:2020jmh,
    author = "Abbasi, R. and others",
    collaboration = "IceCube",
    title = "{The IceCube high-energy starting event sample: Description and flux characterization with 7.5 years of data}",
    eprint = "2011.03545",
    archivePrefix = "arXiv",
    primaryClass = "astro-ph.HE",
    month = "11",
    year = "2020"
}

@article{Super-Kamiokande:2021acd,
    author = "Abe, K. and others",
    collaboration = "Super-Kamiokande",
    title = "{Diffuse Supernova Neutrino Background Search at Super-Kamiokande}",
    eprint = "2109.11174",
    archivePrefix = "arXiv",
    primaryClass = "astro-ph.HE",
    month = "9",
    year = "2021"
}

@article{PierreAuger:2015eyc,
    author = "Aab, Alexander and others",
    collaboration = "Pierre Auger",
    title = "{The Pierre Auger Cosmic Ray Observatory}",
    eprint = "1502.01323",
    archivePrefix = "arXiv",
    primaryClass = "astro-ph.IM",
    reportNumber = "FERMILAB-PUB-15-034-AD-AE-CD-TD",
    doi = "10.1016/j.nima.2015.06.058",
    journal = "Nucl. Instrum. Meth. A",
    volume = "798",
    pages = "172--213",
    year = "2015"
}

@article{PierreAuger:2007kus,
    author = "Allekotte, I. and others",
    collaboration = "Pierre Auger",
    title = "{The Surface Detector System of the Pierre Auger Observatory}",
    eprint = "0712.2832",
    archivePrefix = "arXiv",
    primaryClass = "astro-ph",
    reportNumber = "FERMILAB-PUB-07-664-E-TD",
    doi = "10.1016/j.nima.2007.12.016",
    journal = "Nucl. Instrum. Meth. A",
    volume = "586",
    pages = "409--420",
    year = "2008"
}

@article{PierreAuger:2022aty,
    author = "Abreu, P. and others",
    collaboration = "Pierre Auger",
    title = "{Search for photons above 10$^{19}$ eV with the surface detector of the Pierre Auger Observatory}",
    eprint = "2209.05926",
    archivePrefix = "arXiv",
    primaryClass = "astro-ph.HE",
    reportNumber = "FERMILAB-PUB-22-722-ND-PPD-TD",
    month = "9",
    year = "2022"
}

@article{Cao:2022myt,
    author = "Cao, S. and Zhu, Z-H and Liang, N.",
    collaboration = "LHAASO",
    title = "{Constraints on heavy decaying dark matter from 570 days of LHAASO observations}",
    eprint = "2210.15989",
    archivePrefix = "arXiv",
    primaryClass = "astro-ph.HE",
    month = "10",
    year = "2022"
}

@phdthesis{Linyan:2018rmj,
    author = "Linyan, Wan",
    title = "{Experimental Studies on Low Energy Electron Antineutrinos and Related Physics}",
    school = "Tsinghua U., Beijing",
    year = "2018"
}

@article{Borexino:2019wln,
    author = "Agostini, M. and others",
    collaboration = "Borexino",
    title = "{Search for low-energy neutrinos from astrophysical sources with Borexino}",
    eprint = "1909.02422",
    archivePrefix = "arXiv",
    primaryClass = "hep-ex",
    reportNumber = "FERMILAB-PUB-21-152-AE",
    doi = "10.1016/j.astropartphys.2020.102509",
    journal = "Astropart. Phys.",
    volume = "125",
    pages = "102509",
    year = "2021"
}

@article{Ahmed:2021fvt,
    author = "Ahmed, Aqeel and Grz\k{a}dkowski, Bohdan and Socha, Anna",
    title = "{Implications of time-dependent inflaton decay on reheating and dark matter production}",
    eprint = "2111.06065",
    archivePrefix = "arXiv",
    primaryClass = "hep-ph",
    month = "11",
    year = "2021"
}

@article{Cohen:2016uyg,
    author = "Cohen, Timothy and Murase, Kohta and Rodd, Nicholas L. and Safdi, Benjamin R. and Soreq, Yotam",
    title = "{\ensuremath{\gamma} -ray Constraints on Decaying Dark Matter and Implications for IceCube}",
    eprint = "1612.05638",
    archivePrefix = "arXiv",
    primaryClass = "hep-ph",
    reportNumber = "MIT-CTP-4863, MIT-CTP 4863",
    doi = "10.1103/PhysRevLett.119.021102",
    journal = "Phys. Rev. Lett.",
    volume = "119",
    number = "2",
    pages = "021102",
    year = "2017"
}

@article{Aartsen:2013vja,
    author = "Aartsen, M.G. and others",
    collaboration = "IceCube",
    title = "{Energy Reconstruction Methods in the IceCube Neutrino Telescope}",
    eprint = "1311.4767",
    archivePrefix = "arXiv",
    primaryClass = "physics.ins-det",
    doi = "10.1088/1748-0221/9/03/P03009",
    journal = "JINST",
    volume = "9",
    pages = "P03009",
    year = "2014"
}

@article{Abi:2020wmh,
    author = "Abi, Babak and others",
    collaboration = "DUNE",
    title = "{Deep Underground Neutrino Experiment (DUNE), Far Detector Technical Design Report, Volume I Introduction to DUNE}",
    eprint = "2002.02967",
    archivePrefix = "arXiv",
    primaryClass = "physics.ins-det",
    reportNumber = "FERMILAB-PUB-20-024-ND, FERMILAB-DESIGN-2020-01",
    doi = "10.1088/1748-0221/15/08/T08008",
    journal = "JINST",
    volume = "15",
    number = "08",
    pages = "T08008",
    year = "2020"
}

@article{Murase:2015xka,
    author = "Murase, Kohta and Guetta, Dafne and Ahlers, Markus",
    title = "{Hidden Cosmic-Ray Accelerators as an Origin of TeV-PeV Cosmic Neutrinos}",
    eprint = "1509.00805",
    archivePrefix = "arXiv",
    primaryClass = "astro-ph.HE",
    doi = "10.1103/PhysRevLett.116.071101",
    journal = "Phys. Rev. Lett.",
    volume = "116",
    number = "7",
    pages = "071101",
    year = "2016"
}

@article{Aartsen:2014muf,
    author = "Aartsen, M.G. and others",
    collaboration = "IceCube",
    title = "{Atmospheric and astrophysical neutrinos above 1 TeV interacting in IceCube}",
    eprint = "1410.1749",
    archivePrefix = "arXiv",
    primaryClass = "astro-ph.HE",
    doi = "10.1103/PhysRevD.91.022001",
    journal = "Phys. Rev. D",
    volume = "91",
    number = "2",
    pages = "022001",
    year = "2015"
}

@article{Wissel:2020sec,
    author = "Wissel, S. and others",
    title = "{Prospects for High-Elevation Radio Detection of >100 PeV Tau Neutrinos}",
    eprint = "2004.12718",
    archivePrefix = "arXiv",
    primaryClass = "astro-ph.IM",
    month = "4",
    year = "2020"
}

@article{Abe:2020sbr,
    author = "Abe, K. and others",
    collaboration = "Super-Kamiokande",
    title = "{Indirect search for dark matter from the Galactic Center and halo with the Super-Kamiokande detector}",
    eprint = "2005.05109",
    archivePrefix = "arXiv",
    primaryClass = "hep-ex",
    doi = "10.1103/PhysRevD.102.072002",
    journal = "Phys. Rev. D",
    volume = "102",
    number = "7",
    pages = "072002",
    year = "2020"
}

@article{Barwick:2014pca,
    author = "Barwick, S.W. and others",
    collaboration = "ARIANNA",
    title = "{A First Search for Cosmogenic Neutrinos with the ARIANNA Hexagonal Radio Array}",
    eprint = "1410.7352",
    archivePrefix = "arXiv",
    primaryClass = "astro-ph.HE",
    doi = "10.1016/j.astropartphys.2015.04.002",
    journal = "Astropart. Phys.",
    volume = "70",
    pages = "12--26",
    year = "2015"
}

@ARTICLE{2012APh....35..457A,
    author = "Allison, P. and others",
    collaboration = "Ara Collaboration",
    title = "{Design and initial performance of the Askaryan Radio Array prototype EeV neutrino detector at the South Pole}",
      journal = {Astroparticle Physics},
     keywords = {Astrophysics - Instrumentation and Methods for Astrophysics, Astrophysics - High Energy Astrophysical Phenomena},
         year = 2012,
        month = feb,
       volume = {35},
       number = {7},
        pages = {457-477},
          doi = {10.1016/j.astropartphys.2011.11.010},
archivePrefix = {arXiv},
       eprint = {1105.2854},
 primaryClass = {astro-ph.IM},
       adsurl = {https://ui.adsabs.harvard.edu/abs/2012APh....35..457A},
      adsnote = {Provided by the SAO/NASA Astrophysics Data System}
}

@article{Aguilar:2019jay,
    author = "Aguilar, J.A. and others",
    title = "{The Next-Generation Radio Neutrino Observatory -- Multi-Messenger Neutrino Astrophysics at Extreme Energies}",
    eprint = "1907.12526",
    archivePrefix = "arXiv",
    primaryClass = "astro-ph.HE",
    month = "7",
    year = "2019"
}

@article{Agostini:2020aar,
    author = "Agostini, M. and others",
    title = "{The Pacific Ocean Neutrino Experiment}",
    eprint = "2005.09493",
    archivePrefix = "arXiv",
    primaryClass = "astro-ph.HE",
    month = "5",
    year = "2020"
}

@article{Murase:2012xs,
      author         = "Murase, Kohta and Beacom, John F.",
      title          = "{Constraining Very Heavy Dark Matter Using Diffuse
                        Backgrounds of Neutrinos and Cascaded Gamma Rays}",
      journal        = "JCAP",
      volume         = "1210",
      year           = "2012",
      pages          = "043",
      doi            = "10.1088/1475-7516/2012/10/043",
      eprint         = "1206.2595",
      archivePrefix  = "arXiv",
      primaryClass   = "hep-ph",
      SLACcitation   = "
}

@article{Murase:2019xqi,
      author         = "Murase, Kohta and Shoemaker, Ian M.",
      title          = "{Neutrino Echoes from Multimessenger Transient Sources}",
      journal        = "Phys. Rev. Lett.",
      volume         = "123",
      year           = "2019",
      number         = "24",
      pages          = "241102",
      doi            = "10.1103/PhysRevLett.123.241102",
      eprint         = "1903.08607",
      archivePrefix  = "arXiv",
      primaryClass   = "hep-ph",
      SLACcitation   = "
}

@inproceedings{Ishihara:2019uei,
      author         = "Ishihara, Aya and Kiriki, Ayumi",
      title          = "{Calibration LEDs in the IceCube Upgrade D-Egg Modules}",
      booktitle      = "{HAWC Contributions to the 36th International Cosmic Ray
                        Conference (ICRC2019)}",
      collaboration  = "IceCube",
      year           = "2019",
      eprint         = "1908.10780",
      archivePrefix  = "arXiv",
      primaryClass   = "astro-ph.HE",
      reportNumber   = "PoS-ICRC2019-923",
      SLACcitation   = "
}

@article{Primulando:2017kxf,
      author         = "Primulando, Reinard and Uttayarat, Patipan",
      title          = "{Dark Matter-Neutrino Interaction in Light of Collider
                        and Neutrino Telescope Data}",
      journal        = "JHEP",
      volume         = "06",
      year           = "2018",
      pages          = "026",
      doi            = "10.1007/JHEP06(2018)026",
      eprint         = "1710.08567",
      archivePrefix  = "arXiv",
      primaryClass   = "hep-ph",
      SLACcitation   = "
}

@article{Acharya:2017ttl,
      author         = "Acharya, B. S. and others",
      title          = "Science with the Cherenkov Telescope Array",
      publisher      = "WSP",
      collaboration  = "CTA Consortium",
      year           = "2018",
      doi            = "10.1142/10986",
      eprint         = "1709.07997",
      archivePrefix  = "arXiv",
      primaryClass   = "astro-ph.IM",
      ISBN           = "9789813270084",
      SLACcitation   = "
}

@article{Aguilar:2016kjl,
      author         = "Aguilar, M. and others",
      title          = "{Antiproton Flux, Antiproton-to-Proton Flux Ratio, and
                        Properties of Elementary Particle Fluxes in Primary Cosmic
                        Rays Measured with the Alpha Magnetic Spectrometer on the
                        International Space Station}",
      collaboration  = "AMS",
      journal        = "Phys. Rev. Lett.",
      volume         = "117",
      year           = "2016",
      number         = "9",
      pages          = "091103",
      doi            = "10.1103/PhysRevLett.117.091103",
      SLACcitation   = "
}

@article{Daylan:2014rsa,
      author         = "Daylan, Tansu and Finkbeiner, Douglas P. and Hooper, Dan
                        and Linden, Tim and Portillo, Stephen K. N. and Rodd,
                        Nicholas L. and Slatyer, Tracy R.",
      title          = "{The characterization of the gamma-ray signal from the
                        central Milky Way: A case for annihilating dark matter}",
      journal        = "Phys. Dark Univ.",
      volume         = "12",
      year           = "2016",
      pages          = "1-23",
      doi            = "10.1016/j.dark.2015.12.005",
      eprint         = "1402.6703",
      archivePrefix  = "arXiv",
      primaryClass   = "astro-ph.HE",
      reportNumber   = "FERMILAB-PUB-14-032-A, MIT-CTP-4533",
      SLACcitation   = "
}

@article{Cuoco:2016eej,
      author         = "Cuoco, Alessandro and Kr{\"a}mer, Michael and Korsmeier,
                        Michael",
      title          = "{Novel Dark Matter Constraints from Antiprotons in Light
                        of AMS-02}",
      journal        = "Phys. Rev. Lett.",
      volume         = "118",
      year           = "2017",
      number         = "19",
      pages          = "191102",
      doi            = "10.1103/PhysRevLett.118.191102",
      eprint         = "1610.03071",
      archivePrefix  = "arXiv",
      primaryClass   = "astro-ph.HE",
      SLACcitation   = "
}

@article{Calore:2014xka,
      author         = "Calore, Francesca and Cholis, Ilias and Weniger,
                        Christoph",
      title          = "{Background Model Systematics for the Fermi GeV Excess}",
      journal        = "JCAP",
      volume         = "1503",
      year           = "2015",
      pages          = "038",
      doi            = "10.1088/1475-7516/2015/03/038",
      eprint         = "1409.0042",
      archivePrefix  = "arXiv",
      primaryClass   = "astro-ph.CO",
      reportNumber   = "FERMILAB-PUB-14-289-A",
      SLACcitation   = "
}

@inproceedings{Nagai:2019uaz,
      author         = "Nagai, Ryo and Ishihara, Aya",
      title          = "{Electronics Development for the New Photo-Detectors
                        (PDOM and D-Egg) for IceCube-Upgrade}",
      booktitle      = "{HAWC Contributions to the 36th International Cosmic Ray
                        Conference (ICRC2019)}",
      collaboration  = "IceCube",
      year           = "2019",
      eprint         = "1908.11564",
      archivePrefix  = "arXiv",
      primaryClass   = "astro-ph.IM",
      reportNumber   = "PoS-ICRC2019-966",
      SLACcitation   = "
}

@inproceedings{Ishihara:2019aao,
      author         = "Ishihara, Aya",
      title          = "{The IceCube Upgrade -- Design and Science Goals}",
      booktitle      = "{HAWC Contributions to the 36th International Cosmic Ray
                        Conference (ICRC2019)}",
      collaboration  = "IceCube",
      year           = "2019",
      eprint         = "1908.09441",
      archivePrefix  = "arXiv",
      primaryClass   = "astro-ph.HE",
      reportNumber   = "PoS-ICRC2019-1031",
      SLACcitation   = "
}

@misc{nusquids,
	title="{nuSQuIDS}",
  author = {Arg{\"u}elles, C. A. and Salvado, J. and Weaver, C. N.},
	howpublished={\url{https://github.com/Arguelles/nuSQuIDS}},
	year            = "2015"
}

@article{Delgado:2014kpa,
  author = {Arg\"uelles, Carlos Alberto and Salvado, Jordi and
    Weaver, Christopher N.},
  title = {A Simple Quantum Integro-Differential Solver (SQuIDS)},
  year = "2014",
  eprint = "1412.3832",
  archivePrefix = "arXiv",
  primaryClass = "hep-ph",
  SLACcitation = "
}

@article{Arguelles:2020hss,
    author = "Arg{\"u}elles, Carlos A. and Salvado, Jordi and Weaver, Christopher N.",
    title = "{A Simple Quantum Integro-Differential Solver (SQuIDS)}",
    doi = "10.1016/j.cpc.2020.107405",
    journal = "Comput. Phys. Commun.",
    volume = "255",
    pages = "107405",
    year = "2020"
}

@article{Capozzi:2018dat,
      author         = "Capozzi, Francesco and Li, Shirley Weishi and Zhu,
                        Guanying and Beacom, John F.",
      title          = "{DUNE as the Next-Generation Solar Neutrino Experiment}",
      journal        = "Phys. Rev. Lett.",
      volume         = "123",
      year           = "2019",
      number         = "13",
      pages          = "131803",
      doi            = "10.1103/PhysRevLett.123.131803",
      eprint         = "1808.08232",
      archivePrefix  = "arXiv",
      primaryClass   = "hep-ph",
      reportNumber   = "MPP-2018-214, SLAC-PUB-17199",
      SLACcitation   = "
}

@article{Bauer_2021,
	doi = {10.1007/JHEP06(2021)121},
  
	url = {https://doi.org/10.1007
  
	year = 2021,
	month = {jun},
  
	publisher = {Springer Science and Business Media {LLC}
},
  
	volume = {2021},
  
	number = {6},
  
	author = {Christian W. Bauer and Nicholas L. Rodd and Bryan R. Webber},
  
	title = {Dark matter spectra from the electroweak to the Planck scale},
  
	journal = {Journal of High Energy Physics}
}

@article{Zhu:2018rwc,
      author         = "Zhu, Guanying and Li, Shirley Weishi and Beacom, John F.",
      title          = "{Developing the MeV potential of DUNE: Detailed
                        considerations of muon-induced spallation and other
                        backgrounds}",
      journal        = "Phys. Rev.",
      volume         = "C99",
      year           = "2019",
      number         = "5",
      pages          = "055810",
      doi            = "10.1103/PhysRevC.99.055810",
      eprint         = "1811.07912",
      archivePrefix  = "arXiv",
      primaryClass   = "hep-ph",
      reportNumber   = "SLAC-PUB-17355",
      SLACcitation   = "
}

@article{Arguelles:2019xgp,
      author         = {Arg{\"u}elles, C. A. and others},
      title          = "{White Paper on New Opportunities at the Next-Generation
                        Neutrino Experiments (Part 1: BSM Neutrino Physics and
                        Dark Matter)}",
      year           = "2019",
      eprint         = "1907.08311",
      archivePrefix  = "arXiv",
      primaryClass   = "hep-ph",
      reportNumber   = "FERMILAB-FN-1079-T",
      SLACcitation   = "
}

@article{Akita:2022lit,
    author = "Akita, Kensuke and Lambiase, Gaetano and Niibo, Michiru and Yamaguchi, Masahide",
    title = "{Neutrino lines from MeV dark matter annihilation and decay in JUNO}",
    eprint = "2206.06755",
    archivePrefix = "arXiv",
    primaryClass = "hep-ph",
    reportNumber = "CTPU-PTC-22-14",
    month = "6",
    year = "2022"
}

@article{An:2015jdp,
      author         = "An, Fengpeng and others",
      title          = "{Neutrino Physics with JUNO}",
      collaboration  = "JUNO",
      journal        = "J. Phys.",
      volume         = "G43",
      year           = "2016",
      number         = "3",
      pages          = "030401",
      doi            = "10.1088/0954-3899/43/3/030401",
      eprint         = "1507.05613",
      archivePrefix  = "arXiv",
      primaryClass   = "physics.ins-det",
      SLACcitation   = "
}

@unpublished{nickroddtalk,
title= {Spectra for Heavy Dark Matter},
author = {Nick Rodd},
year = {2020},
note= {TRIUMF Theory Workshop: New Techniques for Dark Matter Discovery},
URL= {https://particletheory.triumf.ca/2020/Talks/rodd.pdf},
}

@article{Liu:2020ckq,
    author = {Liu, Qinrui and Lazar, Jeffrey and Arg\"uelles, Carlos A. and Kheirandish, Ali},
    title = "{$\chi$aro$\nu$: a tool for neutrino flux generation from WIMPs}",
    eprint = "2007.15010",
    archivePrefix = "arXiv",
    primaryClass = "hep-ph",
    doi = "10.1088/1475-7516/2020/10/043",
    journal = "JCAP",
    volume = "10",
    pages = "043",
    year = "2020"
}

@article{Bell:2017irk,
    author = "Bell, Nicole F. and Cai, Yi and Dent, James B. and Leane, Rebecca K. and Weiler, Thomas J.",
    title = "{Enhancing Dark Matter Annihilation Rates with Dark Bremsstrahlung}",
    eprint = "1705.01105",
    archivePrefix = "arXiv",
    primaryClass = "hep-ph",
    doi = "10.1103/PhysRevD.96.023011",
    journal = "Phys. Rev. D",
    volume = "96",
    number = "2",
    pages = "023011",
    year = "2017"
}

@article{Freedman:1973yd,
    author = "Freedman, Daniel Z.",
    title = "{Coherent Neutrino Nucleus Scattering as a Probe of the Weak Neutral Current}",
    reportNumber = "NAL-PUB-73-76-THY, FERMILAB-PUB-73-076-T",
    doi = "10.1103/PhysRevD.9.1389",
    journal = "Phys. Rev. D",
    volume = "9",
    pages = "1389--1392",
    year = "1974"
}

@article{Gando:2013nba,
      author         = "Gando, A. and others",
      title          = "{Reactor On-Off Antineutrino Measurement with KamLAND}",
      collaboration  = "KamLAND",
      journal        = "Phys. Rev.",
      volume         = "D88",
      year           = "2013",
      number         = "3",
      pages          = "033001",
      doi            = "10.1103/PhysRevD.88.033001",
      eprint         = "1303.4667",
      archivePrefix  = "arXiv",
      primaryClass   = "hep-ex",
      SLACcitation   = "
}

@article{Abe:2011em,
      author         = "Abe, S. and others",
      title          = "{Measurement of the 8B Solar Neutrino Flux with the
                        KamLAND Liquid Scintillator Detector}",
      collaboration  = "KamLAND",
      journal        = "Phys. Rev.",
      volume         = "C84",
      year           = "2011",
      pages          = "035804",
      doi            = "10.1103/PhysRevC.84.035804",
      eprint         = "1106.0861",
      archivePrefix  = "arXiv",
      primaryClass   = "hep-ex",
      SLACcitation   = "
}

@article{Chianese:2019kyl,
    author = "Chianese, Marco and Fiorillo, Damiano F. G. and Miele, Gennaro and Morisi, Stefano and Pisanti, Ofelia",
    title = "{Decaying dark matter at IceCube and its signature on High Energy gamma experiments}",
    eprint = "1907.11222",
    archivePrefix = "arXiv",
    primaryClass = "hep-ph",
    doi = "10.1088/1475-7516/2019/11/046",
    journal = "JCAP",
    volume = "11",
    pages = "046",
    year = "2019"
}
@article{Murase:2015gea,
    author = "Murase, Kohta and Laha, Ranjan and Ando, Shin'ichiro and Ahlers, Markus",
    title = "{Testing the Dark Matter Scenario for PeV Neutrinos Observed in IceCube}",
    eprint = "1503.04663",
    archivePrefix = "arXiv",
    primaryClass = "hep-ph",
    doi = "10.1103/PhysRevLett.115.071301",
    journal = "Phys. Rev. Lett.",
    volume = "115",
    number = "7",
    pages = "071301",
    year = "2015"
}

@article{Collaboration:2011jza,
      author         = "Gando, A. and others",
      title          = "{A study of extraterrestrial antineutrino sources with
                        the KamLAND detector}",
      collaboration  = "KamLAND",
      journal        = "Astrophys. J.",
      volume         = "745",
      year           = "2012",
      pages          = "193",
      doi            = "10.1088/0004-637X/745/2/193",
      eprint         = "1105.3516",
      archivePrefix  = "arXiv",
      primaryClass   = "astro-ph.HE",
      SLACcitation   = "
}

@article{Adrian-Martinez:2016fdl,
      author         = "Adrian-Martinez, S. and others",
      title          = "{Letter of intent for KM3NeT 2.0}",
      collaboration  = "KM3Net",
      journal        = "J. Phys.",
      volume         = "G43",
      year           = "2016",
      number         = "8",
      pages          = "084001",
      doi            = "10.1088/0954-3899/43/8/084001",
      eprint         = "1601.07459",
      archivePrefix  = "arXiv",
      primaryClass   = "astro-ph.IM",
      SLACcitation   = "
}

@article{Aartsen:2014njl,
      author         = "Aartsen, M. G. and others",
      title          = "{IceCube-Gen2: A Vision for the Future of Neutrino
                        Astronomy in Antarctica}",
      collaboration  = "IceCube",
      year           = "2014",
      eprint         = "1412.5106",
      archivePrefix  = "arXiv",
      primaryClass   = "astro-ph.HE",
      SLACcitation   = "
}

@article{Karukes:2019jxv,
      author         = "Karukes, Ekaterina V. and Benito, Maria and Iocco, Fabio
                        and Trotta, Roberto and Geringer-Sameth, Alex",
      title          = "{Bayesian reconstruction of the Milky Way dark matter
                        distribution}",
      year           = "2019",
      eprint         = "1901.02463",
      archivePrefix  = "arXiv",
      primaryClass   = "astro-ph.GA",
      SLACcitation   = "
}

@article{Benito:2019ngh,
      author         = "Benito, Maria and Cuoco, Alessandro and Iocco, Fabio",
      title          = "{Handling the Uncertainties in the Galactic Dark Matter
                        Distribution for Particle Dark Matter Searches}",
      journal        = "JCAP",
      volume         = "1903",
      year           = "2019",
      number         = "03",
      pages          = "033",
      doi            = "10.1088/1475-7516/2019/03/033",
      eprint         = "1901.02460",
      archivePrefix  = "arXiv",
      primaryClass   = "astro-ph.GA",
      SLACcitation   = "
}

@article{Ellis:2017tkh,
      author         = "Ellis, John and Fairbairn, Malcolm and Tunney, Patrick",
      title          = "{Anomaly-Free Dark Matter Models are not so Simple}",
      journal        = "JHEP",
      volume         = "08",
      year           = "2017",
      pages          = "053",
      doi            = "10.1007/JHEP08(2017)053",
      eprint         = "1704.03850",
      archivePrefix  = "arXiv",
      primaryClass   = "hep-ph",
      reportNumber   = "KCL-PH-TH-2017-21, CERN-TH-2017-084",
      SLACcitation   = "
}

@article{Bakhti:2017jhm,
      author         = "Bakhti, Pouya and Farzan, Yasaman",
      title          = "{Constraining secret gauge interactions of neutrinos by
                        meson decays}",
      journal        = "Phys. Rev.",
      volume         = "D95",
      year           = "2017",
      number         = "9",
      pages          = "095008",
      doi            = "10.1103/PhysRevD.95.095008",
      eprint         = "1702.04187",
      archivePrefix  = "arXiv",
      primaryClass   = "hep-ph",
      SLACcitation   = "
}

@article{Beacom:2006tt,
      author         = "Beacom, John F. and Bell, Nicole F. and Mack, Gregory D.",
      title          = "{General Upper Bound on the Dark Matter Total
                        Annihilation Cross Section}",
      journal        = "Phys. Rev. Lett.",
      volume         = "99",
      year           = "2007",
      pages          = "231301",
      doi            = "10.1103/PhysRevLett.99.231301",
      eprint         = "astro-ph/0608090",
      archivePrefix  = "arXiv",
      primaryClass   = "astro-ph",
      reportNumber   = "KRL-MAP-322",
      SLACcitation   = "
}

@article{Lopez-Honorez:2013lcm,
      author         = "Lopez-Honorez, Laura and Mena, Olga and Palomares-Ruiz,
                        Sergio and Vincent, Aaron C.",
      title          = "{Constraints on dark matter annihilation from CMB
                        observationsbefore Planck}",
      journal        = "JCAP",
      volume         = "1307",
      year           = "2013",
      pages          = "046",
      doi            = "10.1088/1475-7516/2013/07/046",
      eprint         = "1303.5094",
      archivePrefix  = "arXiv",
      primaryClass   = "astro-ph.CO",
      reportNumber   = "IFIC-13-016, CFTP-13-007",
      SLACcitation   = "
}

@article{Diamanti:2013bia,
      author         = "Diamanti, Roberta and Lopez-Honorez, Laura and Mena, Olga
                        and Palomares-Ruiz, Sergio and Vincent, Aaron C.",
      title          = "{Constraining Dark Matter Late-Time Energy Injection:
                        Decays and P-Wave Annihilations}",
      journal        = "JCAP",
      volume         = "1402",
      year           = "2014",
      pages          = "017",
      doi            = "10.1088/1475-7516/2014/02/017",
      eprint         = "1308.2578",
      archivePrefix  = "arXiv",
      primaryClass   = "astro-ph.CO",
      reportNumber   = "IFIC-13-54",
      SLACcitation   = "
}

@article{Aartsen:2015xup,
      author         = "Aartsen, M. G. and others",
      title          = "{Measurement of the Atmospheric $\nu_e$ Spectrum with
                        IceCube}",
      collaboration  = "IceCube",
      journal        = "Phys. Rev.",
      volume         = "D91",
      year           = "2015",
      pages          = "122004",
      doi            = "10.1103/PhysRevD.91.122004",
      eprint         = "1504.03753",
      archivePrefix  = "arXiv",
      primaryClass   = "astro-ph.HE",
      SLACcitation   = "
}
@misc{iovine2021indirect,
      title={Indirect search for dark matter in the Galactic Centre with IceCube}, 
      author={Nadège Iovine and Juan A. Aguilar},
      year={2021},
      eprint={2107.11224},
      archivePrefix={arXiv},
      primaryClass={astro-ph.HE}
}
@misc{frankiewicz2015searching,
      title={Searching for Dark Matter Annihilation into Neutrinos with Super-Kamiokande}, 
      author={Katarzyna Frankiewicz},
      year={2015},
      eprint={1510.07999},
      archivePrefix={arXiv},
      primaryClass={hep-ex}
}
@article{2017b,
   title={Results from the search for dark matter in the Milky Way with 9 years of data of the ANTARES neutrino telescope},
   volume={769},
   ISSN={0370-2693},
   url={http://dx.doi.org/10.1016/j.physletb.2017.03.063},
   DOI={10.1016/j.physletb.2017.03.063},
   journal={Physics Letters B},
   publisher={Elsevier BV},
   author={Albert, A. and André, M. and Anghinolfi, M. and Anton, G. and Ardid, M. and Aubert, J.-J. and Avgitas, T. and Baret, B. and Barrios-Martí, J. and Basa, S. and et al.},
   year={2017},
   month={Jun},
   pages={249–254}
}

@article{2017,
   title={Search for neutrinos from dark matter self-annihilations in the center of the Milky Way with 3 years of IceCube/DeepCore},
   volume={77},
   ISSN={1434-6052},
   DOI={10.1140/epjc/s10052-017-5213-y},
   number={9},
   journal={The European Physical Journal C},
   publisher={Springer Science and Business Media LLC},
   author={Aartsen, M. G. and Ackermann, M. and Adams, J. and Aguilar, J. A. and Ahlers, M. and Ahrens, M. and Al Samarai, I. and Altmann, D. and Andeen, K. and Anderson, T. and et al.},
   year={2017},
   month={Sep}
}

@article{2016,
   title={All-flavour search for neutrinos from dark matter annihilations in the Milky Way with IceCube/DeepCore},
   volume={76},
   ISSN={1434-6052},
   url={http://dx.doi.org/10.1140/epjc/s10052-016-4375-3},
   DOI={10.1140/epjc/s10052-016-4375-3},
   number={10},
   journal={The European Physical Journal C},
   publisher={Springer Science and Business Media LLC},
   author={Aartsen, M. G. and Abraham, K. and Ackermann, M. and Adams, J. and Aguilar, J. A. and Ahlers, M. and Ahrens, M. and Altmann, D. and Andeen, K. and Anderson, T. and et al.},
   year={2016},
   month={Sep}
}

@article{2018,
   title={Differential limit on the extremely-high-energy cosmic neutrino flux in the presence of astrophysical background from nine years of IceCube data},
   volume={98},
   ISSN={2470-0029},
   url={http://dx.doi.org/10.1103/PhysRevD.98.062003},
   DOI={10.1103/physrevd.98.062003},
   number={6},
   journal={Physical Review D},
   publisher={American Physical Society (APS)},
   author={Aartsen, M. G. and Ackermann, M. and Adams, J. and Aguilar, J. A. and Ahlers, M. and Ahrens, M. and Al Samarai, I. and Altmann, D. and Andeen, K. and Anderson, T. and et al.},
   year={2018},
   month={Sep}
}
@article{Aartsen_2016,
	doi = {10.3847/0004-637x/833/1/3},
  
	url = {https://doi.org/10.3847
  
	year = 2016,
	month = {dec},
  
	publisher = {American Astronomical Society},
  
	volume = {833},
  
	number = {1},
  
	pages = {3},
  
	author = {M. G. Aartsen and others},
  
	title = {{OBSERVATION} {AND} {CHARACTERIZATION} {OF} A {COSMIC} {MUON} {NEUTRINO} {FLUX} {FROM} {THE} {NORTHERN} {HEMISPHERE} {USING} {SIX} {YEARS} {OF} {ICECUBE} {DATA}},
  
	journal = {The Astrophysical Journal}
}

@article{Richard:2015aua,
      author         = "Richard, E. and others",
      title          = "{Measurements of the atmospheric neutrino flux by
                        Super-Kamiokande: energy spectra, geomagnetic effects, and
                        solar modulation}",
      collaboration  = "Super-Kamiokande",
      journal        = "Phys. Rev.",
      volume         = "D94",
      year           = "2016",
      number         = "5",
      pages          = "052001",
      doi            = "10.1103/PhysRevD.94.052001",
      eprint         = "1510.08127",
      archivePrefix  = "arXiv",
      primaryClass   = "hep-ex",
      SLACcitation   = "
}

@article{Andringa:2015tza,
      author         = "Andringa, S. and others",
      title          = "{Current Status and Future Prospects of the SNO+
                        Experiment}",
      collaboration  = "SNO+",
      journal        = "Adv. High Energy Phys.",
      volume         = "2016",
      year           = "2016",
      pages          = "6194250",
      doi            = "10.1155/2016/6194250",
      eprint         = "1508.05759",
      archivePrefix  = "arXiv",
      primaryClass   = "physics.ins-det",
      SLACcitation   = "
}

@article{Aartsen:2017ulx,
      author         = "Aartsen, M. G. and others",
      title          = "{Search for Neutrinos from Dark Matter Self-Annihilations
                        in the center of the Milky Way with 3 years of
                        IceCube/DeepCore}",
      collaboration  = "IceCube",
      journal        = "Eur. Phys. J.",
      volume         = "C77",
      year           = "2017",
      number         = "9",
      pages          = "627",
      doi            = "10.1140/epjc/s10052-017-5213-y",
      eprint         = "1705.08103",
      archivePrefix  = "arXiv",
      primaryClass   = "hep-ex",
      SLACcitation   = "
}

@article{Aartsen:2017nbu,
      author         = "Aartsen, M. G. and others",
      title          = "{Measurement of the $\nu _{\mu}$ energy spectrum with
                        IceCube-79}",
      collaboration  = "IceCube",
      journal        = "Eur. Phys. J.",
      volume         = "C77",
      year           = "2017",
      number         = "10",
      pages          = "692",
      doi            = "10.1140/epjc/s10052-017-5261-3",
      eprint         = "1705.07780",
      archivePrefix  = "arXiv",
      primaryClass   = "astro-ph.HE",
      SLACcitation   = "
}

@article{Aartsen:2016xlq,
      author         = "Aartsen, M. G. and others",
      title          = "{Observation and Characterization of a Cosmic Muon
                        Neutrino Flux from the Northern Hemisphere using six years
                        of IceCube data}",
      collaboration  = "IceCube",
      journal        = "Astrophys. J.",
      volume         = "833",
      year           = "2016",
      number         = "1",
      pages          = "3",
      doi            = "10.3847/0004-637X/833/1/3",
      eprint         = "1607.08006",
      archivePrefix  = "arXiv",
      primaryClass   = "astro-ph.HE",
      SLACcitation   = "
}

@article{Lees:2014xha,
      author         = "Lees, J. P. and others",
      title          = "{Search for a Dark Photon in $e^+e^-$ Collisions at
                        BaBar}",
      collaboration  = "BaBar",
      journal        = "Phys. Rev. Lett.",
      volume         = "113",
      year           = "2014",
      number         = "20",
      pages          = "201801",
      doi            = "10.1103/PhysRevLett.113.201801",
      eprint         = "1406.2980",
      archivePrefix  = "arXiv",
      primaryClass   = "hep-ex",
      reportNumber   = "BABAR-PUB-14-002, SLAC-PUB-15979",
      SLACcitation   = "
}

@article{Abrahamyan:2011gv,
      author         = "Abrahamyan, S. and others",
      title          = "{Search for a New Gauge Boson in Electron-Nucleus
                        Fixed-Target Scattering by the APEX Experiment}",
      collaboration  = "APEX",
      journal        = "Phys. Rev. Lett.",
      volume         = "107",
      year           = "2011",
      pages          = "191804",
      doi            = "10.1103/PhysRevLett.107.191804",
      eprint         = "1108.2750",
      archivePrefix  = "arXiv",
      primaryClass   = "hep-ex",
      reportNumber   = "JLAB-PHY-11-1406, SLAC-PUB-14491",
      SLACcitation   = "
}

@article{Carena:2004xs,
      author         = "Carena, Marcela and Daleo, Alejandro and Dobrescu, Bogdan
                        A. and Tait, Timothy M. P.",
      title          = "{$Z^\prime$ gauge bosons at the Tevatron}",
      journal        = "Phys. Rev.",
      volume         = "D70",
      year           = "2004",
      pages          = "093009",
      doi            = "10.1103/PhysRevD.70.093009",
      eprint         = "hep-ph/0408098",
      archivePrefix  = "arXiv",
      primaryClass   = "hep-ph",
      reportNumber   = "FERMILAB-PUB-04-129-T",
      SLACcitation   = "
}

@article{Blanco:2019hah,
      author         = "Blanco, Carlos and Escudero, Miguel and Hooper, Dan and
                        Witte, Samuel J.",
      title          = "{$Z'$ Mediated WIMPs: Dead, Dying, or Soon to be
                        Detected?}",
      year           = "2019",
      eprint         = "1907.05893",
      archivePrefix  = "arXiv",
      primaryClass   = "hep-ph",
      SLACcitation   = "
}

@inproceedings{Campo:2018dfh,
      author         = "Olivares-Del Campo, Andres and Palomares-Ruiz, Sergio and
                        Pascoli, Silvia",
      title          = "{Implications of a Dark Matter-Neutrino Coupling at
                        Hyper-Kamiokande}",
      booktitle      = "{53rd Rencontres de Moriond on Electroweak Interactions and Unified Theories (Moriond EW 2018) La Thuile, Italy, March 10-17, 2018}",
      year           = "2018",
      eprint         = "1805.09830",
      archivePrefix  = "arXiv",
      primaryClass   = "hep-ph",
      SLACcitation   = "
}

@article{wandkowsky_nancy_2018_1301088,
  author       = {Wandkowsky, Nancy},
  title        = "{Latest results on astrophysical neutrinos using high-energy events with contained vertices}",
  month        = jun,
  year         = 2018,
  doi          = {10.5281/zenodo.1301088},
  url          = {https://doi.org/10.5281/zenodo.1301088}
}

@article{Bernal:2012qh,
      author         = "Bernal, Nicolás and Mart{\'i}n-Albo, Justo and
                        Palomares-Ruiz, Sergio",
      title          = "{A novel way of constraining WIMPs annihilations in the
                        Sun: MeV neutrinos}",
      journal        = "JCAP",
      volume         = "1308",
      year           = "2013",
      pages          = "011",
      doi            = "10.1088/1475-7516/2013/08/011",
      eprint         = "1208.0834",
      archivePrefix  = "arXiv",
      primaryClass   = "hep-ph",
      reportNumber   = "BONN-TH-2012-19, IFIC-12-57, CFTP-12-012",
      SLACcitation   = "
}

@article{Cornell:2013rza,
      author         = "Cornell, Jonathan M. and Profumo, Stefano and Shepherd,
                        William",
      title          = "{Kinetic Decoupling and Small-Scale Structure in
                        Effective Theories of Dark Matter}",
      journal        = "Phys. Rev.",
      volume         = "D88",
      year           = "2013",
      number         = "1",
      pages          = "015027",
      doi            = "10.1103/PhysRevD.88.015027",
      eprint         = "1305.4676",
      archivePrefix  = "arXiv",
      primaryClass   = "hep-ph",
      SLACcitation   = "
}

@article{Shoemaker:2013tda,
      author         = "Shoemaker, Ian M.",
      title          = "{Constraints on Dark Matter Protohalos in Effective
                        Theories and Neutrinophilic Dark Matter}",
      journal        = "Phys. Dark Univ.",
      volume         = "2",
      year           = "2013",
      number         = "3",
      pages          = "157-162",
      doi            = "10.1016/j.dark.2013.07.002",
      eprint         = "1305.1936",
      archivePrefix  = "arXiv",
      primaryClass   = "hep-ph",
      reportNumber   = "LA-UR-13-22524",
      SLACcitation   = "
}

@ARTICLE{2013MNRAS.433.1230W,
       author = {{Watson}, William A. and {Iliev}, Ilian T. and {D'Aloisio}, Anson and
         {Knebe}, Alexander and {Shapiro}, Paul R. and {Yepes}, Gustavo},
        title = "{The halo mass function through the cosmic ages}",
      journal = {\mnras},
     keywords = {methods: numerical, galaxies: haloes, galaxies: high-redshift, cosmology: theory, dark matter, large-scale structure of Universe, Astrophysics - Cosmology and Nongalactic Astrophysics},
         year = "2013",
        month = "Aug",
       volume = {433},
        pages = {1230-1245},
          doi = {10.1093/mnras/stt791},
archivePrefix = {arXiv},
       eprint = {1212.0095},
 primaryClass = {astro-ph.CO},
       adsurl = {https://ui.adsabs.harvard.edu/\#abs/2013MNRAS.433.1230W},
      adsnote = {Provided by the SAO/NASA Astrophysics Data System}
}

@ARTICLE{2012MNRAS.423.3018P,
       author = {{Prada}, Francisco and {Klypin}, Anatoly A. and {Cuesta}, Antonio J. and
         {Betancort-Rijo}, Juan E. and {Primack}, Joel},
        title = "{Halo concentrations in the standard {\ensuremath{\Lambda}} cold dark matter cosmology}",
      journal = {\mnras},
     keywords = {galaxies: haloes, cosmology: theory, dark matter, Astrophysics - Cosmology and Nongalactic Astrophysics},
         year = "2012",
        month = "Jul",
       volume = {423},
        pages = {3018-3030},
          doi = {10.1111/j.1365-2966.2012.21007.x},
archivePrefix = {arXiv},
       eprint = {1104.5130},
 primaryClass = {astro-ph.CO},
       adsurl = {https://ui.adsabs.harvard.edu/\#abs/2012MNRAS.423.3018P},
      adsnote = {Provided by the SAO/NASA Astrophysics Data System}
}

@article{Pato:2015dua,
      author         = "Pato, Miguel and Iocco, Fabio and Bertone, Gianfranco",
      title          = "{Dynamical constraints on the dark matter distribution in
                        the Milky Way}",
      journal        = "JCAP",
      volume         = "1512",
      year           = "2015",
      number         = "12",
      pages          = "001",
      doi            = "10.1088/1475-7516/2015/12/001",
      eprint         = "1504.06324",
      archivePrefix  = "arXiv",
      primaryClass   = "astro-ph.GA",
      SLACcitation   = "
}

@article{Griest:1989wd,
      author         = "Griest, Kim and Kamionkowski, Marc",
      title          = "{Unitarity Limits on the Mass and Radius of Dark Matter
                        Particles}",
      journal        = "Phys. Rev. Lett.",
      volume         = "64",
      year           = "1990",
      pages          = "615",
      doi            = "10.1103/PhysRevLett.64.615",
      reportNumber   = "CFPA-TH-89-013, FERMILAB-PUB-89-205-A",
      SLACcitation   = "
}

@article{Anderson:2018byx,
      author         = "Anderson, M. and others",
      title          = "{Search for invisible modes of nucleon decay in water
                        with the SNO+ detector}",
      collaboration  = "SNO+",
      journal        = "Phys. Rev.",
      volume         = "D99",
      year           = "2019",
      number         = "3",
      pages          = "032008",
      doi            = "10.1103/PhysRevD.99.032008",
      eprint         = "1812.05552",
      archivePrefix  = "arXiv",
      primaryClass   = "hep-ex",
      SLACcitation   = "
}

@article{Farzan:2014gza,
    author = "Farzan, Yasaman and Palomares-Ruiz, Sergio",
    title = "{Dips in the Diffuse Supernova Neutrino Background}",
    eprint = "1401.7019",
    archivePrefix = "arXiv",
    primaryClass = "hep-ph",
    reportNumber = "IFIC-14-02",
    doi = "10.1088/1475-7516/2014/06/014",
    journal = "JCAP",
    volume = "06",
    pages = "014",
    year = "2014"
}

@article{Gorham:2016zah,
    author = "Gorham, P.W. and others",
    collaboration = "ANITA",
    title = "{Characteristics of Four Upward-pointing Cosmic-ray-like Events Observed with ANITA}",
    eprint = "1603.05218",
    archivePrefix = "arXiv",
    primaryClass = "astro-ph.HE",
    doi = "10.1103/PhysRevLett.117.071101",
    journal = "Phys. Rev. Lett.",
    volume = "117",
    number = "7",
    pages = "071101",
    year = "2016"
}

@article{Dudas:2020sbq,
    author = "Dudas, Emilian and Heurtier, Lucien and Mambrini, Yann and Olive, Keith A. and Pierre, Mathias",
    title = "{A Model of Metastable EeV Dark Matter}",
    eprint = "2003.02846",
    archivePrefix = "arXiv",
    primaryClass = "hep-ph",
    reportNumber = "CPHT-RR011.032020, UMN--TH--3913/20, FTPI--MINN--20/03,
  IFT-UAM/CSIC-20-23",
    month = "3",
    year = "2020"
}

@article{Anchordoqui:2019utb,
    author = "Anchordoqui, L.A. and others",
    title = "{The pros and cons of beyond standard model interpretations of ANITA events}",
    eprint = "1907.06308",
    archivePrefix = "arXiv",
    primaryClass = "hep-ph",
    doi = "10.22323/1.358.0884",
    journal = "PoS",
    volume = "ICRC2019",
    number = "884",
    pages = "884",
    year = "2020"
}

@article{Gorham:2018ydl,
    author = "Gorham, P.W. and others",
    collaboration = "ANITA",
    title = "{Observation of an Unusual Upward-going Cosmic-ray-like Event in the Third Flight of ANITA}",
    eprint = "1803.05088",
    archivePrefix = "arXiv",
    primaryClass = "astro-ph.HE",
    doi = "10.1103/PhysRevLett.121.161102",
    journal = "Phys. Rev. Lett.",
    volume = "121",
    number = "16",
    pages = "161102",
    year = "2018"
}

@article{Esmaili:2019pcy,
    author = "Esmaili, Arman and Farzan, Yasaman",
    title = "{Explaining the ANITA events by a $L_e-L_\tau$ gauge model}",
    eprint = "1909.07995",
    archivePrefix = "arXiv",
    primaryClass = "hep-ph",
    doi = "10.1088/1475-7516/2019/12/017",
    journal = "JCAP",
    volume = "12",
    pages = "017",
    year = "2019"
}

@article{Hooper:2019ytr,
    author = "Hooper, Dan and Wegsman, Shalma and Deaconu, Cosmin and Vieregg, Abigail",
    title = "{Superheavy dark matter and ANITA's anomalous events}",
    eprint = "1904.12865",
    archivePrefix = "arXiv",
    primaryClass = "astro-ph.HE",
    reportNumber = "FERMILAB-PUB-19-178-A",
    doi = "10.1103/PhysRevD.100.043019",
    journal = "Phys. Rev. D",
    volume = "100",
    number = "4",
    pages = "043019",
    year = "2019"
}

@article{Cline:2019snp,
    author = "Cline, James M. and Gross, Christian and Xue, Wei",
    title = "{Can the ANITA anomalous events be due to new physics?}",
    eprint = "1904.13396",
    archivePrefix = "arXiv",
    primaryClass = "hep-ph",
    reportNumber = "CERN-TH-2019-057",
    doi = "10.1103/PhysRevD.100.015031",
    journal = "Phys. Rev. D",
    volume = "100",
    number = "1",
    pages = "015031",
    year = "2019"
}

@article{Bradley:2018eev,
    author = "Bradley, Richard F. and Tauscher, Keith and Rapetti, David and Burns, Jack O.",
    title = "{A Ground Plane Artifact that Induces an Absorption Profile in Averaged Spectra from Global 21-cm Measurements - with Possible Application to EDGES}",
    eprint = "1810.09015",
    archivePrefix = "arXiv",
    primaryClass = "astro-ph.IM",
    doi = "10.3847/1538-4357/ab0d8b",
    journal = "Astrophys. J.",
    volume = "874",
    number = "2",
    pages = "153",
    year = "2019"
}

@article{Hoof:2018hyn,
      author         = "Hoof, Sebastian and Geringer-Sameth, Alex and Trotta,
                        Roberto",
      title          = "{A Global Analysis of Dark Matter Signals from 27 Dwarf
                        Spheroidal Galaxies using Ten Years of Fermi-LAT
                        Observations}",
      year           = "2018",
      eprint         = "1812.06986",
      archivePrefix  = "arXiv",
      primaryClass   = "astro-ph.CO",
      SLACcitation   = "
}

@article{Fermi-LAT:2016uux,
      author         = "Albert, A. and others",
      title          = "{Searching for Dark Matter Annihilation in Recently
                        Discovered Milky Way Satellites with Fermi-LAT}",
      collaboration  = "Fermi-LAT, DES",
      journal        = "Astrophys. J.",
      volume         = "834",
      year           = "2017",
      number         = "2",
      pages          = "110",
      doi            = "10.3847/1538-4357/834/2/110",
      eprint         = "1611.03184",
      archivePrefix  = "arXiv",
      primaryClass   = "astro-ph.HE",
      reportNumber   = "FERMILAB-PUB-16-073-AE",
      SLACcitation   = "
}

@article{Honda:2006qj,
      author         = "Honda, Morihiro and Kajita, T. and Kasahara, K. and
                        Midorikawa, S. and Sanuki, T.",
      title          = "{Calculation of atmospheric neutrino flux using the
                        interaction model calibrated with atmospheric muon data}",
      journal        = "Phys. Rev.",
      volume         = "D75",
      year           = "2007",
      pages          = "043006",
      doi            = "10.1103/PhysRevD.75.043006",
      eprint         = "astro-ph/0611418",
      archivePrefix  = "arXiv",
      primaryClass   = "astro-ph",
      SLACcitation   = "
}
@article{deSalas:2019pee,
    author = "de Salas, P. F. and Malhan, K. and Freese, K. and Hattori, K. and Valluri, M.",
    title = "{On the estimation of the Local Dark Matter Density using the rotation curve of the Milky Way}",
    eprint = "1906.06133",
    archivePrefix = "arXiv",
    primaryClass = "astro-ph.GA",
    doi = "10.1088/1475-7516/2019/10/037",
    journal = "JCAP",
    volume = "10",
    pages = "037",
    year = "2019"
}

@article{Fukuda:2002uc,
      author         = "Fukuda, Y. and others",
      title          = "{The Super-Kamiokande detector}",
      booktitle      = "{Advanced computing and analysis techniques in physics
                        research. Proceedings, 8th International Workshop, ACAT
                        2002, Moscow, Russia, June 24-28, 2002}",
      collaboration  = "Super-Kamiokande",
      journal        = "Nucl. Instrum. Meth.",
      volume         = "A501",
      year           = "2003",
      pages          = "418-462",
      doi            = "10.1016/S0168-9002(03)00425-X",
      SLACcitation   = "
}

@article{Frankiewicz:2017trk,
      author         = "Frankiewicz, Katarzyna",
      title          = "{Dark matter searches with the Super-Kamiokande
                        detector}",
      booktitle      = "{Proceedings, 27th International Conference on Neutrino
                        Physics and Astrophysics (Neutrino 2016): London, United
                        Kingdom, July 4-9, 2016}",
      collaboration  = "Super-Kamiokande",
      journal        = "J. Phys. Conf. Ser.",
      volume         = "888",
      year           = "2017",
      number         = "1",
      pages          = "012210",
      doi            = "10.1088/1742-6596/888/1/012210",
      SLACcitation   = "
}

@inproceedings{Adams:2022pbo,
    author = "Adams, C. B. and others",
    title = "{Axion Dark Matter}",
    booktitle = "{2022 Snowmass Summer Study}",
    eprint = "2203.14923",
    archivePrefix = "arXiv",
    primaryClass = "hep-ex",
    month = "3",
    year = "2022"
}

@article{Green:2020jor,
    author = "Green, Anne M. and Kavanagh, Bradley J.",
    title = "{Primordial Black Holes as a dark matter candidate}",
    eprint = "2007.10722",
    archivePrefix = "arXiv",
    primaryClass = "astro-ph.CO",
    doi = "10.1088/1361-6471/abc534",
    journal = "J. Phys. G",
    volume = "48",
    number = "4",
    pages = "043001",
    year = "2021"
}

@phdthesis{FRANKIEWICZ:2018lsh,
      author         = "Frankiewicz, Katarzyna",
      title          = "{Indirect Search for Dark Matter with the
                        Super-Kamiokande Detector}",
      school         = "National Centre for Nuclear Research, Poland",
      year           = "2018",
      url            = "http://www-sk.icrr.u-tokyo.ac.jp/sk/_pdf/articles/2018/Frankiewicz_PhD_final.pdf",
      SLACcitation   = "
}
@article{Abuter:2018drb,
      author         = "Abuter, R. and others",
      title          = "{Detection of the gravitational redshift in the orbit of
                        the star S2 near the Galactic centre massive black hole}",
      collaboration  = "GRAVITY",
      journal        = "Astron. Astrophys.",
      volume         = "615",
      year           = "2018",
      pages          = "L15",
      doi            = "10.1051/0004-6361/201833718",
      eprint         = "1807.09409",
      archivePrefix  = "arXiv",
      primaryClass   = "astro-ph.GA",
      SLACcitation   = "
}

@inproceedings{Lisanti:2016jxe,
    author = "Lisanti, Mariangela",
    title = "{Lectures on Dark Matter Physics}",
    booktitle = "{Theoretical Advanced Study Institute in Elementary Particle Physics}: {New Frontiers in Fields and Strings}",
    eprint = "1603.03797",
    archivePrefix = "arXiv",
    primaryClass = "hep-ph",
    doi = "10.1142/9789813149441_0007",
    pages = "399--446",
    year = "2017"
}

@article{Aghanim:2018eyx,
    author = "Aghanim, N. and others",
    collaboration = "Planck",
    title = "{Planck 2018 results. VI. Cosmological parameters}",
    eprint = "1807.06209",
    archivePrefix = "arXiv",
    primaryClass = "astro-ph.CO",
    doi = "10.1051/0004-6361/201833910",
    journal = "Astron. Astrophys.",
    volume = "641",
    pages = "A6",
    year = "2020",
    note = "[Erratum: Astron.Astrophys. 652, C4 (2021)]"
}

@article{Boddy:2018ike,
      author         = "Boddy, Kimberly K. and Kumar, Jason and Strigari, Louis
                        E.",
      title          = "{The Effective J-Factor of the Galactic Center for
                        Velocity-Dependent Dark Matter Annihilation}",
      year           = "2018",
      eprint         = "1805.08379",
      archivePrefix  = "arXiv",
      primaryClass   = "astro-ph.HE",
      SLACcitation   = "
}

@article{Yuksel:2007ac,
      author         = "Y{\"u}ksel, Hasan and Horiuchi, Shunsaku and Beacom, John F.
                        and Ando, Shin'ichiro",
      title          = "{Neutrino Constraints on the Dark Matter Total
                        Annihilation Cross Section}",
      journal        = "Phys. Rev.",
      volume         = "D76",
      year           = "2007",
      pages          = "123506",
      doi            = "10.1103/PhysRevD.76.123506",
      eprint         = "0707.0196",
      archivePrefix  = "arXiv",
      primaryClass   = "astro-ph",
      SLACcitation   = "
}

@article{PalomaresRuiz:2007eu,
      author         = "Palomares-Ruiz, Sergio and Pascoli, Silvia",
      title          = "{Testing MeV dark matter with neutrino detectors}",
      journal        = "Phys. Rev.",
      volume         = "D77",
      year           = "2008",
      pages          = "025025",
      doi            = "10.1103/PhysRevD.77.025025",
      eprint         = "0710.5420",
      archivePrefix  = "arXiv",
      primaryClass   = "astro-ph",
      reportNumber   = "IPPP-07-81, DCPT-07-162",
      SLACcitation   = "
}

@inproceedings{Frankiewicz:2015zma,
      author         = "Frankiewicz, Katarzyna",
      title          = "{Searching for Dark Matter Annihilation into Neutrinos
                        with Super-Kamiokande}",
      booktitle      = "{Proceedings, Meeting of the APS Division of Particles
                        and Fields (DPF 2015): Ann Arbor, Michigan, USA, 4-8 Aug
                        2015}",
      collaboration  = "Super-Kamiokande",
      year           = "2015",
      eprint         = "1510.07999",
      archivePrefix  = "arXiv",
      primaryClass   = "hep-ex",
      SLACcitation   = "
}

@article{Frankiewicz:2016,
      author         = "Frankiewicz, K.",
      title          = "{Indirect searches for dark matter particles with the
                        Super-Kamiokande detector}",
      booktitle      = "{Proceedings, 29th Rencontres de Physique de La Vallée
                        d'Aoste: La Thuile, Aosta, Italy, March 1-7, 2015}",
      collaboration  = "Super-Kamiokande",
      journal        = "Nuovo Cim.",
      volume         = "C38",
      year           = "2016",
      number         = "4",
      pages          = "125",
      doi            = "10.1393/ncc/i2015-15125-y",
      SLACcitation   = "
}

@article{Navarro:1995iw,
      author         = "Navarro, Julio F. and Frenk, Carlos S. and White, Simon
                        D. M.",
      title          = "{The Structure of cold dark matter halos}",
      journal        = "Astrophys. J.",
      volume         = "462",
      year           = "1996",
      pages          = "563-575",
      doi            = "10.1086/177173",
      eprint         = "astro-ph/9508025",
      archivePrefix  = "arXiv",
      primaryClass   = "astro-ph",
      SLACcitation   = "
}

@article{Kravtsov:1997dp,
      author         = "Kravtsov, Andrey V. and Klypin, Anatoly A. and Bullock,
                        James S. and Primack, Joel R.",
      title          = "{The Cores of dark matter dominated galaxies: Theory
                        versus observations}",
      journal        = "Astrophys. J.",
      volume         = "502",
      year           = "1998",
      pages          = "48",
      doi            = "10.1086/305884",
      eprint         = "astro-ph/9708176",
      archivePrefix  = "arXiv",
      primaryClass   = "astro-ph",
      SLACcitation   = "
}

@article{Moore:1999gc,
      author         = "Moore, Ben and Quinn, Thomas R. and Governato, Fabio and
                        Stadel, Joachim and Lake, George",
      title          = "{Cold collapse and the core catastrophe}",
      journal        = "Mon. Not. Roy. Astron. Soc.",
      volume         = "310",
      year           = "1999",
      pages          = "1147-1152",
      doi            = "10.1046/j.1365-8711.1999.03039.x",
      eprint         = "astro-ph/9903164",
      archivePrefix  = "arXiv",
      primaryClass   = "astro-ph",
      SLACcitation   = "
}

@article{Bellini:2010gn,
      author         = "Bellini, G. and others",
      title          = "{Study of solar and other unknown anti-neutrino fluxes
                        with Borexino at LNGS}",
      collaboration  = "Borexino",
      journal        = "Phys. Lett.",
      volume         = "B696",
      year           = "2011",
      pages          = "191-196",
      doi            = "10.1016/j.physletb.2010.12.030",
      eprint         = "1010.0029",
      archivePrefix  = "arXiv",
      primaryClass   = "hep-ex",
      SLACcitation   = "
}

@article{PalomaresRuiz:2007ry,
      author         = "Palomares-Ruiz, Sergio",
      title          = "{Model-Independent Bound on the Dark Matter Lifetime}",
      journal        = "Phys. Lett.",
      volume         = "B665",
      year           = "2008",
      pages          = "50-53",
      doi            = "10.1016/j.physletb.2008.05.040",
      eprint         = "0712.1937",
      archivePrefix  = "arXiv",
      primaryClass   = "astro-ph",
      reportNumber   = "IPPP-07-96, DCPT-07-192",
      SLACcitation   = "
}

@article{Steigman:2012nb,
      author         = "Steigman, Gary and Dasgupta, Basudeb and Beacom, John F.",
      title          = "{Precise Relic WIMP Abundance and its Impact on Searches
                        for Dark Matter Annihilation}",
      journal        = "Phys. Rev.",
      volume         = "D86",
      year           = "2012",
      pages          = "023506",
      doi            = "10.1103/PhysRevD.86.023506",
      eprint         = "1204.3622",
      archivePrefix  = "arXiv",
      primaryClass   = "hep-ph",
      SLACcitation   = "
}

@article{Aartsen:2016nxy,
      author         = "Aartsen, M. G. and others",
      title          = "{The IceCube Neutrino Observatory: Instrumentation and
                        Online Systems}",
      collaboration  = "IceCube",
      journal        = "JINST",
      volume         = "12",
      year           = "2017",
      number         = "03",
      pages          = "P03012",
      doi            = "10.1088/1748-0221/12/03/P03012",
      eprint         = "1612.05093",
      archivePrefix  = "arXiv",
      primaryClass   = "astro-ph.IM",
      SLACcitation   = "
}

@article{Bellini:2013lnn,
      author         = "Bellini, G. and others",
      title          = "{Final results of Borexino Phase-I on low energy solar
                        neutrino spectroscopy}",
      collaboration  = "Borexino",
      journal        = "Phys. Rev.",
      volume         = "D89",
      year           = "2014",
      number         = "11",
      pages          = "112007",
      doi            = "10.1103/PhysRevD.89.112007",
      eprint         = "1308.0443",
      archivePrefix  = "arXiv",
      primaryClass   = "hep-ex",
      SLACcitation   = "
}

@inproceedings{Arguelles:2019rbn,
      author         = "Argüelles, Carlos A. and Bustamante, Mauricio and
                        Kheirandish, Ali and Palomares-Ruiz, Sergio and Salvado,
                        Jordi and Vincent, Aaron C.",
      title          = "{Fundamental physics with high-energy cosmic neutrinos
                        today and in the future}",
      booktitle      = "{36th International Cosmic Ray Conference (ICRC 2019)
                        Madison, Wisconsin, USA, July 24-August 1, 2019}",
      year           = "2019",
      eprint         = "1907.08690",
      archivePrefix  = "arXiv",
      primaryClass   = "astro-ph.HE",
      SLACcitation   = "
}

@article{Baerwald:2012kc,
      author         = "Baerwald, Philipp and Bustamante, Mauricio and Winter,
                        Walter",
      title          = "{Neutrino Decays over Cosmological Distances and the
                        Implications for Neutrino Telescopes}",
      journal        = "JCAP",
      volume         = "1210",
      year           = "2012",
      pages          = "020",
      doi            = "10.1088/1475-7516/2012/10/020",
      eprint         = "1208.4600",
      archivePrefix  = "arXiv",
      primaryClass   = "astro-ph.CO",
      SLACcitation   = "
}

@article{Collaboration:2011nsa,
      author         = "Ageron, M. and others",
      title          = "{ANTARES: the first undersea neutrino telescope}",
      collaboration  = "ANTARES",
      journal        = "Nucl. Instrum. Meth.",
      volume         = "A656",
      year           = "2011",
      pages          = "11-38",
      doi            = "10.1016/j.nima.2011.06.103",
      eprint         = "1104.1607",
      archivePrefix  = "arXiv",
      primaryClass   = "astro-ph.IM",
      SLACcitation   = "
}

@article{Adrian-Martinez:2015wey,
      author         = "Adrian-Martinez, S. and others",
      title          = "{Search of Dark Matter Annihilation in the Galactic Centre using the ANTARES Neutrino Telescope}",
      collaboration  = "ANTARES",
      journal        = "JCAP",
      volume         = "1510",
      year           = "2015",
      number         = "10",
      pages          = "068",
      doi            = "10.1088/1475-7516/2015/10/068",
      eprint         = "1505.04866",
      archivePrefix  = "arXiv",
      primaryClass   = "astro-ph.HE",
      SLACcitation   = "
}

@article{Aartsen:2016pfc,
      author         = "Aartsen, M. G. and others",
      title          = "{All-flavour Search for Neutrinos from Dark Matter
                        Annihilations in the Milky Way with IceCube/DeepCore}",
      collaboration  = "IceCube",
      journal        = "Eur. Phys. J.",
      volume         = "C76",
      year           = "2016",
      number         = "10",
      pages          = "531",
      doi            = "10.1140/epjc/s10052-016-4375-3",
      eprint         = "1606.00209",
      archivePrefix  = "arXiv",
      primaryClass   = "astro-ph.HE",
      SLACcitation   = "
}

@article{IceCube,
author = {Halzen, Francis and R Klein, Spencer},
year = {2010},
month = {08},
pages = {081101},
title = {Invited Review Article: IceCube: An instrument for neutrino astronomy},
volume = {81},
journal = {The Review of scientific instruments},
doi = {10.1063/1.3480478}
}

@article{Klop:2018ltd,
      author         = "Klop, Niki and Ando, Shin'ichiro",
      title          = "{Constraints on MeV dark matter using neutrino detectors
                        and their implication for the 21-cm results}",
      journal        = "Phys. Rev.",
      volume         = "D98",
      year           = "2018",
      number         = "10",
      pages          = "103004",
      doi            = "10.1103/PhysRevD.98.103004",
      eprint         = "1809.00671",
      archivePrefix  = "arXiv",
      primaryClass   = "hep-ph",
      SLACcitation   = "
}

@article{Agostini:2019yuq,
    author = "Agostini, M. and others",
    collaboration = "Borexino",
    title = "{Search for low-energy neutrinos from astrophysical sources with Borexino}",
    eprint = "1909.02422",
    archivePrefix = "arXiv",
    primaryClass = "hep-ex",
    month = "9",
    year = "2019"
}

@InProceedings{INOproc,
author="Indumathi, D.",
editor="Giri, Anjan
and Mohanta, Rukmani",
title="India-Based Neutrino Observatory (INO): Physics and Status Report",
booktitle="16th Conference on Flavor Physics and CP Violation",
year="2019",
publisher="Springer International Publishing",
address="Cham",
pages="309--314",
abstract="We discuss the physics reach and current status of the India-based Neutrino Observatory (INO) project. We set this in the context of the proposed magnetised iron calorimeter (ICAL) detector at INO, whose main goal is the determination of the neutrino mass hierarchy using atmospheric neutrinos. We also discuss various possible synergies with other planned and upcoming experiments. We also mention the status of the mini-ICAL prototype that has been set up at IICHEP in Madurai.",
isbn="978-3-030-29622-3"
}

@article{Khatun:2017adx,
    author = "Khatun, Amina and Laha, Ranjan and Agarwalla, Sanjib Kumar",
    title = "{Indirect searches of Galactic diffuse dark matter in INO-MagICAL detector}",
    eprint = "1703.10221",
    archivePrefix = "arXiv",
    primaryClass = "hep-ph",
    reportNumber = "IP-BBSR-2017-7",
    doi = "10.1007/JHEP06(2017)057",
    journal = "JHEP",
    volume = "06",
    pages = "057",
    year = "2017"
}

@article{Kumar:2017sdq,
    author = "Ahmed, Shakeel and others",
    collaboration = "ICAL",
    title = "{Physics Potential of the ICAL detector at the India-based Neutrino Observatory (INO)}",
    eprint = "1505.07380",
    archivePrefix = "arXiv",
    primaryClass = "physics.ins-det",
    reportNumber = "INO-ICAL-PHY-NOTE-2015-01",
    doi = "10.1007/s12043-017-1373-4",
    journal = "Pramana",
    volume = "88",
    number = "5",
    pages = "79",
    year = "2017"
}

@article{Acciarri:2015uup,
      author         = "Acciarri, R. and others",
      title          = "{Long-Baseline Neutrino Facility (LBNF) and Deep
                        Underground Neutrino Experiment (DUNE)}",
      collaboration  = "DUNE",
      year           = "2015",
      eprint         = "1512.06148",
      archivePrefix  = "arXiv",
      primaryClass   = "physics.ins-det",
      reportNumber   = "FERMILAB-DESIGN-2016-02",
      SLACcitation   = "
}

@article{Aalbers:2020gsn,
    author = "Aalbers, J. and others",
    collaboration = "DARWIN",
    title = "{Solar Neutrino Detection Sensitivity in DARWIN via Electron Scattering}",
    eprint = "2006.03114",
    archivePrefix = "arXiv",
    primaryClass = "physics.ins-det",
    month = "6",
    year = "2020"
}

@article{McKeen:2018pbb,
      author         = "McKeen, David and Raj, Nirmal",
      title          = "{Monochromatic dark neutrinos and boosted dark matter in
                        noble liquid direct detection}",
      year           = "2018",
      eprint         = "1812.05102",
      archivePrefix  = "arXiv",
      primaryClass   = "hep-ph",
      SLACcitation   = "
}

@article{Aalbers:2016jon,
      author         = "Aalbers, J. and others",
      title          = "{DARWIN: towards the ultimate dark matter detector}",
      collaboration  = "DARWIN",
      journal        = "JCAP",
      volume         = "1611",
      year           = "2016",
      pages          = "017",
      doi            = "10.1088/1475-7516/2016/11/017",
      eprint         = "1606.07001",
      archivePrefix  = "arXiv",
      primaryClass   = "astro-ph.IM",
      SLACcitation   = "
}

@article{Abe:2018uyc,
      author         = "Abe, K. and others",
      title          = "{Hyper-Kamiokande Design Report}",
      collaboration  = "Hyper-Kamiokande",
      year           = "2018",
      eprint         = "1805.04163",
      archivePrefix  = "arXiv",
      primaryClass   = "physics.ins-det",
      SLACcitation   = "
}

@article{Cirelli:2010xx,
      author         = "Cirelli, Marco and Corcella, Gennaro and Hektor, Andi and
                        Hutsi, Gert and Kadastik, Mario and Panci, Paolo and
                        Raidal, Martti and Sala, Filippo and Strumia, Alessandro",
      title          = "{PPPC 4 DM ID: A Poor Particle Physicist Cookbook for
                        Dark Matter Indirect Detection}",
      journal        = "JCAP",
      volume         = "1103",
      year           = "2011",
      pages          = "051",
      doi            = "10.1088/1475-7516/2012/10/E01,
                        10.1088/1475-7516/2011/03/051",
      note           = "[Erratum: JCAP1210,E01(2012)]",
      eprint         = "1012.4515",
      archivePrefix  = "arXiv",
      primaryClass   = "hep-ph",
      reportNumber   = "CERN-PH-TH-2010-057, SACLAY-T10-025, IFUP-TH-2010-44",
      SLACcitation   = "
}

@article{Queiroz:2016zwd,
      author         = "Queiroz, Farinaldo S. and Yaguna, Carlos E. and Weniger,
                        Christoph",
      title          = "{Gamma-ray Limits on Neutrino Lines}",
      journal        = "JCAP",
      volume         = "1605",
      year           = "2016",
      number         = "05",
      pages          = "050",
      doi            = "10.1088/1475-7516/2016/05/050",
      eprint         = "1602.05966",
      archivePrefix  = "arXiv",
      primaryClass   = "hep-ph",
      SLACcitation   = "
}

@article{ElAisati:2017ppn,
      author         = "El Aisati, Chaimae and Garcia-Cely, Camilo and Hambye,
                        Thomas and Vanderheyden, Laurent",
      title          = "{Prospects for discovering a neutrino line induced by
                        dark matter annihilation}",
      journal        = "JCAP",
      volume         = "1710",
      year           = "2017",
      number         = "10",
      pages          = "021",
      doi            = "10.1088/1475-7516/2017/10/021",
      eprint         = "1706.06600",
      archivePrefix  = "arXiv",
      primaryClass   = "hep-ph",
      SLACcitation   = "
}

@ARTICLE{2011ExA....32..193A,
       author = {Actis, M. and others },
        title = "{Design concepts for the Cherenkov Telescope Array CTA: an advanced facility for ground-based high-energy gamma-ray astronomy}",
      journal = {Experimental Astronomy},
     keywords = {Ground based gamma ray astronomy, Next generation Cherenkov telescopes, Design concepts, Astrophysics - Instrumentation and Methods for Astrophysics, Astrophysics - High Energy Astrophysical Phenomena},
         year = 2011,
        month = Dec,
       volume = {32},
        pages = {193-316},
          doi = {10.1007/s10686-011-9247-0},
archivePrefix = {arXiv},
       eprint = {1008.3703},
 primaryClass = {astro-ph.IM},
}

@ARTICLE{2018PhRvD..97g5039O,
       author = {{Olivares-Del Campo}, Andr{\'e}s and {B{\r{A}}`hm}, C{\'e}line and {Palomares-Ruiz}, Sergio and {Pascoli}, Silvia},
        title = "{Dark matter-neutrino interactions through the lens of their cosmological implications}",
      journal = {\prd},
     keywords = {High Energy Physics - Phenomenology},
         year = 2018,
        month = apr,
       volume = {97},
       number = {7},
          eid = {075039},
        pages = {075039},
          doi = {10.1103/PhysRevD.97.075039},
archivePrefix = {arXiv},
       eprint = {1711.05283},
 primaryClass = {hep-ph},
       adsurl = {https://ui.adsabs.harvard.edu/abs/2018PhRvD..97g5039O},
      adsnote = {Provided by the SAO/NASA Astrophysics Data System}
}

@article{Albert:2016emp,
      author         = "Albert, A. and others",
      title          = "{Results from the search for dark matter in the Milky Way
                        with 9 years of data of the ANTARES neutrino telescope}",
      journal        = "Phys. Lett.",
      volume         = "B769",
      year           = "2017",
      pages          = "249-254",
      doi            = "10.1016/j.physletb.2019.05.022,
                        10.1016/j.physletb.2017.03.063",
      note           = "[Erratum: Phys. Lett.B796,253(2019)]",
      eprint         = "1612.04595",
      archivePrefix  = "arXiv",
      primaryClass   = "astro-ph.HE",
      SLACcitation   = "
}

@article{ALBERT2017249,
	author = {ANTARES Collaboration},
	doi = "10.1016/j.physletb.2017.03.063",
	issn = "0370-2693",
	journal = "Physics Letters B",
	keywords = "Dark matter; WIMP; Indirect detection; Neutrino telescope; Galactic Centre; ANTARES",
	title = "{Results from the search for dark matter in the Milky Way with 9 years of data of the ANTARES neutrino telescope}",
	url = "http://www.sciencedirect.com/science/article/pii/S037026931730254X",
	volume = "769",
	year = "2017"
}

@article{Blennow:2019fhy,
      author         = "Blennow, M. and Fernandez-Martinez, E. and Campo, A.
                        Olivares-Del and Pascoli, S. and Rosauro-Alcaraz, S. and
                        Titov, A. V.",
      title          = "{Neutrino Portals to Dark Matter}",
      year           = "2019",
      eprint         = "1903.00006",
      archivePrefix  = "arXiv",
      primaryClass   = "hep-ph",
      reportNumber   = "FTUAM-19-5, IFT-UAM/CSIC-19-19, IPPP/19/17",
      SLACcitation   = "
}

@article{Aartsen:2018vtx,
      author         = "Aartsen, M. G. and others",
      title          = "{Differential limit on the extremely-high-energy cosmic
                        neutrino flux in the presence of astrophysical background
                        from nine years of IceCube data}",
      collaboration  = "IceCube",
      journal        = "Phys. Rev.",
      volume         = "D98",
      year           = "2018",
      number         = "6",
      pages          = "062003",
      doi            = "10.1103/PhysRevD.98.062003",
      eprint         = "1807.01820",
      archivePrefix  = "arXiv",
      primaryClass   = "astro-ph.HE",
      SLACcitation   = "
}

@article{Zas:2017xdj,
      author         = "Zas, Enrique",
      title          = "{Searches for neutrino fluxes in the EeV regime with the
                        Pierre Auger Observatory}",
      booktitle      = "{The Pierre Auger Observatory: Contributions to the 35th
                        International Cosmic Ray Conference (ICRC 2017)}",
      collaboration  = "Pierre Auger",
      journal        = "PoS",
      volume         = "ICRC2017",
      year           = "2018",
      pages          = "972",
      doi            = "10.22323/1.301.0972",
      note           = "[,64(2017)]",
      SLACcitation   = "
}

@article{Alvarez-Muniz:2018bhp,
      author         = "Alvarez-Muniz, Jaime and others",
      title          = "{The Giant Radio Array for Neutrino Detection (GRAND):
                        Science and Design}",
      collaboration  = "GRAND",
      year           = "2018",
      eprint         = "1810.09994",
      archivePrefix  = "arXiv",
      primaryClass   = "astro-ph.HE",
      SLACcitation   = "
}

@article{Gorham:2019guw,
      author         = "Gorham, P. W. and others",
      title          = "{Constraints on the ultra-high energy cosmic neutrino
                        flux from the fourth flight of ANITA}",
      collaboration  = "ANITA",
      year           = "2019",
      eprint         = "1902.04005",
      archivePrefix  = "arXiv",
      primaryClass   = "astro-ph.HE",
      SLACcitation   = "
}

@article{Bays:2011si,
      author         = "Bays, K. and others",
      title          = "{Supernova Relic Neutrino Search at Super-Kamiokande}",
      collaboration  = "Super-Kamiokande",
      journal        = "Phys. Rev.",
      volume         = "D85",
      year           = "2012",
      pages          = "052007",
      doi            = "10.1103/PhysRevD.85.052007",
      eprint         = "1111.5031",
      archivePrefix  = "arXiv",
      primaryClass   = "hep-ex",
      SLACcitation   = "
}

@article{Hosaka:2005um,
      author         = "Hosaka, J. and others",
      title          = "{Solar neutrino measurements in super-Kamiokande-I}",
      collaboration  = "Super-Kamiokande",
      journal        = "Phys. Rev.",
      volume         = "D73",
      year           = "2006",
      pages          = "112001",
      doi            = "10.1103/PhysRevD.73.112001",
      eprint         = "hep-ex/0508053",
      archivePrefix  = "arXiv",
      primaryClass   = "hep-ex",
      SLACcitation   = "
}

@article{Cravens:2008aa,
      author         = "Cravens, J. P. and others",
      title          = "{Solar neutrino measurements in Super-Kamiokande-II}",
      collaboration  = "Super-Kamiokande",
      journal        = "Phys. Rev.",
      volume         = "D78",
      year           = "2008",
      pages          = "032002",
      doi            = "10.1103/PhysRevD.78.032002",
      eprint         = "0803.4312",
      archivePrefix  = "arXiv",
      primaryClass   = "hep-ex",
      SLACcitation   = "
}

@article{Abe:2010hy,
      author         = "Abe, K. and others",
      title          = "{Solar neutrino results in Super-Kamiokande-III}",
      collaboration  = "Super-Kamiokande",
      journal        = "Phys. Rev.",
      volume         = "D83",
      year           = "2011",
      pages          = "052010",
      doi            = "10.1103/PhysRevD.83.052010",
      eprint         = "1010.0118",
      archivePrefix  = "arXiv",
      primaryClass   = "hep-ex",
      SLACcitation   = "
}

@article{ThePierreAuger:2015rma,
      author         = "Aab, Alexander and others",
      title          = "{The Pierre Auger Cosmic Ray Observatory}",
      collaboration  = "Pierre Auger",
      journal        = "Nucl. Instrum. Meth.",
      volume         = "A798",
      year           = "2015",
      pages          = "172-213",
      doi            = "10.1016/j.nima.2015.06.058",
      eprint         = "1502.01323",
      archivePrefix  = "arXiv",
      primaryClass   = "astro-ph.IM",
      reportNumber   = "FERMILAB-PUB-15-034-AD-AE-CD-TD",
      SLACcitation   = "
}

@article{Gorham:2008dv,
      author         = "Gorham, P. W. and others",
      title          = "{The Antarctic Impulsive Transient Antenna Ultra-high
                        Energy Neutrino Detector Design, Performance, and
                        Sensitivity for 2006-2007 Balloon Flight}",
      collaboration  = "ANITA",
      journal        = "Astropart. Phys.",
      volume         = "32",
      year           = "2009",
      pages          = "10-41",
      doi            = "10.1016/j.astropartphys.2009.05.003",
      eprint         = "0812.1920",
      archivePrefix  = "arXiv",
      primaryClass   = "astro-ph",
      SLACcitation   = "
}

@article{Frigerio:2012uc,
      author         = "Frigerio, Michele and Pomarol, Alex and Riva, Francesco
                        and Urbano, Alfredo",
      title          = "{Composite Scalar Dark Matter}",
      journal        = "JHEP",
      volume         = "07",
      year           = "2012",
      pages          = "015",
      doi            = "10.1007/JHEP07(2012)015",
      eprint         = "1204.2808",
      archivePrefix  = "arXiv",
      primaryClass   = "hep-ph",
      SLACcitation   = "
}

@article{Benito:2016kyp,
      author         = "Benito, Maria and Bernal, Nicolas and Bozorgnia, Nassim
                        and Calore, Francesca and Iocco, Fabio",
      title          = "{Particle Dark Matter Constraints: the Effect of Galactic
                        Uncertainties}",
      journal        = "JCAP",
      volume         = "1702",
      year           = "2017",
      number         = "02",
      pages          = "007",
      doi            = "10.1088/1475-7516/2017/02/007,
                        10.1088/1475-7516/2018/06/E01",
      note           = "[Erratum: JCAP1806,no.06,E01(2018)]",
      eprint         = "1612.02010",
      archivePrefix  = "arXiv",
      primaryClass   = "hep-ph",
      SLACcitation   = "
}

@article{Bellini:2010hy,
      author         = "Bellini, G. and others",
      title          = "{Observation of Geo-Neutrinos}",
      collaboration  = "Borexino",
      journal        = "Phys. Lett.",
      volume         = "B687",
      year           = "2010",
      pages          = "299-304",
      doi            = "10.1016/j.physletb.2010.03.051",
      eprint         = "1003.0284",
      archivePrefix  = "arXiv",
      primaryClass   = "hep-ex",
      SLACcitation   = "
}

@article{Alimonti:2008gc,
      author         = "Alimonti, G. and others",
      title          = "{The Borexino detector at the Laboratori Nazionali del
                        Gran Sasso}",
      collaboration  = "Borexino",
      journal        = "Nucl. Instrum. Meth.",
      volume         = "A600",
      year           = "2009",
      pages          = "568-593",
      doi            = "10.1016/j.nima.2008.11.076",
      eprint         = "0806.2400",
      archivePrefix  = "arXiv",
      primaryClass   = "physics.ins-det",
      SLACcitation   = "
}

@article{Formaggio:2013kya,
      author         = "Formaggio, J. A. and Zeller, G. P.",
      title          = "{From eV to EeV: Neutrino Cross Sections Across Energy
                        Scales}",
      journal        = "Rev. Mod. Phys.",
      volume         = "84",
      year           = "2012",
      pages          = "1307-1341",
      doi            = "10.1103/RevModPhys.84.1307",
      eprint         = "1305.7513",
      archivePrefix  = "arXiv",
      primaryClass   = "hep-ex",
      reportNumber   = "FERMILAB-PUB-12-785-E",
      SLACcitation   = "
}

@article{Aartsen:2013jdh,
      author         = "Aartsen, M. G. and others",
      title          = "{Evidence for High-Energy Extraterrestrial Neutrinos at
                        the IceCube Detector}",
      collaboration  = "IceCube",
      journal        = "Science",
      volume         = "342",
      year           = "2013",
      pages          = "1242856",
      doi            = "10.1126/science.1242856",
      eprint         = "1311.5238",
      archivePrefix  = "arXiv",
      primaryClass   = "astro-ph.HE",
      SLACcitation   = "
}

@article{Avrorin:2014vca,
      author         = "Avrorin, A. D. and others",
      title          = "{Sensitivity of the Baikal-GVD neutrino telescope to
                        neutrino emission toward the center of the galactic dark
                        matter halo}",
      journal        = "JETP Lett.",
      volume         = "101",
      year           = "2015",
      number         = "5",
      pages          = "289-294",
      doi            = "10.1134/S0021364015050021",
      eprint         = "1412.3672",
      archivePrefix  = "arXiv",
      primaryClass   = "astro-ph.HE",
      SLACcitation   = "
}

@inproceedings{Avrorin:2019vfc,
      author         = "Avrorin, A. D. and others",
      title          = "{The Baikal-GVD neutrino telescope: First results of
                        multi-messenger studies}",
      booktitle      = "{HAWC Contributions to the 36th International Cosmic Ray
                        Conference (ICRC2019)}",
      collaboration  = "Baikal-GVD",
      year           = "2019",
      eprint         = "1908.05450",
      archivePrefix  = "arXiv",
      primaryClass   = "astro-ph.HE",
      SLACcitation   = "
}

@article{Allison:2014kha,
      author         = "Allison, P. and others",
      title          = "{First Constraints on the Ultra-High Energy Neutrino Flux
                        from a Prototype Station of the Askaryan Radio Array}",
      collaboration  = "ARA",
      journal        = "Astropart. Phys.",
      volume         = "70",
      year           = "2015",
      pages          = "62-80",
      doi            = "10.1016/j.astropartphys.2015.04.006",
      eprint         = "1404.5285",
      archivePrefix  = "arXiv",
      primaryClass   = "astro-ph.HE",
      SLACcitation   = "
}

@article{Barwick:2015ica,
      author         = "Barwick, S. W. and others",
      title          = "{Performance of the ARIANNA Hexagonal Radio Array}",
      booktitle      = "{Proceedings, 34th International Cosmic Ray Conference
                        (ICRC 2015): The Hague, The Netherlands, July 30-August 6,
                        2015}",
      collaboration  = "ARIANNA",
      journal        = "PoS",
      volume         = "ICRC2015",
      year           = "2016",
      pages          = "1149",
      doi            = "10.22323/1.236.1149",
      eprint         = "1509.00109",
      archivePrefix  = "arXiv",
      primaryClass   = "astro-ph.IM",
      SLACcitation   = "
}

@article{Alvey_2019,
	doi = {10.1088/1475-7516/2019/07/041},
  
	url = {https://doi.org/10.1088
  
	year = 2019,
	month = {jul},
  
	publisher = {{IOP} Publishing},
  
	volume = {2019},
  
	number = {07},
  
	pages = {041--041},
  
	author = {J.B.G. Alvey and M. Fairbairn},
  
	title = {Linking scalar dark matter and neutrino masses with {IceCube} 170922A},
  
	journal = {Journal of Cosmology and Astroparticle Physics}
}
@article{Patel:2019zky,
    author = "Patel, Hiren H. and Profumo, Stefano and Shakya, Bibhushan",
    title = "{Loop dominated signals from neutrino portal dark matter}",
    eprint = "1912.05581",
    archivePrefix = "arXiv",
    primaryClass = "hep-ph",
    reportNumber = "CERN-TH-2019-203",
    doi = "10.1103/PhysRevD.101.095001",
    journal = "Phys. Rev. D",
    volume = "101",
    number = "9",
    pages = "095001",
    year = "2020"
}

@article{Farzan:2012sa,
      author         = "Farzan, Yasaman and Ma, Ernest",
      title          = "{Dirac neutrino mass generation from dark matter}",
      journal        = "Phys. Rev.",
      volume         = "D86",
      year           = "2012",
      pages          = "033007",
      doi            = "10.1103/PhysRevD.86.033007",
      eprint         = "1204.4890",
      archivePrefix  = "arXiv",
      primaryClass   = "hep-ph",
      reportNumber   = "UCRHEP-T518",
      SLACcitation   = "
}

@misc{yuan_tianlu_2018_1300506,
  author       = {YUAN, Tianlu},
  title        = {{New measurements with high-energy neutrinos in 
                   IceCube}},
  month        = jun,
  year         = 2018,
  doi          = {10.5281/zenodo.1300506},
  url          = {https://doi.org/10.5281/zenodo.1300506}
}

@article{Escudero:2016tzx,
      author         = "Escudero, Miguel and Rius, Nuria and Sanz, Verónica",
      title          = "{Sterile neutrino portal to Dark Matter I: The
                        $U(1)_{B-L}$ case}",
      journal        = "JHEP",
      volume         = "02",
      year           = "2017",
      pages          = "045",
      doi            = "10.1007/JHEP02(2017)045",
      eprint         = "1606.01258",
      archivePrefix  = "arXiv",
      primaryClass   = "hep-ph",
      reportNumber   = "FTUV-16-0419, IFIC-16-32",
      SLACcitation   = "
}

@article{Escudero:2016ksa,
      author         = "Escudero, Miguel and Rius, Nuria and Sanz, Verónica",
      title          = "{Sterile Neutrino portal to Dark Matter II: Exact Dark
                        symmetry}",
      journal        = "Eur. Phys. J.",
      volume         = "C77",
      year           = "2017",
      number         = "6",
      pages          = "397",
      doi            = "10.1140/epjc/s10052-017-4963-x",
      eprint         = "1607.02373",
      archivePrefix  = "arXiv",
      primaryClass   = "hep-ph",
      SLACcitation   = "
}

@article{Tanabashi:2018oca,
      author         = "Tanabashi, M. and others",
      title          = "{Review of Particle Physics}",
      collaboration  = "Particle Data Group",
      journal        = "Phys. Rev.",
      volume         = "D98",
      year           = "2018",
      number         = "3",
      pages          = "030001",
      doi            = "10.1103/PhysRevD.98.030001",
      SLACcitation   = "
}

@article{Bertuzzo:2018itn,
      author         = "Bertuzzo, Enrico and Jana, Sudip and Machado, Pedro A. N.
                        and Zukanovich Funchal, Renata",
      title          = "{Dark Neutrino Portal to Explain MiniBooNE excess}",
      journal        = "Phys. Rev. Lett.",
      volume         = "121",
      year           = "2018",
      number         = "24",
      pages          = "241801",
      doi            = "10.1103/PhysRevLett.121.241801",
      eprint         = "1807.09877",
      archivePrefix  = "arXiv",
      primaryClass   = "hep-ph",
      reportNumber   = "FERMILAB-PUB-18-336-T, OSU-HEP-18-04",
      SLACcitation   = "
}

@article{Ballett:2018ynz,
      author         = "Ballett, Peter and Pascoli, Silvia and Ross-Lonergan,
                        Mark",
      title          = "{U(1)' mediated decays of heavy sterile neutrinos in
                        MiniBooNE}",
      journal        = "Phys. Rev.",
      volume         = "D99",
      year           = "2019",
      pages          = "071701",
      doi            = "10.1103/PhysRevD.99.071701",
      eprint         = "1808.02915",
      archivePrefix  = "arXiv",
      primaryClass   = "hep-ph",
      reportNumber   = "IPPP/18/70",
      SLACcitation   = "
}

@article{Ballett:2019cqp,
      author         = "Ballett, Peter and Hostert, Matheus and Pascoli, Silvia",
      title          = "{Neutrino Masses from a Dark Neutrino Sector below the
                        Electroweak Scale}",
      journal        = "Phys. Rev.",
      volume         = "D99",
      year           = "2019",
      number         = "9",
      pages          = "091701",
      doi            = "10.1103/PhysRevD.99.091701",
      eprint         = "1903.07590",
      archivePrefix  = "arXiv",
      primaryClass   = "hep-ph",
      reportNumber   = "IPPP/19/21",
      SLACcitation   = "
}

@article{Ballett:2019pyw,
      author         = "Ballett, Peter and Hostert, Matheus and Pascoli, Silvia",
      title          = "{Dark Neutrinos and a Three Portal Connection to the
                        Standard Model}",
      year           = "2019",
      eprint         = "1903.07589",
      archivePrefix  = "arXiv",
      primaryClass   = "hep-ph",
      reportNumber   = "IPPP/19/19",
      SLACcitation   = "
}

@article{Boehm:2006mi,
      author         = "Boehm, Celine and Farzan, Yasaman and Hambye, Thomas and
                        Palomares-Ruiz, Sergio and Pascoli, Silvia",
      title          = "{Is it possible to explain neutrino masses with scalar
                        dark matter?}",
      journal        = "Phys. Rev.",
      volume         = "D77",
      year           = "2008",
      pages          = "043516",
      doi            = "10.1103/PhysRevD.77.043516",
      eprint         = "hep-ph/0612228",
      archivePrefix  = "arXiv",
      primaryClass   = "hep-ph",
      SLACcitation   = "
}

@inproceedings{Baur:2019jwm,
      author         = "Baur, Sebastian",
      title          = "{Dark matter searches with the IceCube Upgrade}",
      booktitle      = "{HAWC Contributions to the 36th International Cosmic Ray
                        Conference (ICRC2019)}",
      collaboration  = "IceCube",
      year           = "2019",
      eprint         = "1908.08236",
      archivePrefix  = "arXiv",
      primaryClass   = "astro-ph.HE",
      reportNumber   = "PoS-ICRC2019-506",
      SLACcitation   = "
}

@article{Agostini:2018uly,
      author         = "Agostini, M. and others",
      title          = "{Comprehensive measurement of $pp$-chain solar
                        neutrinos}",
      collaboration  = "BOREXINO",
      journal        = "Nature",
      volume         = "562",
      year           = "2018",
      number         = "7728",
      pages          = "505-510",
      doi            = "10.1038/s41586-018-0624-y",
      reportNumber   = "FERMILAB-PUB-18-592-ND",
      SLACcitation   = "
}

@article{Aartsen:2014gkd,
      author         = "Aartsen, M. G. and others",
      title          = "{Observation of High-Energy Astrophysical Neutrinos in
                        Three Years of IceCube Data}",
      collaboration  = "IceCube",
      journal        = "Phys. Rev. Lett.",
      volume         = "113",
      year           = "2014",
      pages          = "101101",
      doi            = "10.1103/PhysRevLett.113.101101",
      eprint         = "1405.5303",
      archivePrefix  = "arXiv",
      primaryClass   = "astro-ph.HE",
      SLACcitation   = "
}

@inproceedings{Schneider:2019ayi,
      author         = "Schneider, Austin",
      title          = "{Characterization of the Astrophysical Diffuse Neutrino
                        Flux with IceCube High-Energy Starting Events}",
      booktitle      = "{36th International Cosmic Ray Conference (ICRC 2019)
                        Madison, Wisconsin, USA, July 24-August 1, 2019}",
      year           = "2019",
      eprint         = "1907.11266",
      archivePrefix  = "arXiv",
      primaryClass   = "astro-ph.HE",
      reportNumber   = "PoS-ICRC2019-1004",
      SLACcitation   = "
}

@article{Tinker:2008ff,
      author         = "Tinker, Jeremy L. and Kravtsov, Andrey V. and Klypin,
                        Anatoly and Abazajian, Kevork and Warren, Michael S. and
                        Yepes, Gustavo and Gottlober, Stefan and Holz, Daniel E.",
      title          = "{Toward a halo mass function for precision cosmology: The
                        Limits of universality}",
      journal        = "Astrophys. J.",
      volume         = "688",
      year           = "2008",
      pages          = "709-728",
      doi            = "10.1086/591439",
      eprint         = "0803.2706",
      archivePrefix  = "arXiv",
      primaryClass   = "astro-ph",
      SLACcitation   = "
}

@article{Sheth:1999su,
      author         = "Sheth, Ravi K. and Mo, H. J. and Tormen, Giuseppe",
      title          = "{Ellipsoidal collapse and an improved model for the
                        number and spatial distribution of dark matter haloes}",
      journal        = "Mon. Not. Roy. Astron. Soc.",
      volume         = "323",
      year           = "2001",
      pages          = "1",
      doi            = "10.1046/j.1365-8711.2001.04006.x",
      eprint         = "astro-ph/9907024",
      archivePrefix  = "arXiv",
      primaryClass   = "astro-ph",
      SLACcitation   = "
}

@article{Sheth:1999mn,
      author         = "Sheth, Ravi K. and Tormen, Giuseppe",
      title          = "{Large scale bias and the peak background split}",
      journal        = "Mon. Not. Roy. Astron. Soc.",
      volume         = "308",
      year           = "1999",
      pages          = "119",
      doi            = "10.1046/j.1365-8711.1999.02692.x",
      eprint         = "astro-ph/9901122",
      archivePrefix  = "arXiv",
      primaryClass   = "astro-ph",
      SLACcitation   = "
}

@ARTICLE{1991ApJ...379..440B,
       author = {{Bond}, J.~R. and {Cole}, S. and {Efstathiou}, G. and {Kaiser}, N.},
        title = "{Excursion Set Mass Functions for Hierarchical Gaussian Fluctuations}",
      journal = {\apj},
     keywords = {Computational Astrophysics, Gauss Equation, Dark Matter, Density Distribution, Many Body Problem, Mass Distribution, Monte Carlo Method, Astrophysics, COSMOLOGY, GALAXIES: CLUSTERING, NUMERICAL METHODS},
         year = "1991",
        month = "Oct",
       volume = {379},
        pages = {440},
          doi = {10.1086/170520},
       adsurl = {https://ui.adsabs.harvard.edu/abs/1991ApJ...379..440B},
      adsnote = {Provided by the SAO/NASA Astrophysics Data System}
}

@ARTICLE{1974ApJ...187..425P,
       author = {{Press}, William H. and {Schechter}, Paul},
        title = "{Formation of Galaxies and Clusters of Galaxies by Self-Similar Gravitational Condensation}",
      journal = {\apj},
         year = "1974",
        month = "Feb",
       volume = {187},
        pages = {425-438},
          doi = {10.1086/152650},
       adsurl = {https://ui.adsabs.harvard.edu/abs/1974ApJ...187..425P},
      adsnote = {Provided by the SAO/NASA Astrophysics Data System}
}

@

@article{Bedard:2018zml,
    author = "Bedard, J. and others",
    collaboration = "STRAW",
    title = "{STRAW (STRings for Absorption length in Water): pathfinder for a neutrino telescope in the deep Pacific Ocean}",
    eprint = "1810.13265",
    archivePrefix = "arXiv",
    primaryClass = "astro-ph.IM",
    doi = "10.1088/1748-0221/14/02/P02013",
    journal = "JINST",
    volume = "14",
    number = "02",
    pages = "P02013",
    year = "2019"
}

@article{Aab:2015kma,
      author         = "Aab, Alexander and others",
      title          = "{Improved limit to the diffuse flux of ultrahigh energy
                        neutrinos from the Pierre Auger Observatory}",
      collaboration  = "Pierre Auger",
      journal        = "Phys. Rev.",
      volume         = "D91",
      year           = "2015",
      number         = "9",
      pages          = "092008",
      doi            = "10.1103/PhysRevD.91.092008",
      eprint         = "1504.05397",
      archivePrefix  = "arXiv",
      primaryClass   = "astro-ph.HE",
      reportNumber   = "FERMILAB-PUB-15-150-AD-AE-CD-TD",
      SLACcitation   = "
}

@article{PhysRevD.92.023004,
  title = {Atmospheric neutrino flux calculation using the NRLMSISE-00 atmospheric model},
  author = {Honda, M. and Athar, M. Sajjad and Kajita, T. and Kasahara, K. and Midorikawa, S.},
  journal = {Phys. Rev. D},
  volume = {92},
  issue = {2},
  pages = {023004},
  numpages = {15},
  year = {2015},
  month = {Jul},
  publisher = {American Physical Society},
  doi = {10.1103/PhysRevD.92.023004},
  url = {https://link.aps.org/doi/10.1103/PhysRevD.92.023004}
}

@article{Wissel:2019alx,
      author         = "Wissel, S. and others",
      title          = "{A New Concept for High-Elevation Radio Detection of Tau
                        Neutrinos}",
      booktitle      = "{Proceedings for ARENA 2018: Catania, Italy, June 12-15,
                        2018}",
      journal        = "EPJ Web Conf.",
      volume         = "216",
      year           = "2019",
      pages          = "04007",
      doi            = "10.1051/epjconf/201921604007",
      SLACcitation   = "
}

@article{Safa:2019ege,
      author         = "Safa, Ibrahim and Pizzuto, Alex and Arg{\"u}elles, Carlos A.
                        and Halzen, Francis and Hussain, Raamis and Kheirandish,
                        Ali and Vandenbroucke, Justin",
      title          = "{Observing EeV neutrinos through the Earth: GZK and the
                        anomalous ANITA events}",
      year           = "2019",
      eprint         = "1909.10487",
      archivePrefix  = "arXiv",
      primaryClass   = "hep-ph",
      SLACcitation   = "
}

@ARTICLE{2014arXiv1409.0477H,
       author = {{Hou}, George W. -S.},
        title = "{Neutrino Telescope Array (NTA) - Towards Survey of Astronomical $\nu_\tau$ Sources}",
      journal = {arXiv e-prints},
     keywords = {Astrophysics - Instrumentation and Methods for Astrophysics},
         year = "2014",
        month = "Aug",
          eid = {arXiv:1409.0477},
        pages = {arXiv:1409.0477},
archivePrefix = {arXiv},
       eprint = {1409.0477},
 primaryClass = {astro-ph.IM},
       adsurl = {https://ui.adsabs.harvard.edu/abs/2014arXiv1409.0477H},
      adsnote = {Provided by the SAO/NASA Astrophysics Data System}
}

@article{Profumo:2017obk,
      author         = "Profumo, Stefano and Queiroz, Farinaldo S. and Silk,
                        Joseph and Siqueira, Clarissa",
      title          = "{Searching for Secluded Dark Matter with H.E.S.S.,
                        Fermi-LAT, and Planck}",
      journal        = "JCAP",
      volume         = "1803",
      year           = "2018",
      number         = "03",
      pages          = "010",
      doi            = "10.1088/1475-7516/2018/03/010",
      eprint         = "1711.03133",
      archivePrefix  = "arXiv",
      primaryClass   = "hep-ph",
      SLACcitation   = "
}

@phdthesis{ElAisati:2018vkn,
      author         = "El Aisati, Chaimae",
      title          = "{Gamma-ray and Neutrino Lines from Dark Matter:
                        multi-messenger and dedicated smoking-gun searches}",
      school         = "U. Brussels (main)",
      year           = "2018-01-18",
      url            = "http://difusion.ulb.ac.be/vufind/Record/ULB-DIPOT:oai:dipot.ulb.ac.be:2013/266180/Holdings",
      SLACcitation   = "
}

@article{Arcadi:2017kky,
      author         = "Arcadi, Giorgio and Dutra, Maíra and Ghosh, Pradipta and
                        Lindner, Manfred and Mambrini, Yann and Pierre, Mathias
                        and Profumo, Stefano and Queiroz, Farinaldo S.",
      title          = "{The waning of the WIMP? A review of models, searches,
                        and constraints}",
      journal        = "Eur. Phys. J.",
      volume         = "C78",
      year           = "2018",
      number         = "3",
      pages          = "203",
      doi            = "10.1140/epjc/s10052-018-5662-y",
      eprint         = "1703.07364",
      archivePrefix  = "arXiv",
      primaryClass   = "hep-ph",
      SLACcitation   = "
}

@article{Horiuchi:2008jz,
      author         = "Horiuchi, Shunsaku and Beacom, John F. and Dwek, Eli",
      title          = "{The Diffuse Supernova Neutrino Background is detectable
                        in Super-Kamiokande}",
      journal        = "Phys. Rev.",
      volume         = "D79",
      year           = "2009",
      pages          = "083013",
      doi            = "10.1103/PhysRevD.79.083013",
      eprint         = "0812.3157",
      archivePrefix  = "arXiv",
      primaryClass   = "astro-ph",
      SLACcitation   = "
}

@phdthesis{Olivares-Del-Campo:2019qwe,
      author         = "Olivares-Del-Campo, Andres",
      title          = "{Dark Matter and Neutrinos: A Love-Hate Relationship}",
      school         = "Durham U.",
      year           = "2019",
      url            = "http://etheses.dur.ac.uk/13142/",
      SLACcitation   = "
}

@article{Belolaptikov:1997ry,
      author         = "Belolaptikov, I. A. and others",
      title          = "{The Baikal underwater neutrino telescope: Design,
                        performance and first results}",
      collaboration  = "BAIKAL",
      journal        = "Astropart. Phys.",
      volume         = "7",
      year           = "1997",
      pages          = "263-282",
      doi            = "10.1016/S0927-6505(97)00022-4",
      reportNumber   = "DESY-97-033, BAIKAL-NOTE-97-01",
      SLACcitation   = "
}

@article{Aynutdinov:2006ca,
      author         = "Aynutdinov, V. M. and others",
      title          = "{Baikal neutrino telescope}",
      booktitle      = "{Third NO-VE International Workshop on Neutrino
                        Oscillations in Venice : Fifty years after the neutrino
                        esperimental discovery : Venezia, February 7-10, 2006,
                        Istituto Veneto di Scienze, Lettere ed Arti, Campo Santo
                        Stefano}",
      journal        = "Phys. Atom. Nucl.",
      volume         = "69",
      year           = "2006",
      pages          = "1914-1921",
      doi            = "10.1134/S1063778806110160",
      note           = "[,449(2006)]",
      SLACcitation   = "
}

@article{Barkana:2018lgd,
      author         = "Barkana, Rennan",
      title          = "{Possible interaction between baryons and dark-matter
                        particles revealed by the first stars}",
      journal        = "Nature",
      volume         = "555",
      year           = "2018",
      number         = "7694",
      pages          = "71-74",
      doi            = "10.1038/nature25791",
      eprint         = "1803.06698",
      archivePrefix  = "arXiv",
      primaryClass   = "astro-ph.CO",
      SLACcitation   = "
}

@article{Avrorin:2015bct,
      author         = "Avrorin, A. D. and others",
      title          = "{A search for neutrino signal from dark matter
                        annihilation in the center of the Milky Way with Baikal
                        NT200}",
      collaboration  = "BAIKAL",
      journal        = "Astropart. Phys.",
      volume         = "81",
      year           = "2016",
      pages          = "12-20",
      doi            = "10.1016/j.astropartphys.2016.04.004",
      eprint         = "1512.01198",
      archivePrefix  = "arXiv",
      primaryClass   = "astro-ph.HE",
      SLACcitation   = "
}

@article{Arguelles:2018mtc,
      author         = "Arg{\"u}elles, Carlos A. and Hostert, Matheus and Tsai,
                        Yu-Dai",
      title          = "{Testing New Physics Explanations of MiniBooNE Anomaly at
                        Neutrino Scattering Experiments}",
      year           = "2018",
      eprint         = "1812.08768",
      archivePrefix  = "arXiv",
      primaryClass   = "hep-ph",
      reportNumber   = "IPPP/18/113/FERMILAB-PUB-18-686-A-ND-PPD-T, IPPP/18/113,
                        FERMILAB-PUB-18-686-A-ND-PPD-T",
      SLACcitation   = "
}

@article{Arguelles:2017atb,
      author         = "Arg{\"u}elles, Carlos A. and Kheirandish, Ali and Vincent,
                        Aaron C.",
      title          = "{Imaging Galactic Dark Matter with High-Energy Cosmic
                        Neutrinos}",
      journal        = "Phys. Rev. Lett.",
      volume         = "119",
      year           = "2017",
      number         = "20",
      pages          = "201801",
      doi            = "10.1103/PhysRevLett.119.201801",
      eprint         = "1703.00451",
      archivePrefix  = "arXiv",
      primaryClass   = "hep-ph",
      SLACcitation   = "
}

@article{Farzan:2018pnk,
      author         = "Farzan, Yasaman and Palomares-Ruiz, Sergio",
      title          = "{Flavor of cosmic neutrinos preserved by ultralight dark
                        matter}",
      journal        = "Phys. Rev.",
      volume         = "D99",
      year           = "2019",
      number         = "5",
      pages          = "051702",
      doi            = "10.1103/PhysRevD.99.051702",
      eprint         = "1810.00892",
      archivePrefix  = "arXiv",
      primaryClass   = "hep-ph",
      reportNumber   = "IFIC/18-36",
      SLACcitation   = "
}

@article{Pandey:2018wvh,
      author         = "Pandey, Sujata and Karmakar, Siddhartha and Rakshit,
                        Subhendu",
      title          = "{Interactions of Astrophysical Neutrinos with Dark
                        Matter: A model building perspective}",
      journal        = "JHEP",
      volume         = "01",
      year           = "2019",
      pages          = "095",
      doi            = "10.1007/JHEP01(2019)095",
      eprint         = "1810.04203",
      archivePrefix  = "arXiv",
      primaryClass   = "hep-ph",
      SLACcitation   = "
}

@article{Choi:2019ixb,
      author         = "Choi, Ki-Young and Kim, Jongkuk and Rott, Carsten",
      title          = "{Constraining dark matter-neutrino interactions with
                        IceCube-170922A}",
      journal        = "Phys. Rev.",
      volume         = "D99",
      year           = "2019",
      number         = "8",
      pages          = "083018",
      doi            = "10.1103/PhysRevD.99.083018",
      eprint         = "1903.03302",
      archivePrefix  = "arXiv",
      primaryClass   = "astro-ph.CO",
      SLACcitation   = "
}

@phdthesis{Hostert:2019iia,
      author         = "Hostert, Matheus.",
      title          = "{Hidden Physics at the Neutrino Frontier: Tridents, Dark
                        Forces, and Hidden Particles}",
      school         = "Durham U.",
      year           = "2019",
      url            = "http://etheses.dur.ac.uk/13289/",
      SLACcitation   = "
}

@article{Barenboim:2019tux,
      author         = "Barenboim, Gabriela and Denton, Peter B. and Oldengott,
                        Isabel M.",
      title          = "{Constraints on inflation with an extended neutrino
                        sector}",
      journal        = "Phys. Rev.",
      volume         = "D99",
      year           = "2019",
      number         = "8",
      pages          = "083515",
      doi            = "10.1103/PhysRevD.99.083515",
      eprint         = "1903.02036",
      archivePrefix  = "arXiv",
      primaryClass   = "astro-ph.CO",
      SLACcitation   = "
}

@article{Kelly:2018tyg,
      author         = "Kelly, Kevin J. and Machado, Pedro A. N.",
      title          = "{Multimessenger Astronomy and New Neutrino Physics}",
      journal        = "JCAP",
      volume         = "1810",
      year           = "2018",
      pages          = "048",
      doi            = "10.1088/1475-7516/2018/10/048",
      eprint         = "1808.02889",
      archivePrefix  = "arXiv",
      primaryClass   = "hep-ph",
      reportNumber   = "FERMILAB-PUB-18-374-T, NUHEP-TH/18-07",
      SLACcitation   = "
}

@article{Capozzi:2018bps,
      author         = "Capozzi, Francesco and Shoemaker, Ian M. and Vecchi,
                        Luca",
      title          = "{Neutrino Oscillations in Dark Backgrounds}",
      journal        = "JCAP",
      volume         = "1807",
      year           = "2018",
      number         = "07",
      pages          = "004",
      doi            = "10.1088/1475-7516/2018/07/004",
      eprint         = "1804.05117",
      archivePrefix  = "arXiv",
      primaryClass   = "hep-ph",
      SLACcitation   = "
}

@article{Capozzi:2017auw,
      author         = "Capozzi, Francesco and Shoemaker, Ian M. and Vecchi,
                        Luca",
      title          = "{Solar Neutrinos as a Probe of Dark Matter-Neutrino
                        Interactions}",
      journal        = "JCAP",
      volume         = "1707",
      year           = "2017",
      number         = "07",
      pages          = "021",
      doi            = "10.1088/1475-7516/2017/07/021",
      eprint         = "1702.08464",
      archivePrefix  = "arXiv",
      primaryClass   = "hep-ph",
      SLACcitation   = "
}

@article{Cherry:2016jol,
      author         = "Cherry, John F. and Friedland, Alexander and Shoemaker,
                        Ian M.",
      title          = "{Short-baseline neutrino oscillations, Planck, and
                        IceCube}",
      year           = "2016",
      eprint         = "1605.06506",
      archivePrefix  = "arXiv",
      primaryClass   = "hep-ph",
      reportNumber   = "SLAC-PUB-16517",
      SLACcitation   = "
}

@article{Davis:2015rza,
      author         = "Davis, Jonathan H. and Silk, Joseph",
      title          = "{Spectral and Spatial Distortions of PeV Neutrinos from
                        Scattering with Dark Matter}",
      year           = "2015",
      eprint         = "1505.01843",
      archivePrefix  = "arXiv",
      primaryClass   = "hep-ph",
      SLACcitation   = "
}

@article{Aguilar-Arevalo:2017mqx,
      author         = "Aguilar-Arevalo, A. A. and others",
      title          = "{Dark Matter Search in a Proton Beam Dump with
                        MiniBooNE}",
      collaboration  = "MiniBooNE",
      journal        = "Phys. Rev. Lett.",
      volume         = "118",
      year           = "2017",
      number         = "22",
      pages          = "221803",
      doi            = "10.1103/PhysRevLett.118.221803",
      eprint         = "1702.02688",
      archivePrefix  = "arXiv",
      primaryClass   = "hep-ex",
      reportNumber   = "FERMILAB-PUB-17-059-AD-E-ND",
      SLACcitation   = "
}

@article{Migenda:2017tej,
      author         = "Migenda, Jost",
      title          = "{Astroparticle Physics in Hyper-Kamiokande}",
      booktitle      = "{Proceedings, 2017 European Physical Society Conference
                        on High Energy Physics (EPS-HEP 2017): Venice, Italy, July
                        5-12, 2017}",
      collaboration  = "Hyper-Kamiokande Proto",
      journal        = "PoS",
      volume         = "EPS-HEP2017",
      year           = "2017",
      pages          = "020",
      doi            = "10.22323/1.314.0020",
      eprint         = "1710.08345",
      archivePrefix  = "arXiv",
      primaryClass   = "physics.ins-det",
      SLACcitation   = "
}

@article{He:1990pn,
      author         = "He, X.-G. and Joshi, Girish C. and Lew, H. and Volkas, R.
                        R.",
      title          = "{NEW Z-prime PHENOMENOLOGY}",
      journal        = "Phys. Rev.",
      volume         = "D43",
      year           = "1991",
      pages          = "22-24",
      doi            = "10.1103/PhysRevD.43.R22",
      reportNumber   = "UM-P-90/42, OZ-P-90/16",
      SLACcitation   = "
}

@article{He:1991qd,
      author         = "He, Xiao-Gang and Joshi, Girish C. and Lew, H. and
                        Volkas, R. R.",
      title          = "{Simplest Z-prime model}",
      journal        = "Phys. Rev.",
      volume         = "D44",
      year           = "1991",
      pages          = "2118-2132",
      doi            = "10.1103/PhysRevD.44.2118",
      reportNumber   = "CERN-TH-6084-91, UM-P-91-32, OZ-91-07",
      SLACcitation   = "
}

@phdthesis{WanLinyan:2018,
  title={{Experimental Studies on Low Energy Electron Antineutrinos and Related Physics}},
  author={Linyan, Wan},
  year={2018},
  school={Tsinghua University},
  url={http://www-sk.icrr.u-tokyo.ac.jp/sk/_pdf/articles/2019/SKver-Linyan.pdf?fbclid=IwAR0EgdX5Clr1tFJI1tXV2W4YjnDl2wfiyZ42OnaRGeJRnvJxGsUu6Mi--dA}
}

@article{Wilkinson:2016gsy,
      author         = "Wilkinson, Ryan J. and Vincent, Aaron C. and Boehm,
                        C\'line and McCabe, Christopher",
      title          = "{Ruling out the light weakly interacting massive particle
                        explanation of the Galactic 511 keV line}",
      journal        = "Phys. Rev.",
      volume         = "D94",
      year           = "2016",
      number         = "10",
      pages          = "103525",
      doi            = "10.1103/PhysRevD.94.103525",
      eprint         = "1602.01114",
      archivePrefix  = "arXiv",
      primaryClass   = "astro-ph.CO",
      reportNumber   = "IPPP-16-09",
      SLACcitation   = "
}
@article{Ho:2012ug,
      author         = "Ho, Chiu Man and Scherrer, Robert J.",
      title          = "{Limits on MeV Dark Matter from the Effective Number of
                        Neutrinos}",
      journal        = "Phys. Rev.",
      volume         = "D87",
      year           = "2013",
      number         = "2",
      pages          = "023505",
      doi            = "10.1103/PhysRevD.87.023505",
      eprint         = "1208.4347",
      archivePrefix  = "arXiv",
      primaryClass   = "astro-ph.CO",
      SLACcitation   = "
}
@article{Steigman:2013yua,
      author         = "Steigman, Gary",
      title          = "{Equivalent Neutrinos, Light WIMPs, and the Chimera of
                        Dark Radiation}",
      journal        = "Phys. Rev.",
      volume         = "D87",
      year           = "2013",
      number         = "10",
      pages          = "103517",
      doi            = "10.1103/PhysRevD.87.103517",
      eprint         = "1303.0049",
      archivePrefix  = "arXiv",
      primaryClass   = "astro-ph.CO",
      SLACcitation   = "
}

@article{Boehm:2012gr,
      author         = "Boehm, Celine and Dolan, Matthew J. and McCabe,
                        Christopher",
      title          = "{Increasing Neff with particles in thermal equilibrium
                        with neutrinos}",
      journal        = "JCAP",
      volume         = "1212",
      year           = "2012",
      pages          = "027",
      doi            = "10.1088/1475-7516/2012/12/027",
      eprint         = "1207.0497",
      archivePrefix  = "arXiv",
      primaryClass   = "astro-ph.CO",
      reportNumber   = "IPPP-12-44, DCPT-12-88",
      SLACcitation   = "
}

@article{Serpico:2004nm,
      author         = "Serpico, Pasquale Dario and Raffelt, Georg G.",
      title          = "{MeV-mass dark matter and primordial nucleosynthesis}",
      journal        = "Phys. Rev.",
      volume         = "D70",
      year           = "2004",
      pages          = "043526",
      doi            = "10.1103/PhysRevD.70.043526",
      eprint         = "astro-ph/0403417",
      archivePrefix  = "arXiv",
      primaryClass   = "astro-ph",
      reportNumber   = "MPP-2004-34",
      SLACcitation   = "
}

@article{Kolb:1986nf,
      author         = "Kolb, Edward W. and Turner, Michael S. and Walker,
                        Terrence P.",
      title          = "{The Effect of Interacting Particles on Primordial
                        Nucleosynthesis}",
      journal        = "Phys. Rev.",
      volume         = "D34",
      year           = "1986",
      pages          = "2197",
      doi            = "10.1103/PhysRevD.34.2197",
      reportNumber   = "FERMILAB-PUB-86-077-A",
      SLACcitation   = "
}

@article{Nollett:2013pwa,
      author         = "Nollett, Kenneth M. and Steigman, Gary",
      title          = "{BBN And The CMB Constrain Light, Electromagnetically
                        Coupled WIMPs}",
      journal        = "Phys. Rev.",
      volume         = "D89",
      year           = "2014",
      number         = "8",
      pages          = "083508",
      doi            = "10.1103/PhysRevD.89.083508",
      eprint         = "1312.5725",
      archivePrefix  = "arXiv",
      primaryClass   = "astro-ph.CO",
      SLACcitation   = "
}

@article{Boehm:2013jpa,
      author         = "Boehm, C\'eline and Dolan, Matthew J. and McCabe,
                        Christopher",
      title          = "{A Lower Bound on the Mass of Cold Thermal Dark Matter
                        from Planck}",
      journal        = "JCAP",
      volume         = "1308",
      year           = "2013",
      pages          = "041",
      doi            = "10.1088/1475-7516/2013/08/041",
      eprint         = "1303.6270",
      archivePrefix  = "arXiv",
      primaryClass   = "hep-ph",
      reportNumber   = "IPPP-13-18, DCPT-13-36",
      SLACcitation   = "
}
@article{Nollett:2014lwa,
      author         = "Nollett, Kenneth M. and Steigman, Gary",
      title          = "{BBN And The CMB Constrain Neutrino Coupled Light WIMPs}",
      journal        = "Phys. Rev.",
      volume         = "D91",
      year           = "2015",
      number         = "8",
      pages          = "083505",
      doi            = "10.1103/PhysRevD.91.083505",
      eprint         = "1411.6005",
      archivePrefix  = "arXiv",
      primaryClass   = "astro-ph.CO",
      SLACcitation   = "
}

@article{Steigman:2014pfa,
      author         = "Steigman, Gary and Nollett, Kenneth M.",
      title          = "{Light WIMPs, Equivalent Neutrinos, BBN, and the CMB}",
      journal        = "Mem. Soc. Ast. It.",
      volume         = "85",
      year           = "2014",
      pages          = "175",
      eprint         = "1401.5488",
      archivePrefix  = "arXiv",
      primaryClass   = "astro-ph.CO",
      SLACcitation   = "
}

@article{Cappiello:2019qsw,
      author         = "Cappiello, Christopher and Beacom, John F.",
      title          = "{Strong New Limits on Light Dark Matter from Neutrino
                        Experiments}",
      journal        = "Phys. Rev.",
      volume         = "D100",
      year           = "2019",
      number         = "10",
      pages          = "103011",
      doi            = "10.1103/PhysRevD.100.103011",
      eprint         = "1906.11283",
      archivePrefix  = "arXiv",
      primaryClass   = "hep-ph",
      SLACcitation   = "
}

@article{Smirnov:2019ngs,
      author         = "Smirnov, Juri and Beacom, John F.",
      title          = "{TeV-Scale Thermal WIMPs: Unitarity and its
                        Consequences}",
      journal        = "Phys. Rev.",
      volume         = "D100",
      year           = "2019",
      number         = "4",
      pages          = "043029",
      doi            = "10.1103/PhysRevD.100.043029",
      eprint         = "1904.11503",
      archivePrefix  = "arXiv",
      primaryClass   = "hep-ph",
      SLACcitation   = "
}

@article{Li:2014sea,
      author         = "Li, Shirley Weishi and Beacom, John F.",
      title          = "{First calculation of cosmic-ray muon spallation
                        backgrounds for MeV astrophysical neutrino signals in
                        Super-Kamiokande}",
      journal        = "Phys. Rev.",
      volume         = "C89",
      year           = "2014",
      pages          = "045801",
      doi            = "10.1103/PhysRevC.89.045801",
      eprint         = "1402.4687",
      archivePrefix  = "arXiv",
      primaryClass   = "hep-ph",
      SLACcitation   = "
}

@article{Laha:2013hva,
      author         = "Laha, Ranjan and Beacom, John F.",
      title          = "{Gadolinium in water Cherenkov detectors improves
                        detection of supernova $\nu_e$}",
      journal        = "Phys. Rev.",
      volume         = "D89",
      year           = "2014",
      pages          = "063007",
      doi            = "10.1103/PhysRevD.89.063007",
      eprint         = "1311.6407",
      archivePrefix  = "arXiv",
      primaryClass   = "astro-ph.HE",
      SLACcitation   = "
}

@article{Beacom:2004jb,
      author         = "Beacom, John F. and Candia, Julian",
      title          = "{Shower power: Isolating the prompt atmospheric neutrino
                        flux using electron neutrinos}",
      journal        = "JCAP",
      volume         = "0411",
      year           = "2004",
      pages          = "009",
      doi            = "10.1088/1475-7516/2004/11/009",
      eprint         = "hep-ph/0409046",
      archivePrefix  = "arXiv",
      primaryClass   = "hep-ph",
      reportNumber   = "FERMILAB-PUB-04-188-A-T",
      SLACcitation   = "
}

@article{Bowman:2018yin,
      author         = "Bowman, Judd D. and Rogers, Alan E. E. and Monsalve, Raul
                        A. and Mozdzen, Thomas J. and Mahesh, Nivedita",
      title          = "{An absorption profile centred at 78 megahertz in the
                        sky-averaged spectrum}",
      journal        = "Nature",
      volume         = "555",
      year           = "2018",
      number         = "7694",
      pages          = "67-70",
      doi            = "10.1038/nature25792",
      eprint         = "1810.05912",
      archivePrefix  = "arXiv",
      primaryClass   = "astro-ph.CO",
      SLACcitation   = "
}

@article{Pospelov:2018kdh,
      author         = "Pospelov, Maxim and Pradler, Josef and Ruderman, Joshua
                        T. and Urbano, Alfredo",
      title          = "{Room for New Physics in the Rayleigh-Jeans Tail of the
                        Cosmic Microwave Background}",
      journal        = "Phys. Rev. Lett.",
      volume         = "121",
      year           = "2018",
      number         = "3",
      pages          = "031103",
      doi            = "10.1103/PhysRevLett.121.031103",
      eprint         = "1803.07048",
      archivePrefix  = "arXiv",
      primaryClass   = "hep-ph",
      reportNumber   = "CERN-TH-2018-059",
      SLACcitation   = "
}

@article{Berlin:2018sjs,
      author         = "Berlin, Asher and Hooper, Dan and Krnjaic, Gordan and
                        McDermott, Samuel D.",
      title          = "{Severely Constraining Dark Matter Interpretations of the
                        21-cm Anomaly}",
      journal        = "Phys. Rev. Lett.",
      volume         = "121",
      year           = "2018",
      number         = "1",
      pages          = "011102",
      doi            = "10.1103/PhysRevLett.121.011102",
      eprint         = "1803.02804",
      archivePrefix  = "arXiv",
      primaryClass   = "hep-ph",
      reportNumber   = "FERMILAB-PUB-18-066-A",
      SLACcitation   = "
}

@article{Munoz:2018pzp,
      author         = "Mu{\~n}oz, Julian B. and Loeb, Abraham",
      title          = "{A small amount of mini-charged dark matter could cool
                        the baryons in the early Universe}",
      journal        = "Nature",
      volume         = "557",
      year           = "2018",
      number         = "7707",
      pages          = "684",
      doi            = "10.1038/s41586-018-0151-x",
      eprint         = "1802.10094",
      archivePrefix  = "arXiv",
      primaryClass   = "astro-ph.CO",
      SLACcitation   = "
}

@article{Hagedorn:2018spx,
      author         = "Hagedorn, Claudia and Herrero-Garc{\'i}a, Juan and Molinaro,
                        Emiliano and Schmidt, Michael A.",
      title          = "{Phenomenology of the Generalised Scotogenic Model with
                        Fermionic Dark Matter}",
      journal        = "JHEP",
      volume         = "11",
      year           = "2018",
      pages          = "103",
      doi            = "10.1007/JHEP11(2018)103",
      eprint         = "1804.04117",
      archivePrefix  = "arXiv",
      primaryClass   = "hep-ph",
      reportNumber   = "ADP-18-9-T1057, CP3-ORIGINS-2018-012, ADP-18-9/T1057,
                        CP3-Origins-2018-012-DNRF90",
      SLACcitation   = "
}

@article{Goodenough:2009gk,
      author         = "Goodenough, Lisa and Hooper, Dan",
      title          = "{Possible Evidence For Dark Matter Annihilation In The
                        Inner Milky Way From The Fermi Gamma Ray Space Telescope}",
      year           = "2009",
      eprint         = "0910.2998",
      archivePrefix  = "arXiv",
      primaryClass   = "hep-ph",
      reportNumber   = "FERMILAB-PUB-09-494-A",
      SLACcitation   = "
}

@article{Hooper:2010mq,
      author         = "Hooper, Dan and Goodenough, Lisa",
      title          = "{Dark Matter Annihilation in The Galactic Center As Seen
                        by the Fermi Gamma Ray Space Telescope}",
      journal        = "Phys. Lett.",
      volume         = "B697",
      year           = "2011",
      pages          = "412-428",
      doi            = "10.1016/j.physletb.2011.02.029",
      eprint         = "1010.2752",
      archivePrefix  = "arXiv",
      primaryClass   = "hep-ph",
      reportNumber   = "FERMILAB-PUB-10-414-A",
      SLACcitation   = "
}

@article{Boudaud:2019efq,
    author = "Boudaud, Mathieu and Génolini, Yoann and Derome, Laurent and Lavalle, Julien and Maurin, David and Salati, Pierre and Serpico, Pasquale D.",
    title = "{AMS-02 antiprotons are consistent with a secondary astrophysical origin}",
    eprint = "1906.07119",
    archivePrefix = "arXiv",
    primaryClass = "astro-ph.HE",
    doi = "10.1103/PhysRevResearch.2.023022",
    journal = "Phys. Rev. Res.",
    volume = "2",
    pages = "023022",
    year = "2020"
}

@article{Heisig:2020nse,
    author = "Heisig, Jan and Korsmeier, Michael and Winkler, Martin Wolfgang",
    title = "{Dark matter or correlated errors? Systematics of the AMS-02 antiproton excess}",
    eprint = "2005.04237",
    archivePrefix = "arXiv",
    primaryClass = "astro-ph.HE",
    reportNumber = "CP3-20-19, TTK-20-14",
    month = "5",
    year = "2020"
}

@article{Leane:2019uhc,
      author         = "Leane, Rebecca K. and Slatyer, Tracy R.",
      title          = "{Revival of the Dark Matter Hypothesis for the Galactic
                        Center Gamma-Ray Excess}",
      journal        = "Phys. Rev. Lett.",
      volume         = "123",
      year           = "2019",
      number         = "24",
      pages          = "241101",
      doi            = "10.1103/PhysRevLett.123.241101",
      SLACcitation   = "
}

@ARTICLE{1937ApJ....86..217Z,
       author = {{Zwicky}, F.},
        title = "{On the Masses of Nebulae and of Clusters of Nebulae}",
      journal = {\apj},
         year = 1937,
        month = oct,
       volume = {86},
        pages = {217},
          doi = {10.1086/143864},
       adsurl = {https://ui.adsabs.harvard.edu/abs/1937ApJ....86..217Z},
      adsnote = {Provided by the SAO/NASA Astrophysics Data System}
}

@ARTICLE{1936ApJ....83...23S,
       author = {{Smith}, Sinclair},
        title = "{The Mass of the Virgo Cluster}",
      journal = {\apj},
         year = 1936,
        month = jan,
       volume = {83},
        pages = {23},
          doi = {10.1086/143697},
       adsurl = {https://ui.adsabs.harvard.edu/abs/1936ApJ....83...23S},
      adsnote = {Provided by the SAO/NASA Astrophysics Data System}
}

@article{Bhattacharya:2019ucd,
      author         = "Bhattacharya, Atri and Esmaili, Arman and Palomares-Ruiz,
                        Sergio and Sarcevic, Ina",
      title          = "{Update on decaying and annihilating heavy dark matter
                        with the 6-year IceCube HESE data}",
      journal        = "JCAP",
      volume         = "1905",
      year           = "2019",
      number         = "05",
      pages          = "051",
      doi            = "10.1088/1475-7516/2019/05/051",
      eprint         = "1903.12623",
      archivePrefix  = "arXiv",
      primaryClass   = "hep-ph",
      reportNumber   = "IFIC/19-19, IFIC-19-19",
      SLACcitation   = "
}

@article{Jee:2008qj,
    author = "Jee, M.J. and Tyson, J.A.",
    title = "{Dark Matter in the Galaxy Cluster CL J1226+3332 at Z=0.89}",
    eprint = "0810.0709",
    archivePrefix = "arXiv",
    primaryClass = "astro-ph",
    doi = "10.1088/0004-637X/691/2/1337",
    journal = "Astrophys. J.",
    volume = "691",
    pages = "1337--1347",
    year = "2009"
}

@article{Bertone:2016nfn,
    author = "Bertone, Gianfranco and Hooper, Dan",
    title = "{History of dark matter}",
    eprint = "1605.04909",
    archivePrefix = "arXiv",
    primaryClass = "astro-ph.CO",
    reportNumber = "FERMILAB-PUB-16-157-A",
    doi = "10.1103/RevModPhys.90.045002",
    journal = "Rev. Mod. Phys.",
    volume = "90",
    number = "4",
    pages = "045002",
    year = "2018"
}

@article{Jee:2007nx,
    author = "Jee, Myungkook James and others",
    title = "{Discovery of a Ringlike Dark Matter Structure in the Core of the Galaxy Cluster Cl 0024+17}",
    eprint = "0705.2171",
    archivePrefix = "arXiv",
    primaryClass = "astro-ph",
    doi = "10.1086/517498",
    journal = "Astrophys. J.",
    volume = "661",
    pages = "728--749",
    year = "2007"
}

@article{Chianese:2017nwe,
      author         = "Chianese, Marco and Miele, Gennaro and Morisi, Stefano",
      title          = "{Interpreting IceCube 6-year HESE data as an evidence for
                        hundred TeV decaying Dark Matter}",
      journal        = "Phys. Lett.",
      volume         = "B773",
      year           = "2017",
      pages          = "591-595",
      doi            = "10.1016/j.physletb.2017.09.016",
      eprint         = "1707.05241",
      archivePrefix  = "arXiv",
      primaryClass   = "hep-ph",
      SLACcitation   = "
}

@article{Chianese:2018ijk,
      author         = "Chianese, Marco and Miele, Gennaro and Morisi, Stefano
                        and Peinado, Eduardo",
      title          = "{Neutrinophilic Dark Matter in the epoch of IceCube and
                        Fermi-LAT}",
      journal        = "JCAP",
      volume         = "1812",
      year           = "2018",
      number         = "12",
      pages          = "016",
      doi            = "10.1088/1475-7516/2018/12/016",
      eprint         = "1808.02486",
      archivePrefix  = "arXiv",
      primaryClass   = "hep-ph",
      SLACcitation   = "
}

@article{Dekker:2019gpe,
      author         = "Dekker, Ariane and Chianese, Marco and Ando, Shin'ichiro",
      title          = "{Probing dark matter signals in neutrino telescopes
                        through angular power spectrum}",
      year           = "2019",
      eprint         = "1910.12917",
      archivePrefix  = "arXiv",
      primaryClass   = "hep-ph",
      SLACcitation   = "
}

@article{Sui:2018bbh,
      author         = "Sui, Yicong and Bhupal Dev, P. S.",
      title          = "{A Combined Astrophysical and Dark Matter Interpretation
                        of the IceCube HESE and Throughgoing Muon Events}",
      journal        = "JCAP",
      volume         = "1807",
      year           = "2018",
      number         = "07",
      pages          = "020",
      doi            = "10.1088/1475-7516/2018/07/020",
      eprint         = "1804.04919",
      archivePrefix  = "arXiv",
      primaryClass   = "hep-ph",
      SLACcitation   = "
}

@article{gozzini2019search,
  title={Search for dark matter with the ANTARES and KM3NeT neutrino telescopes},
  author={Gozzini, Sara Rebecca},
  year={2019},
  url={https://pos.sissa.it/358/552/pdf}
}

@article{Mena:2014sja,
      author         = "Mena, Olga and Palomares-Ruiz, Sergio and Vincent, Aaron
                        C.",
      title          = "{Flavor Composition of the High-Energy Neutrino Events in
                        IceCube}",
      journal        = "Phys. Rev. Lett.",
      volume         = "113",
      year           = "2014",
      pages          = "091103",
      doi            = "10.1103/PhysRevLett.113.091103",
      eprint         = "1404.0017",
      archivePrefix  = "arXiv",
      primaryClass   = "astro-ph.HE",
      reportNumber   = "IFIC-14-22",
      SLACcitation   = "
}

@article{Palomares-Ruiz:2015mka,
      author         = "Palomares-Ruiz, Sergio and Vincent, Aaron C. and Mena,
                        Olga",
      title          = "{Spectral analysis of the high-energy IceCube neutrinos}",
      journal        = "Phys. Rev.",
      volume         = "D91",
      year           = "2015",
      number         = "10",
      pages          = "103008",
      doi            = "10.1103/PhysRevD.91.103008",
      eprint         = "1502.02649",
      archivePrefix  = "arXiv",
      primaryClass   = "astro-ph.HE",
      reportNumber   = "IFIC-15-06, IPPP-15-04, DCPT-15-08",
      SLACcitation   = "
}

@article{Arguelles:2015dca,
      author         = "Arg{\"u}elles, Carlos A. and Katori, Teppei and Salvado,
                        Jordi",
      title          = "{New Physics in Astrophysical Neutrino Flavor}",
      journal        = "Phys. Rev. Lett.",
      volume         = "115",
      year           = "2015",
      pages          = "161303",
      doi            = "10.1103/PhysRevLett.115.161303",
      eprint         = "1506.02043",
      archivePrefix  = "arXiv",
      primaryClass   = "hep-ph",
      SLACcitation   = "
}

@article{Bustamante:2015waa,
      author         = "Bustamante, Mauricio and Beacom, John F. and Winter,
                        Walter",
      title          = "{Theoretically palatable flavor combinations of
                        astrophysical neutrinos}",
      journal        = "Phys. Rev. Lett.",
      volume         = "115",
      year           = "2015",
      number         = "16",
      pages          = "161302",
      doi            = "10.1103/PhysRevLett.115.161302",
      eprint         = "1506.02645",
      archivePrefix  = "arXiv",
      primaryClass   = "astro-ph.HE",
      reportNumber   = "DESY-15-099",
      SLACcitation   = "
}

@article{Gonzalez-Garcia:2016gpq,
      author         = "Gonzalez-Garcia, M. C. and Maltoni, Michele and
                        Martinez-Soler, Ivan and Song, Ningqiang",
      title          = "{Non-standard neutrino interactions in the Earth and the
                        flavor of astrophysical neutrinos}",
      journal        = "Astropart. Phys.",
      volume         = "84",
      year           = "2016",
      pages          = "15-22",
      doi            = "10.1016/j.astropartphys.2016.07.001",
      eprint         = "1605.08055",
      archivePrefix  = "arXiv",
      primaryClass   = "hep-ph",
      reportNumber   = "IFT-UAM-CSIC-16-048, YITP-SB-16-23",
      SLACcitation   = "
}

@article{Ahlers:2018yom,
      author         = "Ahlers, Markus and Bustamante, Mauricio and Mu, Siqiao",
      title          = "{Unitarity Bounds of Astrophysical Neutrinos}",
      journal        = "Phys. Rev.",
      volume         = "D98",
      year           = "2018",
      number         = "12",
      pages          = "123023",
      doi            = "10.1103/PhysRevD.98.123023",
      eprint         = "1810.00893",
      archivePrefix  = "arXiv",
      primaryClass   = "astro-ph.HE",
      SLACcitation   = "
}

@article{Arguelles:2019tum,
      author         = "Argüelles, Carlos A. and Farrag, Kareem and Katori,
                        Teppei and Khandelwal, Rishabh and Mandalia, Shivesh and
                        Salvado, Jordi",
      title          = "{Probe of Sterile Neutrinos Using Astrophysical Neutrino
                        Flavor}",
      year           = "2019",
      eprint         = "1909.05341",
      archivePrefix  = "arXiv",
      primaryClass   = "hep-ph",
      SLACcitation   = "
}

@article{Rasmussen:2017ert,
      author         = "Rasmussen, Rasmus W. and Lechner, Lukas and Ackermann,
                        Markus and Kowalski, Marek and Winter, Walter",
      title          = "{Astrophysical neutrinos flavored with Beyond the
                        Standard Model physics}",
      journal        = "Phys. Rev.",
      volume         = "D96",
      year           = "2017",
      number         = "8",
      pages          = "083018",
      doi            = "10.1103/PhysRevD.96.083018",
      eprint         = "1707.07684",
      archivePrefix  = "arXiv",
      primaryClass   = "hep-ph",
      reportNumber   = "DESY-17-107",
      SLACcitation   = "
}

@inproceedings{Halzen:2005qu,
      author         = "Halzen, F.",
      title          = "{Lectures on high-energy neutrino astronomy}",
      booktitle      = "{International WE - Heraeus Summer School: Physics with
                        Cosmic Accelerators Bad Honnef, Germany, July 5-16, 2004}",
      year           = "2005",
      eprint         = "astro-ph/0506248",
      archivePrefix  = "arXiv",
      primaryClass   = "astro-ph",
      SLACcitation   = "
}
@ARTICLE{1970ApJ...159..379R,
       author = {{Rubin}, Vera C. and {Ford}, W. Kent, Jr.},
        title = "{Rotation of the Andromeda Nebula from a Spectroscopic Survey of Emission Regions}",
      journal = {\apj},
         year = 1970,
        month = feb,
       volume = {159},
        pages = {379},
          doi = {10.1086/150317},
       adsurl = {https://ui.adsabs.harvard.edu/abs/1970ApJ...159..379R},
      adsnote = {Provided by the SAO/NASA Astrophysics Data System}
}
@article{Persic:1995ru,
    author = "Persic, Massimo and Salucci, Paolo and Stel, Fulvio",
    archivePrefix = "arXiv",
    doi = "10.1093/mnras/278.1.27",
    eprint = "astro-ph/9506004",
    journal = "Mon. Not. Roy. Astron. Soc.",
    pages = "27",
    reportNumber = "SISSA-60-95-A",
    title = "{The Universal rotation curve of spiral galaxies: 1. The Dark matter connection}",
    volume = "281",
    year = "1996"
}

   @article{Aguilar_2021,
   title={Design and sensitivity of the Radio Neutrino Observatory in Greenland (RNO-G)},
   volume={16},
   ISSN={1748-0221},
   url={http://dx.doi.org/10.1088/1748-0221/16/03/P03025},
   DOI={10.1088/1748-0221/16/03/p03025},
   number={03},
   journal={Journal of Instrumentation},
   publisher={IOP Publishing},
   author={Aguilar, J.A. and others},
   year={2021},
   month={Mar},
   pages={P03025} }

@article{Bell:2020rkw,
    author = "Bell, Nicole F. and Dolan, Matthew J. and Robles, Sandra",
    title = "{Searching for Sub-GeV Dark Matter in the Galactic Centre using Hyper-Kamiokande}",
    eprint = "2005.01950",
    archivePrefix = "arXiv",
    primaryClass = "hep-ph",
    month = "5",
    year = "2020"
}

@article{Chirkin:2013lya,
    author = "Chirkin, Dmitry",
    title = "{Likelihood description for comparing data with simulation of limited statistics}",
    eprint = "1304.0735",
    archivePrefix = "arXiv",
    primaryClass = "astro-ph.IM",
    month = "4",
    year = "2013"
}

@article{Arguelles:2019izp,
    author = "Arg{\"u}elles, Carlos A. and Schneider, Austin and Yuan, Tianlu",
    title = "{A binned likelihood for stochastic models}",
    eprint = "1901.04645",
    archivePrefix = "arXiv",
    primaryClass = "physics.data-an",
    doi = "10.1007/JHEP06(2019)030",
    journal = "JHEP",
    volume = "06",
    pages = "030",
    year = "2019"
}

@article{Glusenkamp:2017rlp,
    author = "Gl{\"u}senkamp, Thorsten",
    title = "{Probabilistic treatment of the uncertainty from the finite size of weighted Monte Carlo data}",
    eprint = "1712.01293",
    archivePrefix = "arXiv",
    primaryClass = "physics.data-an",
    doi = "10.1140/epjp/i2018-12042-x",
    journal = "Eur. Phys. J. Plus",
    volume = "133",
    number = "6",
    pages = "218",
    year = "2018"
}

@article{Glusenkamp:2019uir,
    author = "Gl{\"u}senkamp, Thorsten",
    title = "{A unified perspective on modified Poisson likelihoods for limited Monte Carlo data}",
    eprint = "1902.08831",
    archivePrefix = "arXiv",
    primaryClass = "astro-ph.IM",
    doi = "10.1088/1748-0221/15/01/P01035",
    journal = "JINST",
    volume = "15",
    number = "01",
    pages = "P01035",
    year = "2020"
}

@article{Bohm:2013gla,
    author = "Bohm, G. and Zech, G.",
    title = "{Statistics of weighted Poisson events and its applications}",
    eprint = "1309.1287",
    archivePrefix = "arXiv",
    primaryClass = "physics.data-an",
    doi = "10.1016/j.nima.2014.02.021",
    journal = "Nucl. Instrum. Meth. A",
    volume = "748",
    pages = "1--6",
    year = "2014"
}

@article{Barlow:1993dm,
    author = "Barlow, Roger J. and Beeston, Christine",
    title = "{Fitting using finite Monte Carlo samples}",
    reportNumber = "MAN-HEP-93-1",
    doi = "10.1016/0010-4655(93)90005-W",
    journal = "Comput. Phys. Commun.",
    volume = "77",
    pages = "219--228",
    year = "1993"
}

@article{Cousins:1991qz,
    author = "Cousins, Robert D. and Highland, Virgil L.",
    title = "{Incorporating systematic uncertainties into an upper limit}",
    reportNumber = "DOE-ER-40389-69-REV, DOE-ER-40389-69",
    doi = "10.1016/0168-9002(92)90794-5",
    journal = "Nucl. Instrum. Meth. A",
    volume = "320",
    pages = "331--335",
    year = "1992"
}

@inproceedings{Trotta:2017wnx,
    author = "Trotta, Roberto",
    title = "{Bayesian Methods in Cosmology}",
    eprint = "1701.01467",
    archivePrefix = "arXiv",
    primaryClass = "astro-ph.CO",
    month = "1",
    year = "2017"
}

@article{Gainer:2014bta,
    author = "Gainer, James S. and Lykken, Joseph and Matchev, Konstantin T. and Mrenna, Stephen and Park, Myeonghun",
    title = "{Exploring Theory Space with Monte Carlo Reweighting}",
    eprint = "1404.7129",
    archivePrefix = "arXiv",
    primaryClass = "hep-ph",
    reportNumber = "FERMILAB-PUB-14-114-CD-T, IPMU-14-0100, PUB-14-114-CD-T",
    doi = "10.1007/JHEP10(2014)078",
    journal = "JHEP",
    volume = "10",
    pages = "078",
    year = "2014"
}

@article{Diaz:2019fwt,
    author = "Diaz, A. and Arguelles, C.A. and Collin, G.H. and Conrad, J.M. and Shaevitz, M.H.",
    title = "{Where Are We With Light Sterile Neutrinos?}",
    eprint = "1906.00045",
    archivePrefix = "arXiv",
    primaryClass = "hep-ex",
    month = "5",
    year = "2019"
}

@article{Akimov:2017ade,
    author = "Akimov, D. and others",
    collaboration = "COHERENT",
    title = "{Observation of Coherent Elastic Neutrino-Nucleus Scattering}",
    eprint = "1708.01294",
    archivePrefix = "arXiv",
    primaryClass = "nucl-ex",
    doi = "10.1126/science.aao0990",
    journal = "Science",
    volume = "357",
    number = "6356",
    pages = "1123--1126",
    year = "2017"
}

@article{Vogel:1999zy,
    author = "Vogel, P. and Beacom, John F.",
    title = "{Angular distribution of neutron inverse beta decay, anti-neutrino(e) + p ---> e+ + n}",
    eprint = "hep-ph/9903554",
    archivePrefix = "arXiv",
    doi = "10.1103/PhysRevD.60.053003",
    journal = "Phys. Rev. D",
    volume = "60",
    pages = "053003",
    year = "1999"
}

@article{Ankowski:2016oyj,
    author = "Ankowski, Artur M.",
    title = "{Improved estimate of the cross section for inverse beta decay}",
    eprint = "1601.06169",
    archivePrefix = "arXiv",
    primaryClass = "hep-ph",
    month = "1",
    year = "2016"
}

@book{leo1994techniques,
  title={Techniques for Nuclear and Particle Physics Experiments: A How-to Approach},
  author={Leo, W.R.},
  isbn={9780387572802},
  lccn={93038494},
  url={https://books.google.com/books?id=W7vHQgAACAAJ},
  year={1994},
  publisher={Springer}
}

@article{Aartsen:2019swn,
    author = "Aartsen, M.G. and others",
    collaboration = "IceCube",
    title = "{Neutrino astronomy with the next generation IceCube Neutrino Observatory}",
    eprint = "1911.02561",
    archivePrefix = "arXiv",
    primaryClass = "astro-ph.HE",
    month = "11",
    year = "2019"
}

@article{Aiello:2018usb,
    author = "Aiello, S. and others",
    collaboration = "KM3NeT",
    title = "{Sensitivity of the KM3NeT/ARCA neutrino telescope to point-like neutrino sources}",
    eprint = "1810.08499",
    archivePrefix = "arXiv",
    primaryClass = "astro-ph.HE",
    doi = "10.1016/j.astropartphys.2019.04.002",
    journal = "Astropart. Phys.",
    volume = "111",
    pages = "100--110",
    year = "2019"
}

@article{Katori:2016yel,
    author = "Katori, Teppei and Martini, Marco",
    title = "{Neutrino--nucleus cross sections for oscillation experiments}",
    eprint = "1611.07770",
    archivePrefix = "arXiv",
    primaryClass = "hep-ph",
    doi = "10.1088/1361-6471/aa8bf7",
    journal = "J. Phys. G",
    volume = "45",
    number = "1",
    pages = "013001",
    year = "2018"
}

@article{Fukuda:1998fd,
    author = "Fukuda, Y. and others",
    collaboration = "Super-Kamiokande",
    title = "{Measurements of the solar neutrino flux from Super-Kamiokande's first 300 days}",
    eprint = "hep-ex/9805021",
    archivePrefix = "arXiv",
    reportNumber = "ICRR-417-98-13, ICRR-REPORT-417-98-13, UCI-98-4, CSUDH-HEP-98-02, KEK-PREPRINT-98-40, UMD-98-116, SBHEP-98-2, NGTHEP-98-02, OULNS-98-01, TKU-PAP-98-02, UWSEA-PUB-98-02, LSU-HEPA-2-98, SBHEP98-2, NGTHEP.98-02, TIT-HPE-98-05",
    doi = "10.1103/PhysRevLett.81.1158",
    journal = "Phys. Rev. Lett.",
    volume = "81",
    pages = "1158--1162",
    year = "1998",
    note = "[Erratum: Phys.Rev.Lett. 81, 4279 (1998)]"
}

@article{Ahmad:2001an,
    author = "Ahmad, Q.R. and others",
    collaboration = "SNO",
    title = "{Measurement of the rate of $\nu_e+d \to p+p+e^-$ interactions produced by $^8B$ solar neutrinos at the Sudbury Neutrino Observatory}",
    eprint = "nucl-ex/0106015",
    archivePrefix = "arXiv",
    reportNumber = "UPR-0240E",
    doi = "10.1103/PhysRevLett.87.071301",
    journal = "Phys. Rev. Lett.",
    volume = "87",
    pages = "071301",
    year = "2001"
}

@article{Alimonti:2000xc,
    author = "Alimonti, G and others",
    collaboration = "Borexino",
    title = "{Science and technology of BOREXINO: A Real time detector for low-energy solar neutrinos}",
    eprint = "hep-ex/0012030",
    archivePrefix = "arXiv",
    doi = "10.1016/S0927-6505(01)00110-4",
    journal = "Astropart. Phys.",
    volume = "16",
    pages = "205--234",
    year = "2002"
}

@article{Arpesella:2008mt,
    author = "Arpesella, C. and others",
    collaboration = "Borexino",
    title = "{Direct Measurement of the Be-7 Solar Neutrino Flux with 192 Days of Borexino Data}",
    eprint = "0805.3843",
    archivePrefix = "arXiv",
    primaryClass = "astro-ph",
    doi = "10.1103/PhysRevLett.101.091302",
    journal = "Phys. Rev. Lett.",
    volume = "101",
    pages = "091302",
    year = "2008"
}

@article{Sasaki:2017zwd,
    author = "Sasaki, Makoto and Kifune, Tadashi",
    title = "{Ashra Neutrino Telescope Array (NTA): Combined Imaging Observation of Astroparticles --- For Clear Identification of Cosmic Accelerators and Fundamental Physics Using Cosmic Beams ---}",
    doi = "10.7566/JPSCP.15.011013",
    journal = "JPS Conf. Proc.",
    volume = "15",
    pages = "011013",
    year = "2017"
}

@article{Sasaki:2017uta,
    author = "Sasaki, Makoto",
    collaboration = "NTA",
    title = "{Neutrino Telescope Array (NTA): Multi-Astroparticle Explorer for PeV-EeV Universe--- For Clear Identification of Cosmic Accelerators and Cosmic Beam Physics-}",
    doi = "10.22323/1.301.0941",
    journal = "PoS",
    volume = "ICRC2017",
    pages = "941",
    year = "2018"
}

@book{rqft74,
    author    = "V. B. Berestetskii, E.M. Lifshitz, L. P. Pitaevskii",
    title     = "Relativistic Quantum Theory, Part I",
    year      = "1974",
    publisher = "Pergamon Press",
    address   = ""
}

@article{Jeong:2017mzv,
    author = "Jeong, Yu Seon and Luu, Minh Vu and Reno, Mary Hall and Sarcevic, Ina",
    title = "{Tau energy loss and ultrahigh energy skimming tau neutrinos}",
    eprint = "1704.00050",
    archivePrefix = "arXiv",
    primaryClass = "hep-ph",
    doi = "10.1103/PhysRevD.96.043003",
    journal = "Phys. Rev. D",
    volume = "96",
    number = "4",
    pages = "043003",
    year = "2017"
}

@article{Reno:2019jtr,
    author = "Reno, Mary Hall and Krizmanic, John F. and Venters, Tonia M.",
    title = "{Cosmic tau neutrino detection via Cherenkov signals from air showers from Earth-emerging taus}",
    eprint = "1902.11287",
    archivePrefix = "arXiv",
    primaryClass = "astro-ph.HE",
    doi = "10.1103/PhysRevD.100.063010",
    journal = "Phys. Rev. D",
    volume = "100",
    number = "6",
    pages = "063010",
    year = "2019"
}

@article{Reno:2019qmk,
    author = "Reno, Mary Hall and Venters, Tonia M. and Krizmanic, John F. and Anchordoqui, Luis A. and Guepin, Claire and Olinto, Angela V.",
    collaboration = "POEMMA",
    title = "{A new calculation of Earth-skimming very- and ultra-high energy tau neutrinos}",
    eprint = "1908.03603",
    archivePrefix = "arXiv",
    primaryClass = "astro-ph.HE",
    doi = "10.22323/1.358.0989",
    journal = "PoS",
    volume = "ICRC2019",
    pages = "989",
    year = "2020"
}

@article{Shoemaker:2019xlt,
    author = "Shoemaker, Ian M. and Kusenko, Alexander and Munneke, Peter Kuipers and Romero-Wolf, Andrew and Schroeder, Dustin M. and Siegert, Martin J.",
    title = "{Reflections On the Anomalous ANITA Events: The Antarctic Subsurface as a Possible Explanation}",
    eprint = "1905.02846",
    archivePrefix = "arXiv",
    primaryClass = "astro-ph.HE",
    doi = "10.1017/aog.2020.19",
    month = "5",
    year = "2019"
}

@article{Romero-Wolf:2018zxt,
    author = "Romero-Wolf, A. and others",
    title = "{Comprehensive analysis of anomalous ANITA events disfavors a diffuse tau-neutrino flux origin}",
    eprint = "1811.07261",
    archivePrefix = "arXiv",
    primaryClass = "astro-ph.HE",
    doi = "10.1103/PhysRevD.99.063011",
    journal = "Phys. Rev. D",
    volume = "99",
    number = "6",
    pages = "063011",
    year = "2019"
}

@article{Kurylov:2002vj,
    author = "Kurylov, A. and Ramsey-Musolf, M.J. and Vogel, P.",
    title = "{Radiative corrections to low-energy neutrino reactions}",
    eprint = "hep-ph/0211306",
    archivePrefix = "arXiv",
    doi = "10.1103/PhysRevC.67.035502",
    journal = "Phys. Rev. C",
    volume = "67",
    pages = "035502",
    year = "2003"
}

@ARTICLE{2012TCD.....6.4695B,
       author = {{Besson}, D. and {Doolin}, N. and {Stockham}, M. and {Kravchenko}, I.},
        title = "{Radio-frequency probes of Antarctic ice birefringence at South Pole vs. East Antarctica; evidence for a changing ice fabric}",
      journal = {The Cryosphere Discussions},
         year = 2012,
        month = nov,
       volume = {6},
       number = {6},
        pages = {4695-4731},
          doi = {10.5194/tcd-6-4695-2012},
       adsurl = {https://ui.adsabs.harvard.edu/abs/2012TCD.....6.4695B},
      adsnote = {Provided by the SAO/NASA Astrophysics Data System}
}

@article{Aliaga:2013uqz,
    author = "Aliaga, L. and others",
    collaboration = "MINERvA",
    title = "{Design, Calibration, and Performance of the MINERvA Detector}",
    eprint = "1305.5199",
    archivePrefix = "arXiv",
    primaryClass = "physics.ins-det",
    reportNumber = "FERMILAB-PUB-13-111-E",
    doi = "10.1016/j.nima.2013.12.053",
    journal = "Nucl. Instrum. Meth. A",
    volume = "743",
    pages = "130--159",
    year = "2014"
}

@article{Backhouse:2015xva,
    author = "Backhouse, C. and Patterson, R.B.",
    title = "{Library Event Matching event classification algorithm for electron neutrino interactions in the NO$\nu$A detectors}",
    eprint = "1501.00968",
    archivePrefix = "arXiv",
    primaryClass = "physics.ins-det",
    doi = "10.1016/j.nima.2015.01.017",
    journal = "Nucl. Instrum. Meth. A",
    volume = "778",
    pages = "31--39",
    year = "2015"
}

@article{Aurisano_2016,
	doi = {10.1088/1748-0221/11/09/p09001},
	url = {https://doi.org/10.1088
	year = 2016,
	month = {sep},
	publisher = {{IOP} Publishing},
	volume = {11},
	number = {09},
	pages = {P09001--P09001},
	author = {A. Aurisano and A. Radovic and D. Rocco and A. Himmel and M.D. Messier and E. Niner and G. Pawloski and F. Psihas and A. Sousa and P. Vahle},
	title = {A convolutional neural network neutrino event classifier},
	journal = {Journal of Instrumentation},
	abstract = {Convolutional neural networks (CNNs) have been widely   applied in the computer vision community to solve complex problems   in image recognition and analysis. We describe an application of the   CNN technology to the problem of identifying particle interactions in sampling   calorimeters used commonly in high energy physics and high energy   neutrino physics in particular. Following a discussion of the core   concepts of CNNs and recent innovations in CNN architectures related   to the field of deep learning, we outline a specific application to   the NOvA neutrino detector. This algorithm, CVN (Convolutional    Visual  Network) identifies neutrino   interactions based on their topology without the need for detailed   reconstruction and outperforms algorithms currently in use by the   NOvA collaboration.}
}

@article{Psihas:2019ksa,
    author = "Psihas, F. and Niner, E. and Groh, M. and Murphy, R. and Aurisano, A. and Himmel, A. and Lang, K. and Messier, M.D. and Radovic, A. and Sousa, A.",
    title = "{Context-Enriched Identification of Particles with a Convolutional Network for Neutrino Events}",
    eprint = "1906.00713",
    archivePrefix = "arXiv",
    primaryClass = "physics.ins-det",
    reportNumber = "FERMILAB-PUB-19-258-PPD",
    doi = "10.1103/PhysRevD.100.073005",
    journal = "Phys. Rev. D",
    volume = "100",
    number = "7",
    pages = "073005",
    year = "2019"
}

@article{osti_925917,
title = {Neutral current interactions in MINOS},
author = {Sousa, Alexandre and /Oxford U.},
abstractNote = {The Main Injector Neutrino Oscillation Search (MINOS) long-baseline experiment has been actively collecting beam data since 2005, having already accumulated 3 x 10{sup 20} protons-on-target (POT). The several million neutrinos per year observed at the Near detector may improve the existing body of knowledge of neutrino cross-sections and the Near-Far comparison of the observed energy spectrum neutral current events constrains oscillations into sterile neutrinos. MINOS capabilities of observing neutral current neutrino events are described and the employed methodology for event selection is discussed, along with preliminary results obtained. An outlook on the expected neutral current related contributions from MINOS is also presented.},
doi = {},
journal = {},
number = {},
volume = {},
place = {United States},
year = {2007},
month = {7}
}

@article{Sousa:2015bxa,
    author = "Sousa, Alexandre B.",
    editor = "Kearns, Ed",
    collaboration = "MINOS, MINOS+",
    title = "{First MINOS+ Data and New Results from MINOS}",
    eprint = "1502.07715",
    archivePrefix = "arXiv",
    primaryClass = "hep-ex",
    doi = "10.1063/1.4915576",
    journal = "AIP Conf. Proc.",
    volume = "1666",
    number = "1",
    pages = "110004",
    year = "2015"
}

@article{Conrad:2010mh,
    author = "Conrad, Janet and de Gouvea, Andre and Shalgar, Shashank and Spitz, Joshua",
    title = "{Atmospheric Tau Neutrinos in a Multi-kiloton Liquid Argon Detector}",
    eprint = "1008.2984",
    archivePrefix = "arXiv",
    primaryClass = "hep-ph",
    reportNumber = "NUHEP-TH-10-11",
    doi = "10.1103/PhysRevD.82.093012",
    journal = "Phys. Rev. D",
    volume = "82",
    pages = "093012",
    year = "2010"
}

@article{Abi:2020evt,
    author = "Abi, Babak and others",
    collaboration = "DUNE",
    title = "{Deep Underground Neutrino Experiment (DUNE), Far Detector Technical Design Report, Volume II DUNE Physics}",
    eprint = "2002.03005",
    archivePrefix = "arXiv",
    primaryClass = "hep-ex",
    reportNumber = "FERMILAB-PUB-20-025-ND, FERMILAB-DESIGN-2020-02",
    month = "2",
    year = "2020"
}

@article{Acciarri:2016smi,
    author = "Acciarri, R. and others",
    collaboration = "MicroBooNE",
    title = "{Design and Construction of the MicroBooNE Detector}",
    eprint = "1612.05824",
    archivePrefix = "arXiv",
    primaryClass = "physics.ins-det",
    reportNumber = "FERMILAB-PUB-16-613-ND",
    doi = "10.1088/1748-0221/12/02/P02017",
    journal = "JINST",
    volume = "12",
    number = "02",
    pages = "P02017",
    year = "2017"
}

@article{Ali-Mohammadzadeh:2020fbd,
    author = "Ali-Mohammadzadeh, B. and others",
    title = "{Design and implementation of the new scintillation light detection system of ICARUS T600}",
    eprint = "2006.05261",
    archivePrefix = "arXiv",
    primaryClass = "physics.ins-det",
    reportNumber = "FERMILAB-PUB-20-230-ND",
    month = "6",
    year = "2020"
}

@article{Cavanna:2018yfk,
    author = "Cavanna, F. and Ereditato, A. and Fleming, B.T.",
    title = "{Advances in liquid argon detectors}",
    doi = "10.1016/j.nima.2018.07.010",
    journal = "Nucl. Instrum. Meth. A",
    volume = "907",
    pages = "1--8",
    year = "2018"
}

@article{Adams:2018bvi,
    author = "Adams, C. and others",
    collaboration = "MicroBooNE",
    title = "{Deep neural network for pixel-level electromagnetic particle identification in the MicroBooNE liquid argon time projection chamber}",
    eprint = "1808.07269",
    archivePrefix = "arXiv",
    primaryClass = "hep-ex",
    reportNumber = "FERMILAB-PUB-18-231-ND",
    doi = "10.1103/PhysRevD.99.092001",
    journal = "Phys. Rev. D",
    volume = "99",
    number = "9",
    pages = "092001",
    year = "2019"
}

@article{MicroBooNE:2018mfn,
    author = "MicroBooNE",
    collaboration = "MicroBooNE",
    title = "{First Deep Learning based Event Reconstruction for Low-Energy Excess Searches with MicroBooNE}",
    doi = "10.2172/1573220",
    month = "7",
    year = "2018"
}

@article{Acciarri:2017hat,
    author = "Acciarri, R. and others",
    collaboration = "MicroBooNE",
    title = "{The Pandora multi-algorithm approach to automated pattern recognition of cosmic-ray muon and neutrino events in the MicroBooNE detector}",
    eprint = "1708.03135",
    archivePrefix = "arXiv",
    primaryClass = "hep-ex",
    reportNumber = "FERMILAB-PUB-17-306-ND",
    doi = "10.1140/epjc/s10052-017-5481-6",
    journal = "Eur. Phys. J. C",
    volume = "78",
    number = "1",
    pages = "82",
    year = "2018"
}

@article{Zas:1991jv,
    author = "Zas, E. and Halzen, F. and Stanev, T.",
    title = "{Electromagnetic pulses from high-energy showers: Implications for neutrino detection}",
    reportNumber = "MAD-PH-652",
    doi = "10.1103/PhysRevD.45.362",
    journal = "Phys. Rev. D",
    volume = "45",
    pages = "362--376",
    year = "1992"
}

@article{Askaryan:1962hbi,
    author = "Askar'yan, G.A.",
    title = "{Excess negative charge of an electron-photon shower and its coherent radio emission}",
    journal = "Sov. Phys. JETP",
    volume = "14",
    number = "2",
    pages = "441--443",
    year = "1962"
}

@article{Gusev_1984,
	doi = {10.1070/pu1984v027n07abeh004052},
	url = {https://doi.org/10.1070
	year = 1984,
	month = {jul},
	publisher = {{IOP} Publishing},
	volume = {27},
	number = {7},
	pages = {550--552},
	author = {G A Gusev and I M Zheleznykh},
	title = {On the possibility of detection of neutrinos and muons on the basis of radio radiation of cascades in natural dielectric media (antarctic ice sheet and so forth)},
	journal = {Soviet Physics Uspekhi},
	abstract = {}
}

@article{Markov:1986dx,
    author = "Markov, M.A. and Zheleznykh, I.M.",
    title = "{Large Scale Cherenkov Detectors in Ocean, Atmosphere and Ice}",
    doi = "10.1016/0168-9002(86)90522-X",
    journal = "Nucl. Instrum. Meth. A",
    volume = "248",
    pages = "242--251",
    year = "1986"
}

@article{Anker:2020lre,
    author = "Anker, A. and others",
    title = "{White Paper: ARIANNA-200 high energy neutrino telescope}",
    eprint = "2004.09841",
    archivePrefix = "arXiv",
    primaryClass = "astro-ph.IM",
    month = "4",
    year = "2020"
}

@article{Allison:2019xtn,
    author = "Allison, P. and others",
    collaboration = "ARA",
    title = "{Constraints on the Diffuse Flux of Ultra-High Energy Neutrinos from Four Years of Askaryan Radio Array Data in Two Stations}",
    eprint = "1912.00987",
    archivePrefix = "arXiv",
    primaryClass = "astro-ph.HE",
    month = "12",
    year = "2019"
}

@article{Gorham:2010kv,
    author = "Gorham, P.W. and others",
    collaboration = "ANITA",
    title = "{Observational Constraints on the Ultra-high Energy Cosmic Neutrino Flux from the Second Flight of the ANITA Experiment}",
    eprint = "1003.2961",
    archivePrefix = "arXiv",
    primaryClass = "astro-ph.HE",
    doi = "10.1103/PhysRevD.82.022004",
    journal = "Phys. Rev. D",
    volume = "82",
    pages = "022004",
    year = "2010",
    note = "[Erratum: Phys.Rev.D 85, 049901 (2012)]"
}

@inproceedings{Romero-Wolf:2020pzh,
    author = "Romero-Wolf, Andres and others",
    title = "{An Andean Deep-Valley Detector for High-Energy Tau Neutrinos}",
    booktitle = "{Latin American Strategy Forum for Research Infrastructure}",
    eprint = "2002.06475",
    archivePrefix = "arXiv",
    primaryClass = "astro-ph.IM",
    month = "2",
    year = "2020"
}

@article{Alikhanov:2015kla,
    author = "Alikhanov, I.",
    title = "{Hidden Glashow resonance in neutrino--nucleus collisions}",
    eprint = "1503.08817",
    archivePrefix = "arXiv",
    primaryClass = "hep-ph",
    doi = "10.1016/j.physletb.2016.03.009",
    journal = "Phys. Lett. B",
    volume = "756",
    pages = "247--253",
    year = "2016"
}

@article{Seckel:1997kk,
    author = "Seckel, D.",
    title = "{Neutrino photon reactions in astrophysics and cosmology}",
    eprint = "hep-ph/9709290",
    archivePrefix = "arXiv",
    reportNumber = "BA-97-32",
    doi = "10.1103/PhysRevLett.80.900",
    journal = "Phys. Rev. Lett.",
    volume = "80",
    pages = "900--903",
    year = "1998"
}

@article{Beacom:2019pzs,
    author = "Zhou, Bei and Beacom, John F.",
    title = "{W -boson and trident production in TeV--PeV neutrino observatories}",
    eprint = "1910.10720",
    archivePrefix = "arXiv",
    primaryClass = "hep-ph",
    doi = "10.1103/PhysRevD.101.036010",
    journal = "Phys. Rev. D",
    volume = "101",
    number = "3",
    pages = "036010",
    year = "2020"
}

@article{Zhou:2019vxt,
    author = "Zhou, Bei and Beacom, John F.",
    title = "{Neutrino-nucleus cross sections for W-boson and trident production}",
    eprint = "1910.08090",
    archivePrefix = "arXiv",
    primaryClass = "hep-ph",
    doi = "10.1103/PhysRevD.101.036011",
    journal = "Phys. Rev. D",
    volume = "101",
    number = "3",
    pages = "036011",
    year = "2020"
}

@article{Glashow:1960zz,
      author         = "Glashow, Sheldon L.",
      title          = "{Resonant Scattering of Antineutrinos}",
      journal        = "Phys. Rev.",
      volume         = "118",
      year           = "1960",
      pages          = "316-317",
      doi            = "10.1103/PhysRev.118.316",
      SLACcitation   = "
}

@article{Loewy:2014zva,
      author         = "Loewy, Amit and Nussinov, Shmuel and Glashow, Sheldon L.",
      title          = "{The Effect of Doppler Broadening on the $6.3 \ PeV$
                        $W^-$ Resonance in $\bar{\nu}_e e^-$ Collisions}",
      year           = "2014",
      eprint         = "1407.4415",
      archivePrefix  = "arXiv",
      primaryClass   = "hep-ph",
      SLACcitation   = "
}

@article{Cowen:2007ny,
      author         = "Cowen, D. F.",
      title          = "{Tau neutrinos in IceCube}",
      booktitle      = "{TeV particle astrophysics. Proceedings, 2nd Workshop,
                        Madison, USA, August 28-31, 2006}",
      collaboration  = "IceCube",
      journal        = "J. Phys. Conf. Ser.",
      volume         = "60",
      year           = "2007",
      pages          = "227-230",
      doi            = "10.1088/1742-6596/60/1/048",
      SLACcitation   = "
}

@article{Learned:1994wg,
      author         = "Learned, John G. and Pakvasa, Sandip",
      title          = "{Detecting tau-neutrino oscillations at PeV energies}",
      journal        = "Astropart. Phys.",
      volume         = "3",
      year           = "1995",
      pages          = "267-274",
      doi            = "10.1016/0927-6505(94)00043-3",
      eprint         = "hep-ph/9405296",
      archivePrefix  = "arXiv",
      primaryClass   = "hep-ph",
      reportNumber   = "hep-ph/9408296, UH-511-799-94, DUMAND-3-94",
      SLACcitation   = "
}

@INPROCEEDINGS{2019ICRC...36..945L,
       author = {{Lu}, L.},
        title = "{A novel method of rejecting muon backgrounds for the detection of the highest energy neutrinos}",
    booktitle = {36th International Cosmic Ray Conference (ICRC2019)},
         year = 2019,
       series = {International Cosmic Ray Conference},
       volume = {36},
        month = jul,
          eid = {945},
        pages = {945},
       adsurl = {https://ui.adsabs.harvard.edu/abs/2019ICRC...36..945L},
      adsnote = {Provided by the SAO/NASA Astrophysics Data System}
}

@article{Stachurska:2019wfb,
    author = "Stachurska, Juliana",
    collaboration = "IceCube",
    title = "{First Double Cascade Tau Neutrino Candidates in IceCube and a New Measurement of the Flavor Composition}",
    eprint = "1908.05506",
    archivePrefix = "arXiv",
    primaryClass = "astro-ph.HE",
    reportNumber = "PoS-ICRC2019-1015",
    doi = "10.22323/1.358.1015",
    journal = "PoS",
    volume = "ICRC2019",
    pages = "1015",
    year = "2020"
}

@article{stachurska_juliana_2018_1301122,
  author       = {Stachurska, Juliana},
  title        = "{New measurement of the flavor composition of high- 
                   energy neutrino events with contained vertices in
                   IceCube}",
  month        = jun,
  year         = 2018,
  publisher    = {Zenodo},
  doi          = {10.5281/zenodo.1301122},
  url          = {https://doi.org/10.5281/zenodo.1301122}
}

@article{Aartsen:2020tdl,
    author = "Albert, A. and others",
    collaboration = "ANTARES, IceCube",
    title = "{Combined search for neutrinos from dark matter self-annihilation in the Galactic Centre with ANTARES and IceCube}",
    eprint = "2003.06614",
    archivePrefix = "arXiv",
    primaryClass = "astro-ph.HE",
    month = "3",
    year = "2020"
}

@article{doi:10.1146/annurev.nucl.57.090506.123052,
author = {Heinrich, Joel and Lyons, Louis},
title = {Systematic Errors},
journal = {Annual Review of Nuclear and Particle Science},
volume = {57},
number = {1},
pages = {145-169},
year = {2007},
doi = {10.1146/annurev.nucl.57.090506.123052},

URL = { 
        https://doi.org/10.1146/annurev.nucl.57.090506.123052
    
},
eprint = { 
        https://doi.org/10.1146/annurev.nucl.57.090506.123052
    
}
,
    abstract = { To introduce the ideas of statistical and systematic errors, this review first describes a simple pendulum experiment. We follow with a brief discussion of the Bayesian and frequentist approaches. Two widely used applications of statistical techniques in particle physics data include extracting ranges for parameters of interest (e.g., mass of the W boson, cross section for top production, neutrino mixing angles, etc.) and assessing the significance of possible signals (e.g., is there evidence for Higgs boson production?). These two topics are first discussed in the absence of systematics, and then methods of incorporating systematic effects are described. We give a detailed discussion of a Bayesian approach to setting upper limits on a Poisson process in the presence of background and/or acceptance uncertainties. The relevance of the choice of priors and how this affects the coverage properties of the method are described. }
}

@article{Watanabe:2008ru,
    author = "Watanabe, H. and others",
    collaboration = "Super-Kamiokande",
    title = "{First Study of Neutron Tagging with a Water Cherenkov Detector}",
    eprint = "0811.0735",
    archivePrefix = "arXiv",
    primaryClass = "hep-ex",
    doi = "10.1016/j.astropartphys.2009.03.002",
    journal = "Astropart. Phys.",
    volume = "31",
    pages = "320--328",
    year = "2009"
}

@article{Beacom:2003nk,
    author = "Beacom, John F. and Vagins, Mark R.",
    title = "{GADZOOKS! Anti-neutrino spectroscopy with large water Cherenkov detectors}",
    eprint = "hep-ph/0309300",
    archivePrefix = "arXiv",
    reportNumber = "FERMILAB-PUB-03-249-A",
    doi = "10.1103/PhysRevLett.93.171101",
    journal = "Phys. Rev. Lett.",
    volume = "93",
    pages = "171101",
    year = "2004"
}

@INPROCEEDINGS{2019ICRC...36.1003S,
       author = {{Sasaki}, M.},
        title = "{Galactic Bulge Monitor with Ashra-1 and NTA detectors}",
    booktitle = {36th International Cosmic Ray Conference (ICRC2019)},
         year = 2019,
       series = {International Cosmic Ray Conference},
       volume = {36},
        month = jul,
          eid = {1003},
        pages = {1003},
       adsurl = {https://ui.adsabs.harvard.edu/abs/2019ICRC...36.1003S},
      adsnote = {Provided by the SAO/NASA Astrophysics Data System}
}

@article{Anchordoqui:2019omw,
    author = "Anchordoqui, Luis A. and others",
    title = "{Performance and science reach of the Probe of Extreme Multimessenger Astrophysics for ultrahigh-energy particles}",
    eprint = "1907.03694",
    archivePrefix = "arXiv",
    primaryClass = "astro-ph.HE",
    doi = "10.1103/PhysRevD.101.023012",
    journal = "Phys. Rev. D",
    volume = "101",
    number = "2",
    pages = "023012",
    year = "2020"
}

@article{Aartsen:2020vir,
    author = "Aartsen, M.G. and others",
    collaboration = "IceCube",
    title = "{A search for IceCube events in the direction of ANITA neutrino candidates}",
    eprint = "2001.01737",
    archivePrefix = "arXiv",
    primaryClass = "astro-ph.HE",
    doi = "10.3847/1538-4357/ab791d",
    month = "1",
    year = "2020"
}

@article{Fechner:2009aa,
    author = "Fechner, M. and others",
    collaboration = "Super-Kamiokande",
    title = "{Kinematic reconstruction of atmospheric neutrino events in a large water Cherenkov detector with proton identification}",
    eprint = "0901.1645",
    archivePrefix = "arXiv",
    primaryClass = "hep-ex",
    doi = "10.1103/PhysRevD.79.112010",
    journal = "Phys. Rev. D",
    volume = "79",
    pages = "112010",
    year = "2009"
}

@article{Jiang:2019xwn,
    author = "Jiang, M. and others",
    collaboration = "Super-Kamiokande",
    title = "{Atmospheric Neutrino Oscillation Analysis with Improved Event Reconstruction in Super-Kamiokande IV}",
    eprint = "1901.03230",
    archivePrefix = "arXiv",
    primaryClass = "hep-ex",
    doi = "10.1093/ptep/ptz015",
    journal = "PTEP",
    volume = "2019",
    number = "5",
    pages = "053F01",
    year = "2019"
}

@article{Sabti:2019mhn,
    author = "Sabti, Nashwan and Alvey, James and Escudero, Miguel and Fairbairn, Malcolm and Blas, Diego",
    title = "{Refined Bounds on MeV-scale Thermal Dark Sectors from BBN and the CMB}",
    eprint = "1910.01649",
    archivePrefix = "arXiv",
    primaryClass = "hep-ph",
    reportNumber = "KCL-2019-75",
    doi = "10.1088/1475-7516/2020/01/004",
    journal = "JCAP",
    volume = "01",
    pages = "004",
    year = "2020"
}

@article{Escudero:2018mvt,
    author = "Escudero, Miguel",
    title = "{Neutrino decoupling beyond the Standard Model: CMB constraints on the Dark Matter mass with a fast and precise $N_{\rm eff}$ evaluation}",
    eprint = "1812.05605",
    archivePrefix = "arXiv",
    primaryClass = "hep-ph",
    reportNumber = "KCL-2018-76",
    doi = "10.1088/1475-7516/2019/02/007",
    journal = "JCAP",
    volume = "02",
    pages = "007",
    year = "2019"
}

@article{Dutta:2022wuc,
    author = "Dutta, Koushik and Ghosh, Avirup and Kar, Arpan and Mukhopadhyaya, Biswarup",
    title = "{Model-independent constraints on decaying scalar dark matter: existing observations and projected radio signals at the SKA}",
    eprint = "2204.06024",
    archivePrefix = "arXiv",
    primaryClass = "hep-ph",
    month = "4",
    year = "2022"
}

@article{Sjostrand:2014zea,
    author = {Sj\"ostrand, Torbj\"orn and Ask, Stefan and Christiansen, Jesper R. and Corke, Richard and Desai, Nishita and Ilten, Philip and Mrenna, Stephen and Prestel, Stefan and Rasmussen, Christine O. and Skands, Peter Z.},
    title = "{An introduction to PYTHIA 8.2}",
    eprint = "1410.3012",
    archivePrefix = "arXiv",
    primaryClass = "hep-ph",
    reportNumber = "LU-TP-14-36, MCNET-14-22, CERN-PH-TH-2014-190, FERMILAB-PUB-14-316-CD, DESY-14-178, SLAC-PUB-16122",
    doi = "10.1016/j.cpc.2015.01.024",
    journal = "Comput. Phys. Commun.",
    volume = "191",
    pages = "159--177",
    year = "2015"
}

@article{IceCube:2019scr,
    author = "Aartsen, M. G. and others",
    collaboration = "IceCube",
    title = "{Search for PeV Gamma-Ray Emission from the Southern Hemisphere with 5 Years of Data from the IceCube Observatory}",
    eprint = "1908.09918",
    archivePrefix = "arXiv",
    primaryClass = "astro-ph.HE",
    doi = "10.3847/1538-4357/ab6d67",
    journal = "Astrophys. J.",
    volume = "891",
    pages = "9",
    month = "8",
    year = "2019"
}

@article{Sjostrand:2019zhc,
    author = {Sj\"ostrand, Torbj\"orn},
    title = "{The PYTHIA Event Generator: Past, Present and Future}",
    eprint = "1907.09874",
    archivePrefix = "arXiv",
    primaryClass = "hep-ph",
    reportNumber = "LU TP 19-31, MCnet-19-16",
    doi = "10.1016/j.cpc.2019.106910",
    journal = "Comput. Phys. Commun.",
    volume = "246",
    pages = "106910",
    year = "2020"
}

@article{Olinto:2020oky,
    author = "Olinto, A. V. and others",
    title = "{The POEMMA (Probe of Extreme Multi-Messenger Astrophysics) Observatory}",
    eprint = "2012.07945",
    archivePrefix = "arXiv",
    primaryClass = "astro-ph.IM",
    month = "12",
    year = "2020"
}

@article{HAWC:2017udy,
    author = "Abeysekara, A. U. and others",
    collaboration = "HAWC",
    title = "{A Search for Dark Matter in the Galactic Halo with HAWC}",
    eprint = "1710.10288",
    archivePrefix = "arXiv",
    primaryClass = "astro-ph.HE",
    reportNumber = "LA-UR-17-29899, LCTP-17-01, MIT-CTP-4951",
    doi = "10.1088/1475-7516/2018/02/049",
    journal = "JCAP",
    volume = "02",
    pages = "049",
    year = "2018"
}

@misc{cta_sens, 
    title={Performance of the Cherenkov Telescope Array}, 
    url = {https://arxiv.org/abs/1907.08171},
    doi = {10.48550/ARXIV.1907.08171},
    publisher={arXiv}, 
    author={Maier, G. and Arrabito, L. and Bernlöhr, K. and Bregeon, J. and Cumani, P. and Hassan, T. and Hinton, J. and Moralejo, A.},
    year={2019}
    }

@misc{chianese_2021, 
    title={Constraints on decaying dark matter with LHAASO-KM2A}, url={https://video.desy.de/video/Constraints-on-decaying-dark-matter-with-LHAASO-KM2A/8bb5cb0b3ddb5a40c472597c1dc4f2d7}, 
    journal={DESY Webcast}, 
    publisher={ICRC 2021}, 
    author={Chianese, Marco}, 
    year={2021}
    } 
    
@article{Abe_2022,
	doi = {10.3847/1538-4357/ac32c1},
  
	url = {https://doi.org/10.3847
  
	year = 2022,
	month = {jan},
  
	publisher = {American Astronomical Society},
  
	volume = {925},
  
	number = {1},
  
	pages = {14},
  
	author = {S. Abe and S. Asami and A. Gando and Y. Gando and T. Gima and A. Goto and T. Hachiya and K. Hata and S. Hayashida and K. Hosokawa and K. Ichimura and S. Ieki and H. Ikeda and K. Inoue and K. Ishidoshiro and Y. Kamei and N. Kawada and Y. Kishimoto and T. Kinoshita and M. Koga and N. Maemura and T. Mitsui and H. Miyake and K. Nakamura and K. Nakamura and R. Nakamura and H. Ozaki and T. Sakai and H. Sambonsugi and I. Shimizu and Y. Shirahata and J. Shirai and K. Shiraishi and A. Suzuki and Y. Suzuki and A. Takeuchi and K. Tamae and K. Ueshima and Y. Wada and H. Watanabe and Y. Yoshida and S. Obara and A. K. Ichikawa and A. Kozlov and D. Chernyak and Y. Takemoto and S. Yoshida and S. Umehara and K. Fushimi and S. Hirata and K. Z. Nakamura and M. Yoshida and B. E. Berger and B. K. Fujikawa and J. G. Learned and J. Maricic and S. N. Axani and L. A. Winslow and Z. Fu and J. Ouellet and Y. Efremenko and H. J. Karwowski and D. M. Markoff and W. Tornow and A. Li and J. A. Detwiler and S. Enomoto and M. P. Decowski and C. Grant and T. O'Donnell and S. Dell'Oro},
  
	title = {Limits on Astrophysical Antineutrinos with the {KamLAND} Experiment},
  
	journal = {The Astrophysical Journal}
}

@article{Murase:2012df,
    author = "Murase, Kohta and Beacom, John F. and Takami, Hajime",
    title = "{Gamma-Ray and Neutrino Backgrounds as Probes of the High-Energy Universe: Hints of Cascades, General Constraints, and Implications for TeV Searches}",
    eprint = "1205.5755",
    archivePrefix = "arXiv",
    primaryClass = "astro-ph.HE",
    doi = "10.1088/1475-7516/2012/08/030",
    journal = "JCAP",
    volume = "08",
    pages = "030",
    year = "2012"
}

@article{Franceschini:2017iwq,
    author = "Franceschini, Alberto and Rodighiero, Giulia",
    title = "{The extragalactic background light revisited and the cosmic photon-photon opacity}",
    eprint = "1705.10256",
    archivePrefix = "arXiv",
    primaryClass = "astro-ph.HE",
    doi = "10.1051/0004-6361/201629684",
    journal = "Astron. Astrophys.",
    volume = "603",
    pages = "A34",
    year = "2017"
}

@article{Abbasi_2013,
       author = "Abbasi, R. and others",
        title = "{IceTop: The surface component of IceCube. The IceCube Collaboration}",
      journal = {Nuclear Instruments and Methods in Physics Research A},
     keywords = {Astrophysics - Instrumentation and Methods for Astrophysics},
         year = 2013,
        month = feb,
       volume = {700},
        pages = {188-220},
          doi = {10.1016/j.nima.2012.10.067},
archivePrefix = {arXiv},
       eprint = {1207.6326},
 primaryClass = {astro-ph.IM},
       adsurl = {https://ui.adsabs.harvard.edu/abs/2013NIMPA.700..188A},
      adsnote = {Provided by the SAO/NASA Astrophysics Data System}
}

@article{LHAASO:2019qtb,
    author = "Addazi, Andrea and others",
    collaboration = "LHAASO",
    title = "{The Large High Altitude Air Shower Observatory (LHAASO) Science Book (2021 Edition)}",
    eprint = "1905.02773",
    archivePrefix = "arXiv",
    primaryClass = "astro-ph.HE",
    journal = "Chin. Phys. C",
    volume = "46",
    pages = "035001--035007",
    year = "2022"
}

@article{HAWC:2013htd,
    author = "Abeysekara, A. U. and others",
    collaboration = "HAWC",
    title = "{The HAWC Gamma-Ray Observatory: Observations of Cosmic Rays}",
    eprint = "1310.0072",
    archivePrefix = "arXiv",
    primaryClass = "astro-ph.HE",
    month = "9",
    year = "2013"
}

@article{Bigongiari:2016amk,
    author = "Bigongiari, Ciro",
    editor = "Cataldi, Gabriella and De Mitri, Ivan and Martello, Daniele",
    collaboration = "CTA Consortium",
    title = "{The Cherenkov Telescope Array}",
    eprint = "1606.08190",
    archivePrefix = "arXiv",
    primaryClass = "astro-ph.IM",
    doi = "10.1016/j.nuclphysbps.2016.10.025",
    journal = "Nucl. Part. Phys. Proc.",
    volume = "279-281",
    pages = "174--181",
    year = "2016"
}

@article{Atwood_2009,
	doi = {10.1088/0004-637x/697/2/1071},
  
	url = {https://doi.org/10.1088
  
	year = 2009,
	month = {may},
  
	publisher = {American Astronomical Society},
  
	volume = {697},
  
	number = {2},
  
	pages = {1071--1102},
  
	author = "Atwood, W. B. and others",
  
	title = {{THE} {LARGE} {AREA} {TELESCOPE} {ON} {THE}$\less$i$\greater${FERMI} {GAMMA}-{RAY} {SPACE} {TELESCOPE}$\less$/i$\greater${MISSION}},
  
	journal = {The Astrophysical Journal}
}

@article{KLAGES199792,
	author = "Klages, H. O. and others",
	journal = {Nuclear Physics B - Proceedings Supplements},
	number = {3},
	pages = {92-102},
	title = {The Kascade experiment},
	volume = {52},
	year = {1997}}

@article{GIBBS198867,
	author = {Kenneth G Gibbs},
	journal = {Nuclear Instruments and Methods in Physics Research Section A: Accelerators, Spectrometers, Detectors and Associated Equipment},
	number = {1},
	pages = {67-73},
	title = {The Chicago Air Shower Array (CASA)},
	volume = {264},
	year = {1988}}

@ARTICLE{EAS-MSU,
       author = "Vernov, S.~N. and others",
        title = "{The new installation at Moscow State University for the study of extensive air showers with energies up to 10 to the 18th eV}",
      journal = {Akademiia Nauk SSSR Izvestiia Seriia Fizicheskaia},
     keywords = {Cosmic Ray Showers, Particle Energy, Radiation Measuring Instruments, Electrons, Energy Spectra, Geiger Counters, Hodoscopes, Magnetic Spectroscopy, Muons, Particle Density (Concentration), Primary Cosmic Rays, Scintillation Counters, Instrumentation and Photography},
         year = 1980,
        month = mar,
       volume = {44},
        pages = {537-543},
       adsurl = {https://ui.adsabs.harvard.edu/abs/1980IzSSR..44..537V},
      adsnote = {Provided by the SAO/NASA Astrophysics Data System}
}

@article{TASD,
	author = "T. Abu-Zayyad and others",
	journal = {Nuclear Instruments and Methods in Physics Research Section A: Accelerators, Spectrometers, Detectors and Associated Equipment},
	pages = {87-97},
	title = {The surface detector array of the Telescope Array experiment},
	volume = {689},
	year = {2012}}

@article{Navarro_1996,
	doi = {10.1086/177173},
  
	url = {https://doi.org/10.1086
  
	year = 1996,
	month = {may},
  
	publisher = {American Astronomical Society},
  
	volume = {462},
  
	pages = {563},
  
	author = {Julio F. Navarro and Carlos S. Frenk and Simon D. M. White},
  
	title = {The Structure of Cold Dark Matter Halos},
  
	journal = {The Astrophysical Journal}
}

@article{Apel_2017,
	doi = {10.3847/1538-4357/aa8bb7},
  
	url = {https://doi.org/10.3847
  
	year = 2017,
	month = {oct},
  
	publisher = {American Astronomical Society},
  
	volume = {848},
  
	number = {1},
  
	pages = {1},
  
	author = "W. D. Apel and others",
  
	title = {{KASCADE}-Grande Limits on the Isotropic Diffuse Gamma-Ray Flux between 100 {TeV} and 1 {EeV}},
  
	journal = {The Astrophysical Journal}
}

@article{PhysRevLett.79.1805,
  title = {Limits on the Isotropic Diffuse Flux of Ultrahigh Energy $\ensuremath{\gamma}$ Radiation},
  author = {Chantell, M. C. and Covault, C. E. and Cronin, J. W. and Fick, B. E. and Fortson, L. F. and Fowler, J. W. and Green, K. D. and Newport, B. J. and Ong, R. A. and Oser, S. and Catanese, M. A. and Glasmacher, M. A. K. and Matthews, J. and Nitz, D. F. and Sinclair, D. and van der Velde, J. C. and Kieda, D. B.},
  journal = {Phys. Rev. Lett.},
  volume = {79},
  issue = {10},
  pages = {1805--1808},
  numpages = {0},
  year = {1997},
  month = {Sep},
  publisher = {American Physical Society},
  doi = {10.1103/PhysRevLett.79.1805},
  url = {https://link.aps.org/doi/10.1103/PhysRevLett.79.1805}
}

@article{PhysRevD.95.123011,
  title = {Constraints on the flux of $\ensuremath{\sim}({10}^{16}\ensuremath{-}1{0}^{17.5})\text{ }\text{ }\mathrm{eV}$ cosmic photons from the EAS--MSU muon data},
  author = {Fomin, Yu. A. and Kalmykov, N. N. and Karpikov, I. S. and Kulikov, G. V. and Kuznetsov, M. Yu. and Rubtsov, G. I. and Sulakov, V. P. and Troitsky, S. V.},
  journal = {Phys. Rev. D},
  volume = {95},
  issue = {12},
  pages = {123011},
  numpages = {8},
  year = {2017},
  month = {Jun},
  publisher = {American Physical Society},
  doi = {10.1103/PhysRevD.95.123011},
  url = {https://link.aps.org/doi/10.1103/PhysRevD.95.123011}
}

@article{TelescopeArray:2018rbt,
    author = "Abbasi, R. U. and others",
    collaboration = "Telescope Array",
    title = "{Constraints on the diffuse photon flux with energies above $10^{18}$ eV using the surface detector of the Telescope Array experiment}",
    eprint = "1811.03920",
    archivePrefix = "arXiv",
    primaryClass = "astro-ph.HE",
    reportNumber = "INR-TH-2018-026",
    doi = "10.1016/j.astropartphys.2019.03.003",
    journal = "Astropart. Phys.",
    volume = "110",
    pages = "8--14",
    year = "2019"
}

@article{Abbasi_2019,
	doi = {10.1016/j.astropartphys.2019.03.003},
  
	url = {https://doi.org/10.1016\%2Fj.astropartphys.2019.03.003},
  
	year = 2019,
	month = {jul},
  
	publisher = {Elsevier {BV}
},
  
	volume = {110},
  
	pages = {8--14},
  
	author = "R.U. Abbasi and others",
  
	title = {Constraints on the diffuse photon flux with energies above 1018~{eV} using the surface detector of the Telescope Array experiment},
  
	journal = {Astroparticle Physics}
}

@article{Baumholzer_2020,
	doi = {10.1007/jhep09(2020)136},
  
	url = {https://doi.org/10.1007
  
	year = 2020,
	month = {sep},
  
	publisher = {Springer Science and Business Media {LLC}
},
  
	volume = {2020},
  
	number = {9},
  
	author = {Sven Baumholzer and Vedran Brdar and Pedro Schwaller and Alexander Segner},
  
	title = {Shining light on the scotogenic model: interplay of colliders and cosmology},
  
	journal = {Journal of High Energy Physics}
}
@article{Chianese:2021jke,
    author = "Chianese, Marco and Fiorillo, Damiano F. G. and Hajjar, Rasmi and Miele, Gennaro and Saviano, Ninetta",
    title = "{Constraints on heavy decaying dark matter with current gamma-ray measurements}",
    eprint = "2108.01678",
    archivePrefix = "arXiv",
    primaryClass = "hep-ph",
    doi = "10.1088/1475-7516/2021/11/035",
    journal = "JCAP",
    volume = "11",
    pages = "035",
    year = "2021"
}

@article{P-ONE:2020ljt,
    author = "Agostini, Matteo and others",
    collaboration = "P-ONE",
    title = "{The Pacific Ocean Neutrino Experiment}",
    eprint = "2005.09493",
    archivePrefix = "arXiv",
    primaryClass = "astro-ph.HE",
    doi = "10.1038/s41550-020-1182-4",
    journal = "Nature Astron.",
    volume = "4",
    number = "10",
    pages = "913--915",
    year = "2020"
}

@article{Chianese:2021htv,
    author = "Chianese, Marco and Fiorillo, Damiano F. G. and Hajjar, Rasmi and Miele, Gennaro and Morisi, Stefano and Saviano, Ninetta",
    title = "{Heavy decaying dark matter at future neutrino radio telescopes}",
    eprint = "2103.03254",
    archivePrefix = "arXiv",
    primaryClass = "hep-ph",
    doi = "10.1088/1475-7516/2021/05/074",
    journal = "JCAP",
    volume = "05",
    pages = "074",
    year = "2021"
}

@article{PhysRevD.101.115029,
  title = {Model of metastable EeV dark matter},
  author = {Dudas, Emilian and Heurtier, Lucien and Mambrini, Yann and Olive, Keith A. and Pierre, Mathias},
  journal = {Phys. Rev. D},
  volume = {101},
  issue = {11},
  pages = {115029},
  numpages = {12},
  year = {2020},
  month = {Jun},
  publisher = {American Physical Society},
  doi = {10.1103/PhysRevD.101.115029},
  url = {https://link.aps.org/doi/10.1103/PhysRevD.101.115029}
}

@article{PhysRevD.99.095014,
  title = {Dark matter interpretation of the ANITA anomalous events},
  author = {Heurtier, Lucien and Mambrini, Yann and Pierre, Mathias},
  journal = {Phys. Rev. D},
  volume = {99},
  issue = {9},
  pages = {095014},
  numpages = {16},
  year = {2019},
  month = {May},
  publisher = {American Physical Society},
  doi = {10.1103/PhysRevD.99.095014},
  url = {https://link.aps.org/doi/10.1103/PhysRevD.99.095014}
}

@article{PhysRevD.105.L041301,
  title = {Search for dark matter using sub-PeV $\ensuremath{\gamma}$-rays observed by Tibet ${\mathrm{AS}}_{\ensuremath{\gamma}}$},
  author = {Maity, Tarak Nath and Saha, Akash Kumar and Dubey, Abhishek and Laha, Ranjan},
  journal = {Phys. Rev. D},
  volume = {105},
  issue = {4},
  pages = {L041301},
  numpages = {7},
  year = {2022},
  month = {Feb},
  publisher = {American Physical Society},
  doi = {10.1103/PhysRevD.105.L041301},
  url = {https://link.aps.org/doi/10.1103/PhysRevD.105.L041301}
}

@ARTICLE{2011ApJ...743L..14A,
       author = {{Adri{\'a}n-Mart{\'\i}nez} and others},
       collaboration="ANTARES",
        title = "{First Search for Point Sources of High-energy Cosmic Neutrinos with the ANTARES Neutrino Telescope}",
      journal = {\apjl},
     keywords = {astroparticle physics, cosmic rays, neutrinos, Astrophysics - High Energy Astrophysical Phenomena},
         year = 2011,
        month = dec,
       volume = {743},
       number = {1},
          eid = {L14},
        pages = {L14},
          doi = {10.1088/2041-8205/743/1/L14},
archivePrefix = {arXiv},
       eprint = {1108.0292},
 primaryClass = {astro-ph.HE},
       adsurl = {https://ui.adsabs.harvard.edu/abs/2011ApJ...743L..14A},
      adsnote = {Provided by the SAO/NASA Astrophysics Data System}
}

@article{Chantell_1997,
	doi = {10.1103/physrevlett.79.1805},
  
	url = {https://doi.org/10.1103
  
	year = 1997,
	month = {sep},
  
	publisher = {American Physical Society ({APS})},
  
	volume = {79},
  
	number = {10},
  
	pages = {1805--1808},
  
	author = "M. C. Chantell and others",
  
	title = {Limits on the Isotropic Diffuse Flux of Ultrahigh Energy$\less$mml:math xmlns:mml="http://www.w3.org/1998/Math/{MathML}" display="inline"$\greater$$\less$mml:mi$\greater$$\upgamma$$\less$/mml:mi$\greater$$\less$/mml:math$\greater$Radiation},
  
	journal = {Physical Review Letters}
}

@article{Campo:2017nwh,
      author         = "Olivares-Del Campo, Andr{\'e}s and Boehm, Céline and
                        Palomares-Ruiz, Sergio and Pascoli, Silvia",
      title          = "{Dark matter-neutrino interactions through the lens of
                        their cosmological implications}",
      journal        = "Phys. Rev.",
      volume         = "D97",
      year           = "2018",
      number         = "7",
      pages          = "075039",
      doi            = "10.1103/PhysRevD.97.075039",
      eprint         = "1711.05283",
      archivePrefix  = "arXiv",
      primaryClass   = "hep-ph",
      reportNumber   = "IFIC-17-54, IPPP-17-84",
      SLACcitation   = "
}

@article{Fomin_2017,
	doi = {10.1103/physrevd.95.123011},
  
	url = {https://doi.org/10.1103
  
	year = 2017,
	month = {jun},
  
	publisher = {American Physical Society ({APS})},
  
	volume = {95},
  
	number = {12},
  
	author = "Y. A. Fomin and others",
  
	title = {Constraints on the flux of 
		$\less$mml:math xmlns:mml="http://www.w3.org/1998/Math/{MathML}" display="inline"$\greater$$\less$mml:mrow$\greater$$\less$mml:mo$\greater$$\sim$$\less$/mml:mo$\greater$$\less$mml:mo stretchy="false"$\greater$($\less$/mml:mo$\greater$$\less$mml:msup$\greater$$\less$mml:mrow$\greater$$\less$mml:mn$\greater$10$\less$/mml:mn$\greater$$\less$/mml:mrow$\greater$$\less$mml:mrow$\greater$$\less$mml:mn$\greater$16$\less$/mml:mn$\greater$$\less$/mml:mrow$\greater$$\less$/mml:msup$\greater$$\less$mml:mi$\greater$-$\less$/mml:mi$\greater$$\less$mml:mn$\greater$1$\less$/mml:mn$\greater$$\less$mml:msup$\greater$$\less$mml:mrow$\greater$$\less$mml:mn$\greater$0$\less$/mml:mn$\greater$$\less$/mml:mrow$\greater$$\less$mml:mrow$\greater$$\less$mml:mn$\greater$17.5$\less$/mml:mn$\greater$$\less$/mml:mrow$\greater$$\less$/mml:msup$\greater$$\less$mml:mo stretchy="false"$\greater$)$\less$/mml:mo$\greater$$\less$mml:mtext$\greater${\hspace{0.167em}}$\less$/mml:mtext$\greater$$\less$mml:mtext$\greater${\hspace{0.167em}}$\less$/mml:mtext$\greater$$\less$mml:mi$\greater${eV}$\less$/mml:mi$\greater$$\less$/mml:mrow$\greater$$\less$/mml:math$\greater$
		 cosmic photons from the {EAS}{\textendash}{MSU} muon data},
  
	journal = {Physical Review D}
}

%% file: appendix.tex










\onecolumngrid
\appendix


\section{Supplementary Figures and Tables}
\begin{figure}[h]
    \includegraphics[width =0.8\textwidth]{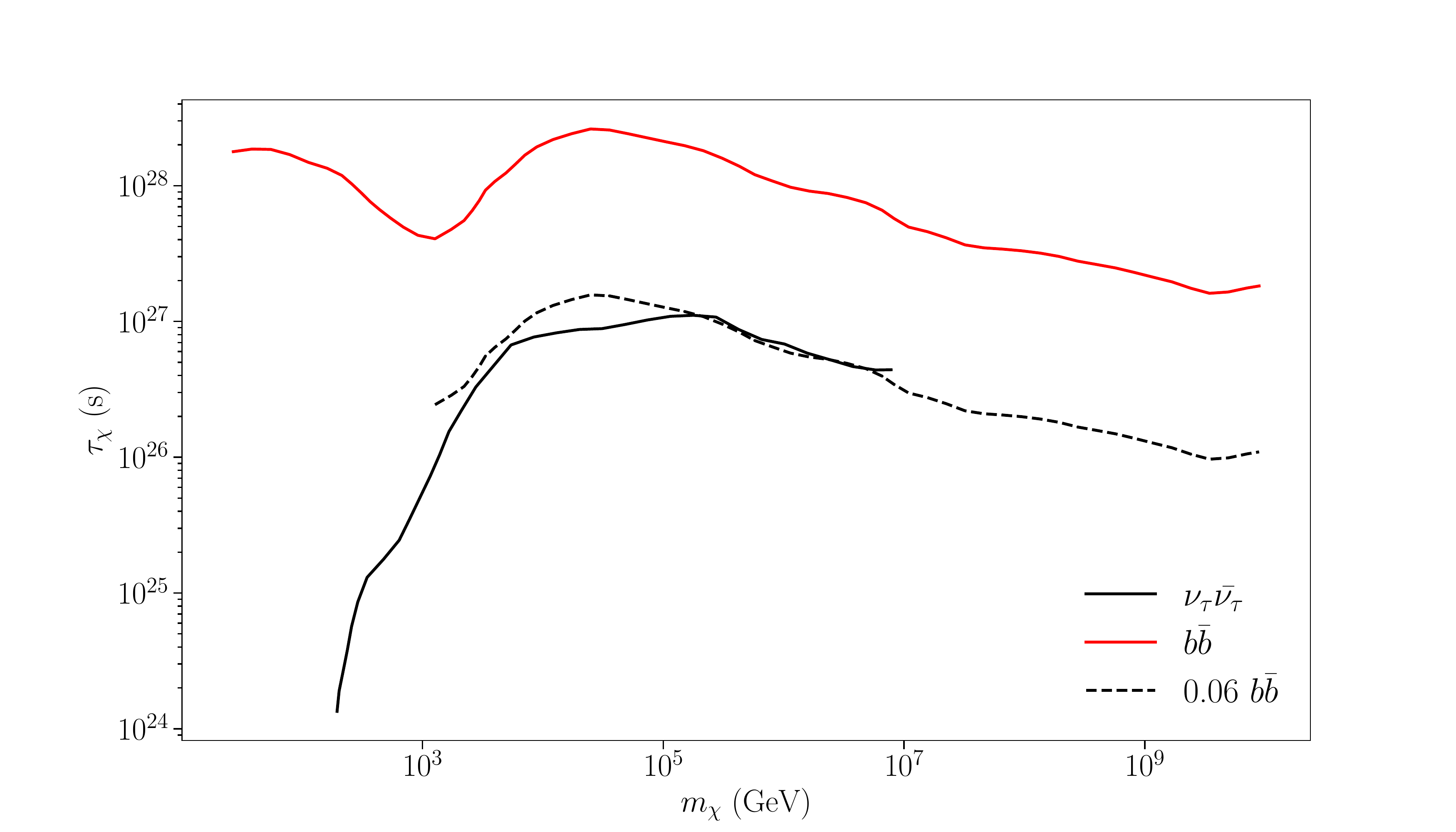}
    \caption{\textbf{\textit{Demonstration of the scaling from dark matter decay to $b \bar b$ constraints to neutrinos.}}
    Solid lines show lifetime limits on dark matter decay to $b\bar{b}$ (red) and $\nu_\tau\bar{\nu}_\tau$ (black) as determined by {\it Cohen et al.} \cite{Cohen:2016uyg} using gamma-ray observation from {\em Fermi}-LAT. Dashed line shows an approximation for extending the $\chi \to \nu_\tau\bar{\nu_\tau}$ limits by rescaling the decay to quark limits, by a factor of 0.06, the ratio between total photon production fractions in each channel.}
    \label{fig:FermiLimits}
\end{figure}

\begin{sidewaysfigure}[ht!]
    \centering
    \includegraphics[width=\textwidth]{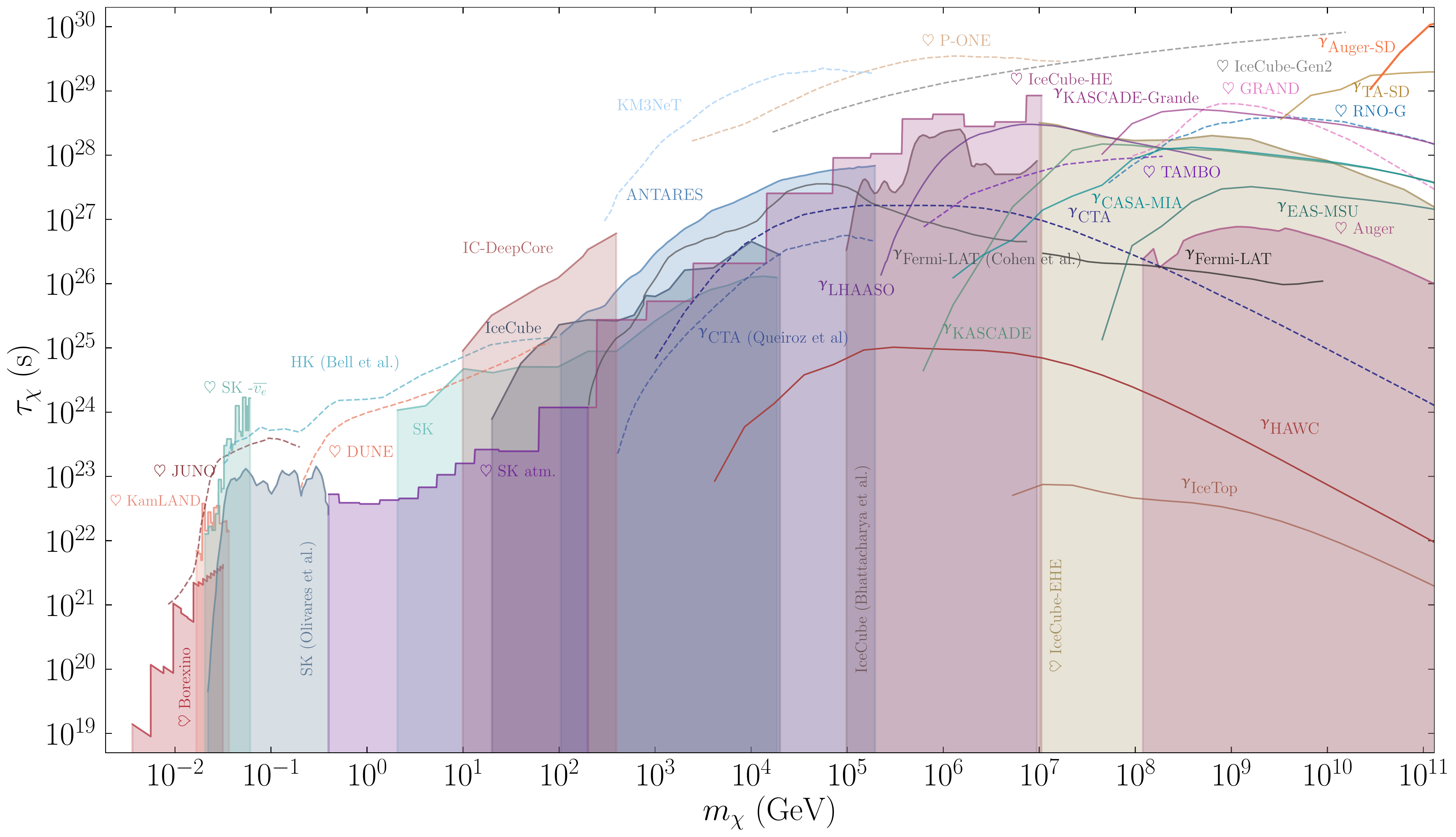}
    \caption{\textbf{\textit{Larger version of Fig.~\ref{fig:DecayLimits}.}}
    Solid lines bordering shaded regions represent limits from existing neutrino telescope data, solid lines without shading correspond to limits from existing gamma-ray observatories (as shown in Fig.~\ref{fig:GammaDecayLimits}), and dashed lines show the reach of future experiments.
    }
    \label{fig:DecayLimitsSideways}
\end{sidewaysfigure}

Here, we include the following tables and figures:
\begin{itemize}
    \item Tab. \ref{tab:lifetime_method} shows the full list of references to data sets or prior analyses used to produce the results of Fig. \ref{fig:DecayLimits}.
    \item Tab. \ref{tab:Jtable} shows the $D$-factors relevant for each experiment for which we recast diffuse limit fluxes into limits on dark matter decay.
    \item Fig. \ref{fig:FermiLimits} illustrates the rescaling of \textit{Fermi}-LAT limits on decays to $\bar b b$ pairs to limits on $\bar \nu \nu$.
    \item Fig. \ref{fig:DecayLimitsSideways} is a larger version of the main results Fig. \ref{fig:DecayLimits}.
\end{itemize}

\vspace*{15mm}

\begin{table}[!ht]
    \begin{center}
    \makebox{

    \begin{tabularx}{0.7\linewidth}{ @{\hskip 0.2in}>{\arraybackslash}c @{\hskip 0.2in} | >{\centering\arraybackslash}X}
    
    \hline \hline 
         \textbf{Analyses} &  \textbf{Data Used to Derive Lifetime Limit}  \\ \hline 
        ANTARES & Annihilation Cross Section Limit \cite{Albert:2016emp}
        \\ \hline
         Auger & Diffuse Neutrino Flux \cite{Zas:2017xdj} \\ 
        \hline
         Borexino & Upper Limit of Neutrino Flux \cite{Agostini:2019yuq} 
         \\ \hline

        DUNE  & Annihilation Cross Section Sensitivity \cite{Arguelles:2019ouk} \\ 
        \hline
        GRAND-200k   & Diffuse Neutrino Flux Sensitivity \cite{Alvarez-Muniz:2018bhp}
        \\ \hline
        
        Hyper-Kamiokande & Annihilation Cross Section Sensitivity \cite{Bell:2020rkw} \\ 
        \hline
        
        IceCube & Annihilation Cross Section Limits \cite{Aartsen:2016pfc,Aartsen:2017ulx}
        \\ \hline
        IC-DeepCore & Annihilation Cross Section Limit \cite{iovine2021indirect}
        \\ \hline
        IC (atm.) & Diffuse Neutrino Flux \cite{Aartsen:2015xup}
        \\ \hline
        IceCube-HE & Diffuse Neutrino Flux \cite{Aartsen_2016,IceCube:2020wum}
        \\ \hline
        IceCube-EHE & Diffuse Neutrino Flux \cite{Aartsen:2018vtx}
        \\ \hline
        IceCube (Bhattacharya) & Lifetime Limit \cite{Bhattacharya:2019ucd}
        \\ \hline
        
        IceCube-Gen2  & Diffuse Neutrino Flux \cite{Aartsen:2019swn} \\ 
        \hline
        
         JUNO & Lifetime Sensitivity \cite{Akita:2022lit}
         \\ \hline
        
        KamLand & Upper Limit of Neutrino Flux  \cite{KamLAND:2021gvi} \cite{Agostini:2019yuq}
         \\ \hline
         
        KM3NET  & Annihilation Cross Section Sensitivity \cite{gozzini2019search} \\
        \hline
        P-ONE  & Projected Effective Areas \cite{Agostini:2020aar}  \\
        \hline
        
        RNO-G   & Diffuse Neutrino Flux Sensitivity \cite{Aguilar_2021}. \\ 
        \hline
         
         SK-$\bar{\nu_e}$ & Upper Limit of Neutrino Flux \cite{WanLinyan:2018}
         \\ \hline
        SK (Olivares) & Annihilation Cross Section \cite{Campo:2017nwh}
        \\ \hline
        SK & Annihilation Cross Section \cite{Frankiewicz:2017trk}
        \\ \hline
        SK (atm.) & Diffuse Neutrino Flux \cite{Richard:2015aua}
        \\ \hline

        TAMBO   & Projected Effective Areas \cite{Romero-Wolf:2020pzh} \\
        \hline \hline
        
        IceTop & Upper Limit of Gamma-Ray Flux \cite{IceCube:2019scr}
         \\ \hline
         CTA (Queiroz et al.)& Annihilation Cross Section Projection \cite{Queiroz:2016zwd} \\ 
        \hline
        CTA & Projected $\gamma$ Sensitivity \cite{cta_sens} \\\hline
        HAWC & Upper Limit of Gamma-Ray Flux  \cite{HAWC:2017udy}
         \\ \hline
        KASCADE & Upper Limit of Gamma-Ray Flux \cite{Apel_2017}
         \\ \hline
        KASCADE-Grande & Upper Limit of Gamma-Ray Flux \cite{Apel_2017}
         \\ \hline
        CASA-MIA & Upper Limit of Gamma-Ray Flux \cite{Chantell_1997}
         \\ \hline
        EAS-MSU & Upper Limit of Gamma-Ray Flux \cite{Fomin_2017}
         \\ \hline    
        TA-SD & Upper Limit of Gamma-Ray Flux \cite{TelescopeArray:2018rbt} \\ \hline
        Auger-SD & Upper Limit of Gamma-Ray Flux \cite{PierreAuger:2022aty}
        \\ \hline \hline
    \end{tabularx}
    }
    \caption{\textbf{\textit{List of data sets used in this work and their references.}}
    Left column names the experiment, while the right one provides a short description of the data set.}
    \label{tab:lifetime_method}
    \end{center}
\end{table}

\newpage
\clearpage


\begin{table*}[!ht]
    \begin{center}
    \makebox[\linewidth]{
    \begin{tabular}{c | c | c }
    \hline \hline 
         Experiment& Exposure & $D/{10^{23}}$  \\ \hline
         All-sky & All-sky & $2.65$  \\\hline
         GRAND  & Elevation-dependent exposure, Fig. 24 of \cite{Alvarez-Muniz:2018bhp} & $0.298$  \\ \hline
         ANITA & dec = $[1.5^{\circ} , 4^{\circ}]$ & $0.052$ \\\hline 
         TAMBO & Elevation-dependent exposure, Fig. 3 \& 4 of \citep{Romero-Wolf:2020pzh} & 0.001 \\ \hline
         Auger & \begin{tabular}{c}${\rm zenith} = [90^{\circ}, 95^{\circ}]$\\${\rm zenith} = [75^{\circ},90^{\circ}]$ \\ ${\rm zenith} = [60^{\circ}, 75^{\circ}]$\end{tabular} & \begin{tabular}{c}0.11\\0.35\\0.33\end{tabular} \\ \hline
         P-ONE  & \begin{tabular}{c}$\cos({\rm zenith}) = [-1, -0.5]$\\  \hspace{6.5mm} $\cos({\rm zenith}) = [-0.5, 0.5]$ \cite{P-ONE:2020ljt} \\ $\cos({\rm zenith}) = [0.5, 1]$\end{tabular} & \begin{tabular}{c}0.83\\1.35\\0.47\end{tabular} \\ \hline 
         ANTARES  & \begin{tabular}{c}${\rm zenith} = [90^{\circ}, 120^{\circ}]$\\ \hspace{6.5mm} ${\rm zenith} = [120^{\circ},150^{\circ}]$ \cite{2005foap.conf..573R} \\ ${\rm zenith} = [150^{\circ}, 180^{\circ}]$\end{tabular} & \begin{tabular}{c}0.51\\0.46\\0.29\end{tabular} \\ \hline
        \hline
         IceTop & All-sky  & $2.63$ \\ \hline
         CTA & Galactic Centre \cite{Queiroz:2016zwd} & $0.003$ \\ \hline
         HAWC & dec = [$-25^\circ$ - $5^\circ $] & $0.0295$ \\ \hline
         KASCADE & dec = [$14^\circ$ - $84^\circ $] & $0.76$ \\ \hline
         CASA-MIA & dec = [$-20^\circ$ - $90^\circ $] & $1.58$ \\ \hline
         EAS-MSU & dec = [$7^\circ$ - $78^\circ $] & $0.92$ \\ \hline
         TA-SD & zen = [$-90^\circ$ - $45^\circ $] & $1.85$ \\ \hline

        \hline
        
    \end{tabular}
    }
    \end{center}
    
    \caption{
    \textbf{\textit{Exposures and D-factors for experiments discussed in this work.}} Exposures are converted to RA-dec and averaged over 24 hour period to obtain the corresponding $D$-factors, given in units of GeV cm$^{-2}$ sr, and computed according to Eq.~\eqref{eq:Dfactordef}.}
    \label{tab:Jtable}
\end{table*}

\newpage
\clearpage
